\documentclass[a4paper]{article}
\usepackage[utf8]{inputenc}
\usepackage[a4paper,
total={7in,9.5in}]{geometry} 
\usepackage{graphicx}
\usepackage{amssymb}
\usepackage{epstopdf}
\usepackage[font=small,labelfont=bf,format=plain]{caption}
\usepackage{soul}
\usepackage{xcolor}

\newcommand{\str}[1]{{\color{teal}\st{#1}}}
\newcommand{\rev}[1]{{\color{black}#1}}

\newcommand{\eps}{\varepsilon}
\newcommand{\R}{\mathbb{R}}
\newcommand{\Var}{\textrm{Var}}
\newcommand{\calM}{\mathcal{M}}

\begin{document}

\title{Thermal Macroeconomics: \\
An axiomatic theory of aggregate economic phenomena}
\author{{Nick Chater}$^a$, R.S.MacKay$^b$\footnote{Order of authorship is alphabetical.}
\\$^a$Warwick\ Business\ School,\ $^b$Mathematics\ Institute, \\ University of Warwick, Coventry CV4 7AL, UK \\
Nick.Chater@wbs.ac.uk, R.S.MacKay@warwick.ac.uk}
\date{\today}                       

\maketitle

\abstract

An axiomatic approach to macroeconomics based on the mathematical structure of thermodynamics is presented. 
We call the resulting theory ``Thermal Macroeconomics'' (TM). Subject to the axioms,
this approach deduces
relations between aggregate properties of an economy, concerning quantities and flows of goods and money, prices, and the value of money, 
without any recourse to `microeconomic' foundations about the preferences and actions of individual economic agents. 

The approach has three important payoffs. Firstly, \rev{it provides a new foundation for some aspects of standard macroeconomic theory, at least in the context of extensive exchange economies (explained below), without relying on implausibly strong rationality assumptions over individual microeconomic agents.  These aspects include} 
the existence of market prices, the value of money, the meaning of inflation, the symmetry and negative-definiteness of a macro-version of the Slutsky matrix, 
and the Le Chatelier-Samuelson principle, without relying on implausibly strong rationality assumptions over individual microeconomic agents. Secondly, the approach generates new results, including implications for money flow and trade when two or more economies are put in contact, in terms of new concepts such as economic entropy, economic temperature, goods' ``values'', and money capacity.  We will see that some of these are related to standard economic concepts (e.g.~marginal utility of money, market prices). \rev{But here they are derived} at a purely macroeconomic level.
Other concepts, such as economic entropy and temperature, have no direct counterparts in standard economics, \rev{though they have important economic interpretations} 
as ``aggregate utility'' and the inverse marginal aggregate utility of money, respectively.
Thirdly, this analysis \rev{provides a natural framework for exploring more complex economic phenomena using thermodynamic ideas, as well as for measuring economic counterparts of thermodynamic properties in real economies, for example, through an economic analogue of calorimetry. Moreover, the present approach may open up}
new frontiers in macroeconomics by building a bridge to ideas from non-equilibrium thermodynamics.  In contrast to the strict economic notion of equilibrium, the present approach starts from a statistical physics notion of equilibrium and thereby has the potential to extend to more realistic non-equilibrium settings.
More broadly, we hope that the economic analogue of entropy (governing the possible transitions between states of economic systems) may prove to be as fruitful for the social sciences as entropy has been in the natural sciences. 

We restrict attention to ``extensive'' economic systems:~those which can be subdivided into parts with the same properties apart from size (for example, we ignore possible ``lumpiness'' that might be caused by large individual companies).  Moreover, following a long tradition in economics, we \rev{focus on} 
exchange economies, in which the total quantities of goods and money remain fixed, but can be redistributed between agents, for example through trade. We
leave production, consumption, labour, \rev{multiple currencies}, finance etc.~for future treatment. 
Building on an axiomatic approach to thermodynamics by Lieb and Yngvason \cite{LY}, we explain how extensivity plus plausible economic assumptions lead to the definition of an economic analogue of entropy, which \rev{can be interpreted as} 
an aggregate utility, and an economic temperature, whose inverse is the marginal (aggregate) utility of money. We obtain a ``second law'' of economics, governing the allowed direction of changes of economic state of a multi-part economy. This approach captures the `value' of nominal money in a way that does not rely on the usual rather arbitrarily chosen ``baskets of goods'' and their prices.  It explains the direction of financial flow in terms of temperature differences.
We illustrate the theory with an economic analogue of the Carnot cycle whereby a trader makes money out of temperature differences. 

We obtain quantitative conditions for trade:~on making contact, nett trade flows occur if and only if they increase the total entropy. 
We \rev{emphasise that it is not necessary for the entropy of all economies to increase, but it does contain} 
the subset of ``mutually beneficial'' trades.
We derive some macroeconomic relations between cross-derivatives, in particular economic ``flexibilities'' (the inverse of the elasticities of standard economics).  We also introduce the concept of ``pure money''. We discuss limitations and possible extensions of our approach, in particular to relaxing the extensivity assumption and incorporating important aspects of real economies, including production, consumption and manufacturing.

\rev{\section*{List of notation}}
\rev{
\begin{itemize}
\item[TM] Thermal Macroeconomics (abstract and sec.~\ref{sec:precursors})
\item[CD] Cobb-Douglas (sec.~\ref{sec:toy})
\item[MB] mutually beneficial (sec.~\ref{sec:trade})
\item[$\preceq$] accessibility relation (sec.~\ref{sec:accessibility})
\item[$\sim$] reversible accessibility (sec.~\ref{sec:accessibility})
\item[$\prec$] irreversible accessibility (sec.~\ref{sec:accessibility})
\item[$\equiv$] financial equilibrium (sec.~\ref{sec:fineqm})
\item[$\theta$] state of financial join (sec.~\ref{sec:fineqm})
\item[$N$] number of agents (sec.~\ref{sec:exchange})
\item[$M$] amount of money (sec.~\ref{sec:eqm})
\item[$Au$, $\odot$] unit of money (sec.~\ref{sec:exchange})
\item[$G$] amount of a good (sec.~\ref{sec:eqm})
\item[$\Gamma$] state space (sec.~\ref{sec:eqm})
\item[$A_X$] forward sector of $X$ (sec.~\ref{sec:convexity})
\item[$S$] economic entropy (sec.~\ref{sec:entropy})
\item[$T$] financial temperature (sec.~\ref{sec:temperature})
\item[$\beta$] financial coolness (sec.~\ref{sec:temperature})
\item[$C$] money capacity (sec.~\ref{sec:moneycap})
\item[$I$] inflation rate (sec.~\ref{sec:moneycap})
\item[$\nu$] marginal value of a good (sec.~\ref{sec:price})
\item[$\mu$] market price of a good (sec.~\ref{sec:price})
\item[$\tau$] tariff (sec.~\ref{sec:tariffs})
\item[$\calM$] flexibilities (sec.~\ref{sec:symm})
\item[$\eth$] compensated derivative (sec.~\ref{sec:symm})
\end{itemize}
}

\section{Introduction}

Economies \rev{show} 
many large-scale regularities, concerning the balance between supply and demand to clear markets, patterns of interdependence between prices and volumes of different types of goods, the value of money, the movement of exchange rates, and many more. Yet economies also have many irregular properties---including crashes and booms, gluts and shortages, herding, fashion, and unpredictability due not just to technological or organizational innovation, but apparently due to endogenous, and little understood, fluctuations at many scales. 

\rev{For much of the twentieth century, macroeconomics was largely independent of assumptions about microfoundations:~that is, the relationships between macroscopic variables were analysed (and often estimated from historical data), without consideration of the decisions of individual consumers and businesses (or indeed the legal, regulatory and institutional frameworks under which they operate). A seminal paper by Lucas \cite{Lu} argued that historical macroscopic relationships were not reliable indicators of the impacts of future policy interventions, because such relationships would change when individuals, households and firms substantially change their behaviour in response to policy changes. In essence, the critique highlights the crucial difference between correlation (in past economic data) and causation (required to infer the impact of interventions on the economy through policy change). This concern led to the development of Dynamic Stochastic General Equilibrium (DSGE) models \cite{KP, LP} to provide a causal underpinning of macroeconomic phenomena. In these models, macroscopic behaviour arises from the interaction of households and firms, where households seek to maximize utility and firms seek to maximise profits. In early models, to make the analysis tractable, the variety of households and firms was not initially considered:~each type of economic actor was modeled by a single so-called ``representative agent,'' which was assumed to operate with high levels of knowledge and complete rationality, in complete and perfectly efficient markets. DSGE models have been the workhorse of central banks and economic policy for nearly half a century. Moreover, macroeconomists have gradually extended these models to include incomplete or inefficient markets (e.g., with economic ``frictions'' such as sticky prices and wages \cite{CEE}), heterogeneous agents with differing levels of wealth and liquidity \cite{KMV}, and rationality that is bounded in various ways \cite{Gab20}.}

\rev{Using this type of model,} economists have tried to explain the mix of regularity and irregularity in aggregate properties of the economy through appeal to micro-foundations,~i.e., explaining aggregate properties as arising from the interaction of economic agents with specified properties. Roughly speaking, the interactions of fully rational and informed consumers and producers, each aiming to maximize their own utility, are presumed to generate regularities at the macroscopic level. 
Thus, the pattern of explanation has been bottom-up:~from the properties of economic agents and their interactions to the large-scale economic phenomena.

There is, however, an interesting complementary approach \rev{to responding to the Lucas critique:~building a causal model of the economy that operates directly at the macro level, without reference to micro-foundations, just requiring specification or measurement of a few macroeconomic functions (our approach depends on only one, which we call the entropy).}  
This approach has the potential advantage that it may be possible to abstract away from detailed ``micro’’ assumptions (e.g., about the psychological properties of consumers, or the behaviour of companies), which are likely to be very difficult to specify and highly variable between contexts. Interestingly, this strategy predates the modern emphasis on the micro-foundations of economics, and is implicit in arbitrage arguments from David Hume and Adam Smith onwards (see discussion in \cite{Su}), and which are the basis for much of financial economics \cite{V}. 

In this paper, we explore this non-micro-foundational approach by applying conceptual and mathematical methods borrowed from 
``classical’’ thermodynamics, developed in the nineteenth century by Carnot, Joule, Clausius, Maxwell and Gibbs (``classical'' means without involving statistical mechanics \cite{LY99}). Classical thermodynamics begins with high-level principles, such as the conservation of energy (the first law of thermodynamics) and the non-decrease of entropy (the second law), and derives systems of relationships between large-scale observable phenomena including work, pressure, volume, temperature, and so on. It requires the existence of a priori
abstract quantities, such as energy and entropy, which provide a concise and elegant framework for expressing these laws. These fundamental laws turn out to be extremely general and to apply to systems with very diverse components, from the earth's climate system to chemical engineering and the functioning of biological cells. Thus, while in specific contexts there are elegant connections to microfoundations (e.g., to statistical properties of interacting gas particles), the principles of classical thermodynamics are more general and more abstract (e.g., \cite{LY,G}) and apply to diverse physical systems, chemical reactions, and biological processes. \rev{With regard to the Lucas critique, notice that classical thermodynamics provides a highly accurate \emph{causal} model (if one knows or has measured the entropy function):~i.e., rather than relying on correlations from prior measurements of temperatures, pressures and so on, classical thermodynamics provides rich and highly accurate predictions concerning the effect of interventions on a physical system---e.g., how temperature and pressure will change if gas in an insulated cylinder is compressed by some factor; the temperature two bodies will reach if they are thermally connected, and so on.} 

Can a similar approach work for economics? Specifically, is there a thermodynamics of economic systems that can derive macro-economic regularities without making detailed assumptions about consumers and companies? We formulate such a theory here, by reinterpreting thermodynamic quantities in economic terms. 

In particular, we map energy to money and deduce economic analogues of entropy and temperature, which in contrast to money are not standard economic quantities. 
Our economic entropy can be viewed as an aggregate utility, but importantly it is a cardinal rather than ordinal utility and applies to an economy as a whole rather than to individual agents; our economic temperature can be viewed as a macro-version of the inverse marginal utility of money.
This provides an interesting new perspective on some well-known economic regularities, captured by equilibrium arguments in conventional mathematical economics. Most importantly, we establish an analogue of the second law of thermodynamics, for economic systems, which specifies possible transitions between states of an economy, ruling out those with non-decreasing entropy. Moreover, this approach provides a principled meaning for the value of nominal currency (e.g., dollars, euros), which does not depend on the prices of somewhat arbitrarily chosen ``baskets of goods.'' It also opens up the possibility of using generalizations of thermodynamics, developed in the natural sciences, to understand more complex economic phenomena (including the analysis of economic systems which are out of equilibrium) without recourse to microfoundational assumptions about lack of information or bounded rationality at the level of individual economic agents. If such a high-level thermodynamic approach to the economy proves fruitful, it has the potential to complement and inform traditional microeconomic analysis, just as in physics and chemistry, classical thermodynamics
has complemented and informed statistical mechanics.

\rev{The paper has two main parts.  The first part (Sections~\ref{sec:precursors}-\ref{sec:convexity}) sets the framework.  The second part (Sections~\ref{sec:entropy}-\ref{sec:cross}) deduces consequences. The paper concludes with three sections on proposed extensions and discussion.}
We begin in Section~\ref{sec:precursors} by reviewing precursors of using thermodynamics in economics.  We restrict attention in this paper to exchange economies, reviewed briefly in Section~\ref{sec:exchange}.  In Sections~\ref{sec:eqm}-\ref{sec:convexity} we state our main assumptions and their justifications.  In Section~\ref{sec:entropy} we deduce the existence of an entropy function for economies and the all-important second law of non-decrease of entropy.  We illustrate the theory with a \rev{concrete, and micro-founded, toy economy in} Section~\ref{sec:toy}. \rev{We check that this toy economy fits with the axioms} in Appendix~\ref{app:axioms}, and \rev{the entropy of this economy} is computed at the micro-level in Section~\ref{sec:toy} and Appendix~\ref{app:CDentropy}. \rev{It is reassuring that the study of the toy economy allows us to confirm that the account given by thermal macroeconomics aligns with what is predicted from micro-foundations where the micro-foundations are sufficiently simple that they can be analysed directly.}
We deduce existence of a temperature function for economies and its relation to financial equilibrium in Section~\ref{sec:temperature}.  A subsidiary concept of money capacity is introduced in Section~\ref{sec:moneycap} and its significance described.  In Section~\ref{sec:price} we deduce the existence of a market price for each good, via a concept of ``value'' of a good, and \rev{obtain} a key relation for reversible trade.
In Section~\ref{sec:puremoney} we introduce the concept of ``pure money'' and discuss some of its consequences.
Section~\ref{sec:thermometer} proposes a way to measure the temperature of an economy \rev{and other quantities, in particular setting out how one could measure its entropy}.  Section~\ref{sec:Price differences} discusses the natural way to make money from price differences between two economies.  Section~\ref{sec:Carnot} describes the less intuitive way to make money out of the temperature ratio between two economies.  Section~\ref{sec:trade} gives quantitative conditions for trade,
highlights the significance of mutually beneficial trades (which increase the entropy of all parties), and explains how to extract the maximum amount of money from a pair of economies.
Section~\ref{sec:cross} derives relations between various partial derivatives, based on the existence and concavity of entropy.
Section~\ref{sec:further} raises some further issues for exchange economies.
Section~\ref{sec:future} discusses potential directions for development of the theory beyond exchange economies.
The paper concludes with a summary and conclusions in Section~\ref{sec:conc}.

\section{Precursors}
\label{sec:precursors}

Many previous researchers have considered possible links between classical thermodynamics and economics. Indeed, the architect of modern mathematical economics, Paul Samuelson, stressed the importance of Gibbs' work as foundational to his own. But Samuelson (and precursors such as Slutsky) applied principles of maximisation at the level of individual agents, rather than aggregate system properties. Samuelson despaired of numerous ill-thought-out proposals to attempt to find an economic analogue of entropy and doubted that such an analogue existed.  

The present approach aims to vindicate those who have intuitively felt the existence of a deep connection between thermodynamics and economics, and to set Samuelson's concerns aside. We begin by reviewing some of the most notable precursors of the approach developed here.

The idea that thermodynamics is relevant to economics goes back more than a century.  Irving Fisher's 1891 PhD  (considered by many to be the first in Economics) was co-supervised (along with the sociologist Sumner) by Gibbs, an early champion of thermodynamics.  Utility theory, as developed by Fisher, Slutsky, Hotelling, Houthakker, Samuelson and others, shares many features with thermodynamics. For example, the theory contains economic analogues of some of the Maxwell relations between partial derivatives (known in economics as the Hotelling conditions \cite{S50}).

Connections between economics and thermodynamics are particularly prominent in Samuelson's work  (e.g.~see the review of Samuelson's legacy by Dixit \cite{Di}).  Specifically, Samuelson notes that \cite{S72} ``Pressure and volume, and for that matter absolute temperature and entropy, have to each other the same conjugate or dualistic relation that the wage rate has to labor or the land rent has to acres of land''  and ``My earlier formulation of the inequality [$\sum_i \Delta p_i \Delta v_i < 0$] owed much to Wilson's lectures on thermodynamics.''\footnote{We will derive an analogue of this inequality in Section~\ref{sec:cross}.}
Yet Samuelson was also scathing about attempts to make connections between the subjects \cite{S72}:~``There is really nothing more pathetic than to have an economist or a retired engineer try to force analogies between the concepts of physics and the concepts of economics''. In the light of Samuelson's own inspiration from thermodynamics, this statement is rather puzzling, but it has certainly deterred researchers from the economic and natural sciences from pursuing such potential analogies further.\footnote{\cite{SF} pick out a particularly clear instruction from Samuelson \cite{S60} to economists and physicists to desist from seeking what he saw as spurious parallels: ``The formal mathematical analogy between classical thermodynamics and
mathematic economic systems has now been explored. This does not warrant
the commonly met attempt to find more exact analogies of physical magnitudes –
such as entropy or energy – in the economic realm. Why should there be laws like
the first or second laws of thermodynamics holding in the economic realm? Why
should ‘utility’ be literally identified with entropy, energy, or anything else? Why
should a failure to make such a successful identification lead anyone to overlook
or deny the mathematical isomorphism that does exist between minimum systems
that arise in different disciplines?''  The present paper shows that there is indeed a powerful economic notion of entropy and a corresponding second law of thermal macroeconomics, which go beyond the link between thermodynamics and maximisation that was so productive in Samuelson's work.} In particular, while Samuelson found important links between thermodynamics and maximisation problems in economics (typically defined for an individual agent, but including an individual company), the present link is more fundamental. Rather than beginning with a presumed economic analogue of entropy, we show that there exists a cardinal quantity such that economies, or systems or economies, can only change in ways that do not decrease that quantity. We call this quantity economic entropy.  

An influential book published by Georgescu-Roegen in 1971 \cite{GR} 
appear\rev{ed} to make a link between entropy and economics.  But our reading of Georgescu-Roegen's claim that entropy governs economics, has, we believe, a rather different aim than the project Samuelson was critiquing (and that we are pursuing here). Georgescu-Roegen's aim was to explore how physical entropy might place constraints on economic activity, rather than attempting to create an analogue of entropy within economics itself. This direction has been pursued under various names including ecological economics, systems ecology (for one take, see \cite{Ku}), and thermoeconomics (on which more shortly). \rev{The volume introduced by \cite{P+} provides a recent selection of approaches to bringing thermodynamics into the social sciences.}

Subsequently, physicists like Jaynes \cite{J} and Farjoun \& Machover \cite{FM}, began again to address economics, but from the viewpoint of statistical mechanics, rather than classical thermodynamics. This developed into a vigorous field of research that was christened ``Econophysics'' by Gene Stanley in 1996.  Particularly relevant articles from this school in our opinion are \cite{Sas} and \cite{RP}; also the book \cite{C+} on econophysics related to the Marxist view of economics, which centres on labour as measure of value.  Some perspectives on relations between economics and thermodynamics have been given in \cite{RMH} and \cite{MM}.
Econophysics has applied a wide variety of ideas and methods developed for analysing complex physical systems to economic phenomena and especially to financial markets. Yet econophysics has not generally found favour with economists. 

Others have attempted to establish links between classical thermodynamics and economics, but like Samuelson, they typically focus on mapping thermodynamics onto the utility function of individual agents, rather than considering economies from a purely macro-perspective as here. For example, \cite{STa,STb} aim to build links founded on the observation that both maximising entropy and maximising utility are problems of constrained optimisation, and draw out a number of detailed formal parallels.\footnote{They also highlight early discussion hinting at analogies between the equations used in economics and thermodynamics including \cite{Can, Da, Lis}.} A particularly sophisticated bridge between the two fields is outlined by \cite{SF}, and our approach has various links with theirs that we will highlight at relevant points. A more recent development by Cooper and Russell \cite{CR01,CR11} uses ideas from Samuelson to help formulate physical thermodynamics, drawing on ideas from symplectic geometry:~thus, their primary focus has the opposite explanatory direction from the account developed here, and has a narrower scope. The present purely macroscopic approach, aiming to map the full axiomatic structure of thermodynamics into a macro-analysis of economics, seems to be new---although exploring potentially fruitful links with prior work is an interesting topic for future research.

The term we use for our approach, ``Thermal Macroeconomics'' (that we abbreviate to TM), aims to express our objective of providing an analysis of economic theory paralleling the mathematical structure of thermodynamics. We distinguish TM  from the variety of loosely connected usages of the term ``thermoeconomics'', which typically point to attempts that, like \cite{GR}, aim to \emph{combine} economic and thermodynamic principles in contexts as diverse as understanding plant growth, the emergence of order in biological evolution, designing commercial energy systems, or clarifying the limits to economic growth.\footnote{For example, in applying economic ideas to understanding living systems \cite{CK}, \cite{Co} states ``The term `thermoeconomics' denotes a paradigm shift in our understanding of the role of energy in living systems, and in evolution. It is based on the proposition that energy in biological evolution can best be defined and understood, not in terms of the Second Law of Thermodynamics but in terms of such economic criteria as productivity, efficiency, and especially the costs and benefits (or `profitability') of various mechanisms for capturing and utilizing energy to build biomass and do work.'' Others, apparently beginning with Myron Tribus in 1962 \cite{Murphy}, have used ``thermoeconomics'' for the exploration of how fundamental thermodynamic constraints in the physical world shape and constrain economic activity, whether at the level of the individual industrial processes \cite{TE} or the entire economy in the tradition of \cite{GR}. According to \cite{DG}, for example, ``Thermoeconomics combines thermodynamic principles with economic analysis and brings some fundamental changes in the economic evaluation, design, and maintenance of processes.'' In sum, no specific usage of the term has yet become well established. We use the new term ``thermal macroeconomics'' to distinguish the present approach from this work. A very recent book \cite{GC} (though building on \cite{Ch}) {uses the term ``entropy economics''}, but is unrelated to the approach developed here: they aim to use Shannon entropy (part of the statistical mechanics, rather than classical thermodynamics, notion of entropy) to create a scarcity theory of value.
{In contrast to these other approaches,} ``thermal economics'' applies the \emph{mathematical structure} of classical thermodynamics, not its physical content nor statistical mechanics, to model economic phenomena. Thus, it has no specific connection to economics questions concerning heat, energy or any other physical quantities.}

The objective of TM is both narrower in scope and has a very different focus than thermoeconomics:~aiming to map the mathematical structure of thermodynamics into economics, in order to provide a purely `macro-level' analysis of economic phenomena. Thus, we aim to construct an economic theory using the mathematical and conceptual machinery of thermodynamics.
Despite various past attempts, and the intuition among many researchers (but not, as we've seen, Samuelson) that such a link might exist, a purely thermodynamic approach to economics 
seems not yet to have reached fruition.\footnote{Our own efforts in building a link between thermodynamics and economics has had a long gestation. We started discussions on formulating a link in 2010, including supervising MSc student Daniel Sprague's simulations of simple economic systems in 2011. We arrived at our present position only after many iterations and false starts.}  

A key ingredient for our thinking was provided by the axiomatic formulation of thermodynamics by Lieb and Yngvason in a series of papers surveyed in \cite{LY}, which clarifies the underlying mathematical structure of thermodynamics, independent of its physical interpretation. Their approach formulates axioms governing abstract relations $\precsim$ and $\equiv$ between macro-states, which we shall introduce in Sections~\ref{sec:accessibility} and \ref{sec:fineqm} respectively. Thus, any analogies between the domains of physics and economics can emerge naturally from the mathematics; equally, where there are disanalogies, these will emerge automatically too.
The present work takes literally their invitation in \cite{LY98}:~``Maybe an ingenious reader will find an application of
this same logical structure to another field of science.''
Indeed, possible links to economics, and in particular to utility theory, are mentioned in \cite{Y}.

Reading \cite{LY} is not a prerequisite for our paper, however. We require only their axioms and consequent results on existence and properties of entropy (which we review here). While our exposition is self-contained, we 
adopt their numbering system for the axioms (though inserting our own slight variants), and largely follow their notation.\footnote{We make slight adjustments to their notation for `adiabatic accessibility' and `adiabatic equivalence'---and we also drop the term `adiabatic' entirely, as it is unfamiliar in economics and is ambiguous even in physics where it can mean without heat flow or alternatively, slow variation, which are not equivalent.} 
There are nuances in their treatment that we skip over; our presentation is stream-lined in an attempt to give the key ideas without overwhelming the reader. There are alternative axiomatic formulations of thermodynamics, in particular by Carath\'eodory (extended by Boyling \cite{Bo}) and by Giles \cite{Gi}, but that of \cite{LY} seemed to us the easiest to adapt.

The approach developed here is purely at the macro-economic level. Many of the topics treated below, such as the formation of market prices, are typically studied in microeconomics starting from fundamental assumptions about the rationality of individual economic agents. We treat these questions from an aggregate point of view, by making assumptions about markets as a whole, rather than their participants. This may create a more natural bridge from micro-economics to macroeconomic questions concerning, for example, the value of money and flows of trade. We see the ability to generate strong theoretical results without specific micro-foundations as a major advantage of this approach, regarding both generality and tractability. Nonetheless, the functional form of the resulting entropy function depends on the micro-foundations, and
it is of course interesting to consider how a macro-analysis can connect with micro-foundations. We show how this can be done in the special case of an exchange economy consisting of agents with a Cobb-Douglas utility function.

The economics community is divided on the value of axiomatic approaches.  On the one hand, the axiomatic approach of Arrow and Debreu \cite{AD} and McKenzie \cite{McK} to general equilibrium is regarded by many theoretical economists as a pinnacle of intellectual achievement.  On the other hand, practical economists tend to be sceptical of axiomatic arguments and of mathematical theories in general, as encapsulated by Krugman's comment on the 2008 financial crisis \cite{Kr}:~``The economics profession went astray because economists, as a group, mistook beauty, clad in impressive-looking mathematics, for truth.''  We take a pragmatic view, justifying the axioms from economic reality as far as we can but identifying their limitations and proposing directions for extension. In the present context, that of building a link between thermodynamics and economics, an axiomatic approach is particularly valuable because it provides a rigorous way of clarifying a common underlying mathematical structure, rather than relying on intuitive analogies between ideas in each domain.

Finally, the economics community is, as we have already indicated, divided on the value of ``imitating'' physics \cite{Mir, Yee}. We aim to set aside such general concerns, and focus on showing how a thermodynamic perspective on economic systems can yield valuable insights for economic analysis. We hope that, especially if the present approach can be extended beyond the exchange economies treated here, concepts of economic temperature, economic entropy, money capacity and so on, may become part of standard economic practice.  Indeed, we suspect that the problematic relationship between economics and physics arises not because economists have been excessively entranced by attempting to imitate physics, but because the connections between the disciplines have not been pushed far enough.

\section{Exchange economies}
\label{sec:exchange}

In this paper we restrict attention to exchange economies. In an exchange economy, agents exchange various sorts  of durable good with each other; no goods are produced or consumed, nor are they combined or disassembled to make other sorts of good. 
This is of course a very limited view of economics, ignoring production, consumption, manufacturing, recycling, labour, interest rates, etc, but we intend to extend the approach in later work. In any case, it is fairly standard in economic and finance theory to reduce much to exchange, for example, exchanging options to consume a good, provide labour, money, or supply a service, on some future date or conditional on a specific state of the world \cite{Ga, Va}.

We suppose that \rev{the number $N$ of agents is large but finite and fixed,} the set of types of good is finite, that goods of the same type are indistinguishable, and that goods can be divided in any ratio.  These are idealisations that we hope to relax in future work.

We suppose there is a particular durable good that we will use as a reference (a ``num\'eraire'' in economics) and will call {\em money}.  We assume that at the aggregate level, more money is always ``preferred'' to less, no matter how much money and other goods are present already. A more precise meaning will become clear in our treatment below---but roughly this means that if a trader (a figure we will introduce in Section~\ref{sec:accessibility}) were to offer freely to give or take money to or from members of an economy, they would in aggregate accept more money than they would give (we do not require the stronger assumption that this holds for every individual).\footnote{The corresponding aggregate property need not hold for other types of goods:~there may be regimes where it is preferable not to have more, or even preferable to have less (e.g., if the good acts as a pollutant or is simply ``surplus to requirements'').} 
Money should also be easy to store and exchange, though in our exchange economies all goods are durable and exchangeable; in particular, we want it to be easily transferable to another economy. To avoid giving a national preference, we measure its amount in ``aurum'', denoted by $Au$ or $\odot$.  The unit is named after gold, in the light of gold's historical role as a num\'eraire for a large part of the world.  Money may or may not have intrinsic value (contrast gold with paper); we explore the consequences of ``pure money'' in Section~\ref{sec:puremoney}. Some of the other types of good might also be considered as currencies.  
We touch briefly on settings in which there is more than one currency in Section~\ref{sec:currencies}, but leave extending the approach for future work.

Some comment is required here on the meaning of conservation of money. 
In the real world, money can be printed, destroyed, \rev{and} created against debt, e.g.~through the sale of bonds as in ``quantitative easing'' after the 2007/8 financial crisis. We regard that as consistent with conservation of money because money enters the system only in identified ways.  This is analogous to the conservation of energy in physics:~conservation of energy does not exclude arrival of sunlight into the earth system, the dissipation of the earth's heat into space or the generation of energy by nuclear fission.

As noted in the previous section, we take a deliberately macroscopic view. Hence we do not have to specify the process by which individual people choose which bundles of goods to exchange nor how frequently they do so.  Nonetheless, we will study one microeconomic example (Section~\ref{sec:toy}) to illustrate how the thermodynamic quantities of our macroeconomic theory emerge from a microeconomic model.

\rev{To aid the reader in understanding what we mean by an exchange economy, we give a rapid preview of this example here.  The economy has $N$ agents, an amount $M$ of money and an amount $G$ of a single type of good.  Each agent $i$ possesses a non-negative amount $m_i$ of money and $g_i$ of good.  Agents $i$ and $j$ make independent pairwise encounters at rate $k_{ij}$.  On each encounter they pool their belongings and redistribute them between themselves with a probability density proportional
to the product of their ``utilities'' for the outcome.  The goal of our analysis is to be able to say what happens if one puts such an economy in contact with a trader offering to buy or sell goods at a given price, or in contact with one or more other economies.}

The literature contains many other examples of microeconomic models of exchange, for which some references can be found in a recent paper \cite{DGMS}. Other recent references on exchange models include the special issue edited by \cite{TSB} and the review \cite{GG}.

An important aspect of our theory is that we allow economic systems with internal barriers.  For example, it might be impossible to exchange some types of good between two islands without external help.  Or the system might consist of several parts under different political regimes, with prohibitions on certain exchanges between the parts.  
We will call an exchange economy ``basic'' if there are no internal barriers, and ``multi-part'' otherwise.\footnote{One might wish to extend the notion of internal barriers to ones that restrict exchange of some linear combinations of different types of good or money, rather than just some types, but we leave that for future work.}

The most direct way to test the theory is by computer simulations of micro-interactions between simple economic agents. This allows direct comparison between the ``macro'' predictions of TM with the measured behaviour of a simulated exchange economy with clear micro-foundations. In \cite{LMC}, we describe a wide range of simulations which are in good quantitative agreement with the predictions of TM.

\section{Equilibrium}
\label{sec:eqm}

Our formulation depends on an assumption\footnote{In their analysis of thermodynamic entropy, Lieb and Yngvason didn't label the corresponding property as an axiom. Nonetheless, it is implicit in their work and we feel it merits a label (A0). \rev{It has been recognised as necessary and dubbed the ``minus first'' law of thermodynamics in \cite{BU}.}} that we call
\vskip 1ex
\rev{\noindent{\bf Axiom A0}:
Any closed exchange economy, with given amounts of each type of good or money in each connected component for that type, settles down to a unique statistical equilibrium state.
\vskip 1ex}
\noindent By a statistical equilibrium state we mean a stationary probability distribution over micro-states.\footnote{It is important to distinguish the notion of micro-state from the macroscopic notion of ``state'' as a statistical state determined by aggregate quantities of goods and money, prices and so on, which is our main concern in TM.} 

To elaborate on this, for a basic economy, i.e., an economy with no internal barriers, A0 says that a unique equilibrium is specified by the total quantities of each type of good and money. For a multi-part economy, i.e.~with some internal barriers to some types of good or money, specifying the unique equilibrium is slightly more complex:~we need to specify not only the total quantity of each good (including money), but also how that quantity is allocated to each of the distinct connected components for that good.
As a straightforward example, the economic system might consist of two unconnected economies $A$ and $B$.  Then the equilibrium states for the joint system are specified by an equilibrium state for $A$ and an equilibrium state for $B$.

Thermal \rev{macro}economics allows us to pose and answer interesting questions, such as what happens \rev{if} a trade barrier between previously separate economies or parts of an economy is removed.   

It is possible to specify micro-level dynamics where A0 fails, in particular by including strong herding effects, e.g.~\cite{GBB}. To apply our theory to such systems would require some additional labels for the state, to indicate which equilibrium it has chosen.\footnote{For example, in a 2D Ising model the possible states are specified by temperature except that below the critical temperature there is an interval of possible states, corresponding to convex combinations of two extremal states at the same temperature, so one has to specify the convex combination as well.}  We avoid such considerations here \rev{but hope that establishing thermodynamic principles for simple cases will, as in physics, be a useful foundation for studying more complex phenomena.}

If we suppose that money is one of the goods in {an economy} and that money can flow between all parts of {an economy}, then Axiom A0 makes an exchange economy ``simple'' in the language of \cite{LY}:~the state is specified by a point in a subset of $\R^n$, for some dimension $n$,
\rev{one of whose coordinates is the total amount $M$ of money. Specifically,}  the coordinates are the amounts \rev{$G_{jc}$} of each type \rev{$j$} of good in each connected component \rev{$c$} for that type of good, and the total amount \rev{$M$} of money. \rev{We denote the set of all possible states of the system by $\Gamma$, called the {\em state space} for the system.}

But even economies with more than one connected component for money can be considered ``simple'' if one of the connected components for money is chosen and its money content is used as 
\rev{a} distinguished money coordinate for the state.  The amounts of money in the other components are treated on a par with those for other types of good.  This will be a useful extension\footnote{So our use of ``simple'' is a slight generalisation of that of \cite{LY}.}, \rev{though a reader may ignore it on first reading by considering money to be able to flow arbitrarily within an economy}.
\rev{We highlight this as
\vskip1ex
\noindent{\bf Definition (simple system)}:~A {\em simple system} is an exchange economy in which there is money and the state can be specified by a point in a subset $\Gamma$ of $\R^n$ for some $n>0$, one of whose coordinates is the amount $M$ of money in a distinguished component for flow of money.
\vskip1ex
}
\noindent When we {introduce} financial contact between simple economies, we will always mean contact between their distinguished money components.  Similarly, the temperature of a simple economy, {introduced below, is relative} to the {the choice of} distinguished money component.
{Note that Axiom A0 makes} a simple system ``financially indecomposable'':~however the specified amount of money is initially distributed in the distinguished money component, it ends up distributed with the same probability distribution. 

One could also consider the number of agents as a coordinate for the state, but for present purposes we will take the number of agents in a system as fixed; {we leave the analysis of} migration for future work.  

We emphasise that statistical equilibrium is not an equilibrium in the sense of Nash, where no player can benefit from a unilateral change \cite{N50}.  Nor is it an equilibrium in the sense of Arrow and Debreu, where prices are such that supply balances demand and then all exchanges stop \cite{A51, D51}.  Instead, it is a dynamic state of continuing exchanges but with a steady probability distribution for the micro-state.  For example, it is entirely possible that people's preferences continually vary (e.g., they become bored with objects they currently have and would rather exchange them for different ones). Perhaps the closest analogue in economics is dynamic stochastic general equilibrium (DSGE) \cite{Ga}, the workhorse of treasury modelling, but there the dynamics arises only as a response to external shocks.  
The idea of equilibrium in a strict sense has been criticised from early days in economics.  For example, Irving Fisher wrote in 1933 \cite{F} that ``It is as absurd to assume that, for any long period of time, the variables in the economic organization, or any part of them, will `stay put,' in perfect equilibrium, as to assume that the Atlantic Ocean can ever be without a wave.''  
A statistical equilibrium with steady (or slowly varying) macro-parameters is more plausible.\footnote{A possible criticism of our equilibrium assumption, especially from finance practioners, is that it cannot account for endogenous business cycles. We note that there is considerable debate in the academic literature concerning the nature and even existence of such cycles \cite{Sl37,FG94, Man89}. But we also note that the TM framework can potentially generate such cycles through the system being held out of equilibrium by fluxes of inputs and outputs, just as a physical system may exhibit oscillations when held out of thermodynamic equilibrium (e.g., in Rayleigh-B\'enard convection or the Belousov-Zhabotinskii reaction \cite{NN}). This is an important direction for future work.
Here, however, we allow only temporary fluxes of inputs and outputs (e.g.~with an external trader).}

The notion of statistical equilibrium is itself, of course, a simplification of real economic activity, in which there are, among other things, continual technological and business innovations. These considerations do not arise in our idealized exchange economies. 
Nonetheless, we believe that the present approach could usefully be extended to deal with \rev{some} such cases, expanding on the intuition that the only innovations that catch on are those that increase economic entropy. For now, we note that entropy is useful in physics even in a time-varying environment and helps explain the formation of complex structure in physics, chemistry and biology. In particular, it is worth making a comparison with biochemistry.  Concepts from equilibrium thermodynamics like changes in Gibbs free energy are very useful in understanding which processes occur in a cell, even if it is a highly structured environment and subject to time-varying conditions. The assumption behind this analysis is local thermodynamic equilibrium.
{Indeed,} the inclusion of ``dynamics'' in the term ``thermodynamics'' reflects that it is relevant to time-dependent systems.

A different concern with the equilibrium assumption is that an economy might have multiple equilibria \cite{Fa}. \rev{For example, \cite{BF} exhibit an exchange economy over assets in which ``public opinion'' is, in a sense, self-fulfilling, that leads to non-ergodic behavior; this type of case is beyond our scope here.} 
\rev{Even worse, it might never settle to a statistical equilibrium at all \cite{D+} (the same is true of evolutionary systems and has been adapted to economics in \cite{RJ}).
These} might well be the case when there are fluxes of inputs and outputs; but without inputs and outputs, it seems to us a plausible working assumption that an economy with specified amounts of each type of good in each component for that type goes to a unique statistical equilibrium.  This type of indecomposability assumption is widely made in analysing physical systems, in particular in statistical mechanics. It is analogous to the property of ergodicity in stochastic dynamics, meaning that there is a unique stationary distribution and it attracts all initial distributions \cite{Li}.  Similarly, it is analogous to the property of mixing for measure-preserving dynamics $\phi_t$, meaning that for all measurable subsets $A$ and $B$, the probability of $A \cap \phi_t B$ converges to the product of the probabilities of $A$ and $B$ as $t\to +\infty$ \cite{Wa}.

To gain an intuition for the significance of the assumption that our exchange economy has a unique statistical equilibrium, it is useful to consider how this assumption can be violated. Suppose that {perhaps unbeknown to the theorist}, the {economy} under study is decomposable into two or more non-interacting parts (e.g., corresponding to islands which are entirely cut off from each other).  
Then there will be a continuum of ways in which goods and money 
\rev{could have been} divided between the parts (and this division will be conserved as the system evolves, because there is no trade between the parts). Thus each of the continuum of possible divisions will correspond to a distinct region of the state space with its own statistical equilibrium. We assume, unless specified as in this example, that this does not occur.\footnote{As already indicated, there is a parallel in classical statistical mechanics:~for example, one assumes that there is no extra conserved quantity, beyond those already identified as labels for the state, that would prevent wide exploration of the space corresponding to the state.  More generally, one assumes that there is no invariant surface separating the space into two parts, or even more generally, one assumes ergodicity of the conditional of Liouville measure on the specified values of conserved quantities. For a simple introductory reference, see \cite{Se}.}

Finally, following \cite{LY}, an unconnected product of simple economies will be called ``compound''. 
This is a particular case of a multi-part economy.
By putting parts of the simple systems of a compound system in contact for various types of good or money, one obtains a multi-part economy with perhaps more than one distinguished money coordinate, but we reserve the use of ``compound'' for unconnected products in order to be clear about using the corresponding results of \cite{LY}.

\section{Accessibility}
\label{sec:accessibility}

The next key notion is the \emph{accessibility} of one state of an economy from another when subject to external influence (for example, through interaction with other economies, or wealthy traders).  Indeed, the main aim of the present approach is to understand which transitions between macro-level states are, or are not, possible. 

\rev{To gain an intuition for the economic significance of the notion of accessibility, we can consider how a trader with unlimited wealth might seek to move the economy from one state to another.} One might \rev{initially} suspect that the trader could rearrange goods between agents in the economy in any way desired, by paying people to buy and sell to each other to order; but to do this, the people in the economy would thereby have to be given additional money by the trader to incentivize them to engage in the desired transactions. \rev{Note though, that} money is itself part of the state of the economy. So the trader does not have complete control after all---some state-transitions are possible and others are not. Indeed, suppose the trader wanted rapidly (rather than arbitrarily slowly) to reverse the allocations of goods to its original state. Doing so would require further financial inducements; so that while goods (apart from money) could be returned to the initial state, the economy would now contain more money than before. The reader may have the intuition that the trader forcibly ``moving'' goods about the economy seems analogous to doing work on a physical system (e.g., compressing a gas by a piston). Just as moving a physical system from one state to another and back again will ``heat'' the system being operated upon (except when the movement is arbitrarily slow), so forcibly moving an economic system from one state to another will increase the economic temperature of the system (by increasing the amount of money in the economy, and thus reducing the value of money). We will make these intuitions rigorous below. 

\rev{To strengthen our intuition further, consider two isolated economies, one with cheap and plentiful apples but few and expensive bananas; and the other with cheap and plentiful bananas but few and expensive apples. Merely by allowing the economies to trade freely, the trader can oversee a nett flow of apples in one direction and a nett flow of bananas in the other; and the prices of apples and bananas will be the same in each economy (assuming no transportation costs); and indeed, there will typically be ``gains of trade'' for both economies. But the reverse transformation, to return to the original state, is not possible. People will only give up the gains of having both apples and bananas if the trader pays them to do so---but then, as in the above case, we have not returned to the original state, because the amount of money in the two isolated economies will be greater than before. Indeed, in general, spontaneous trade between economies will irreversible, in the same way that mixing gases is irreversible.}

\rev{To make these intuitions precise, we} consider a system of one or more exchange economies connected to an external trader who has unlimited goods and money, and access to some other economic system if desired, and ask what effects on the systems the trader can achieve. 

\rev{\vskip1ex
\noindent{\bf Definition (Accessibility, $\preceq$)}: 
We say a state $Y$ of an economic system is {\em accessible} from state $X$, written $X \precsim Y$, if the trader can move the system from state $X$ to state $Y$ with arbitrarily small nett change in the trader's external system but allowing arbitrary changes in the trader's money and goods.
\vskip1ex}
\noindent Note that our notation is slightly different from that of \cite{LY}, who use $\prec$ instead.\footnote{In earlier versions of this paper, we allowed only change in money, but here we allow changes in both money and goods, a choice that fits better with the theory of \cite{LY}.  While \cite{LY} say their analogue of the trader has to return to the same state apart from its energy, in fact they allow the system to change volume too, so implicitly the trader, considered as the outside, also changes volume with the opposite sign.}   
\rev{Note also that \cite{LY} require the external system to return to exactly the same state apart from the ``weight'', but we consider it essential to allow arbitrarily small nett changes (in the physics context too).}

The accessibility relation extends to compare states of different systems if one system can be obtained from the other by action of the trader, e.g.~putting two parts in some form of contact or inserting a new barrier.

Note that directly 
from the definition, we can see that $\precsim$ is a pre-order (it is reflexive:~$X \precsim X$, because the trader does not have to do anything to achieve no change; and it is transitive:~$X\precsim Y \ \&\ Y \precsim Z$ implies $X\precsim Z$, because the trader just has to make one change followed by the other).  Transitivity of $\precsim$ is A2 of \cite{LY} but we get it for free here \rev{(we will not display on separate lines obvious axioms such as this one)}.

It is important to note that $\preceq$ is not assumed to be a {\emph{total} pre-order (called ``preference relation" in economics), i.e., a pre-order such that} for all $X,Y$ then $X\preceq Y$ or $Y \preceq X$ ({or} both).  For example, the states of two economic systems with different numbers of agents can not be compared, because we don't allow the trader to change the number of agents.  The strategy of \cite{LY} is to find conditions on pairs of systems under which $\preceq$ is total (one says then that every pair of states is ``comparable") and then to represent it by $\le$ for a real-valued function of state, which will be called ``entropy".

\rev{\vskip1ex
\noindent{\bf Definition (reversible accessibility, $\sim$)}: 
We say state $Y$ is {\em reversibly accessible} from $X$, and write $X \sim Y$, if $X\precsim Y \ \&\ Y \precsim X$.
\vskip1ex}
\noindent It is an equivalence relation, because it is symmetric:~$X\sim Y$ iff $Y\sim X$, reflexive (which is A1 of \cite{LY} but again we get it for free) and transitive.  

\rev{\vskip1ex
\noindent{\bf Definition (irreversible accessibility, $\prec$)}: State $Y$ is {\em irreversibly accessible} from state $X$, written $X \prec Y$, 
if $X\precsim Y$ \& $X \not\sim Y$.
\vskip1ex}
\noindent For example, we can translate our assumption of desirability of money at the aggregate level into the statement that for a simple system and any $M>0$, $X \prec X+M$, where $X+M$ denotes the state where the distinguished money coordinate is increased by $M$. This justifies 
\rev{\vskip1ex
\noindent{\bf Axiom A8}: For all states $X$ of a simple system there is a state $Y$ with $X\prec Y$.
\vskip1ex}

Denote the state of a (compound) system consisting of two unconnected economies (or systems of economies) $A$ and $B$ in states $X_A, Y_B$ respectively, by $(X_A,Y_B)$.  In particular,  if $X_A\precsim X'_A$ and $Y_B \precsim Y'_B$ then $(X_A,Y_B) \precsim (X'_A,Y'_B)$, because the trader can make the changes to $A$ and $B$ in parallel.  This is A3 of \cite{LY} but we obtain it for free again.

\section{Financial equilibrium}
\label{sec:fineqm}

A further key notion is financial equilibrium.  This will be crucial to making sense of the notion of the ``temperature'' of an economy, which will represent the reciprocal of the marginal utility of money at the macro-level. \rev{Intuitively, the temperature of an economy will determine the nett direction in which, other things being equal, money will flow between two economies:~money will flow from the economy with the higher temperature to that with the lower temperature (i.e., money flows to where it has highest marginal utility, as intuition would suggest). Where there is no nett money flow between economies, they are at the same temperature. We now make these ideas precise.}

Given two simple economies $A$ and $B$, we define their {\em financial join} to be the joint economy where money is allowed to flow freely between the distinguished money-components but no other goods can; {neither can money flow from or to any other money-components if there are any}.  
We say that economies between which money flows freely between their distinguished money-components are in {\em financial contact}.  The financial join of two simple economies is again a simple economy, where the value of the distinguished money coordinate for the joint economy is the sum of the values of the  distinguished money coordinates of the two economies.

There are various reasons why financial flow without goods flow might occur and the present approach is neutral between them. For example, people working in one economy might wish to send money to relatives in the other.  Even more simply, one person could own assets in both economies; when financial contact is allowed, they can choose to send money from their holdings in one economy to their holdings in the other.  It will make sense to send money to the economy in which its value is higher, i.e.,~to the one with the lower temperature (to be formalised in Section~\ref{sec:temperature}).\footnote{A challenge for future work is to analyse some of the forces that drive real-world flows, such as interest rates and investment opportunities. We also leave treating surrogates for money, such as cheques or I-owe-yous (IOUs), and how to deal with credit, for later work.}

\rev{\vskip1ex
\noindent{\bf Definition (financial join $\theta$)}: 
Denote by $\theta(X_A,Y_B)$ the state of the financial join of $A,B$, that is reached from initial states $X_A,Y_B$, after $A$ and $B$ have come to equilibrium.}
\vskip1ex
\noindent The equilibrium is specified by the total amount of money in the new distinguished money component and the individual amounts of the other goods (including possibly money) in each part of each economy.
Because $A$ and $B$ are assumed to be indecomposable, it is reasonable to assume that the equilibrium state of the financial join is to a high approximation independent of the way the join is made.\footnote{As far as we can see, Lieb and Yngvason did not address the corresponding question about thermal joins.} The idea is that if say $A$ were decomposable into $A_1$ and $A_2$ then the effects of joining $B$ to $A_1 $ or $A_2$ or both would certainly be different, depending on the distribution of goods and money between $A_1$ and $A_2$, whereas if $A$ and $B$ are indecomposable then any way of letting money flow between them should produce approximately the same statistical state.

The process of coming to equilibrium will in general lead to a nett transfer of some money from one economy to the other, but in equilibrium there is no nett flow of money between them (though there will be on-going, but zero-mean, exchanges of money between the economies).  Note that we are hereby applying A0 to the financial join and thus assuming that allowing money to flow between the two economies is enough to achieve ergodicity of the joint system.  We deduce that $(X_A,Y_B) \precsim \theta(X_A,Y_B)$, because the trader does not have to do anything except to put the two in financial contact.  This is A11 of \cite{LY} but we get it for free.

We also deduce that there exist states $X'_A,Y'_B$ such that $\theta(X_A,Y_B) \sim (X'_A,Y'_B)$. To see why, let $X'_A,Y'_B$ be the states of $A$ and $B$ after financial contact is removed. Then the only difference between the left- and right-hand sides is that the left-side has financial contact between $A$ and $B$; but the trader can act as a financial contact between $A$ and $B$ on the right-hand side by simply 
\rev{offering a money transfer service}.  So in the presence of the trader, the two sides are the same.
Thus they are reversibly accessible from each other. This is A12 of \cite{LY} and we get it again for free.

\rev{\vskip1ex
\noindent{\bf Definition (financial equilibrium, $\equiv$)}:
If there is no nett flow of money between $A$ and $B$ on bringing them into financial contact, we say their states are in {\em financial equilibrium}, and write $X_A \equiv Y_B$.
\vskip1ex}
\noindent Thus, after nett money flow has occurred, $X'_A \equiv Y'_B$.  From the definition, $\equiv$ is symmetric.\footnote{\cite{LY} deduce symmetry of $\equiv$ in their Lemma 2(ii), from their axioms A4, A5, A7, A11 and A12 but we have not yet introduced A4, A5, A7, and prefer to reduce our dependence on them.}

We deduce also (A13$'$) that $\equiv$ is reflexive:~i.e.,~that each state of a simple economy is in financial equilibrium with itself, or put differently, there are no internal barriers to money flow (within the distinguished money component). Formally, this requires us to be able to make a copy of the system, to make the financial join of a system to itself, but let us suppose that this is possible. Then the unique equilibrium must be symmetric between the copies, because otherwise\ its reflection would be another equilibrium (contradicting uniqueness). Reflection-symmetry implies that there can be no nett direction of money flow.
Thus $\equiv$ is reflexive.\footnote{A13$'$ is not one of the axioms of \cite{LY}; instead they deduce it in their Lemma 2(i) from A4, A5, A7, A11 and A12. We prefer to list it as an axiom, to avoid relying on A4 and A5, and also reflexivity doesn't seem to us as clear a logical consequence as they claim:~it might depend on the way in which the join is made.  So we treat reflexivity as an axiom, A13$'$.}

We 
\rev{make} the plausible assumption\footnote{One consequence to note is that this implies that for the equilibrium of a network of economies in financial contact, there are no nett cyclic flows of money.  This is because one could remove links to leave just a spanning tree, which has no cycles, and so in equilibrium no nett money flows; by transitivity when one puts back the remaining links there is no change to the nett flows because the economies are already in equilibrium. This assumption might not hold exactly for real economies but we hope it is a good enough approximation. One possible line of argument is that any cyclic flows of money would seem to provide arbitrage opportunities, and that such flows would therefore be reduced as those opportunities are exploited.} 
\vskip1ex
\rev{\noindent{\bf Axiom A13}:
Financial equilibrium $\equiv$ is transitive.  
\vskip1ex}
\noindent
Thus, $\equiv$ is symmetric, reflexive and transitive, and hence is an equivalence relation.

We also need 
\rev{\vskip1ex
\noindent{\bf Axiom A15}: For any states $X_A,Y_B$ of two simple systems $A,B$, there exists $M \ge 0$ such that $X_A \equiv Y_B+M$ or $Y_B \equiv X_A+M$.}
\vskip1ex
\noindent In an economic context, this axiom is very natural.  If there is nett money flow from $A$ to $B$ then we should be able to reverse {the direction of flow} by adding enough money to $B$; and {therefore, assuming continuity, there must be an amount $M$ that when added to $B$ leads to zero nett money flow}.

To give some intuition for the significance of $\equiv$, we will see that economies which are in financial equilibrium, i.e.,~such that there is no nett tendency for money to flow from one to the other, are at the same ``temperature'' (although the notion of temperature is only derived later in \rev{our} 
exposition). We will see that this corresponds intuitively to money having the same ``value'' in each economy, so that there is no tendency to generate a nett flow of money from one economy to another.

\section{Accessibility and financial equilibrium}
\label{sec:accfineqm}

A key component of Lieb and Yngvason's approach to thermodynamics is to formulate connections between the two binary relations $\precsim$ and $\equiv$.
In particular, we want\footnote{{Note that A14 is a significant strengthening of A8, mentioned at the end of section~\ref{sec:accessibility}.}} 
\rev{\vskip1ex
\noindent{\bf Axiom A14}:
For each state $X$ of a simple system $A$ there are states $X_0, X_1$ with $X_0 \equiv X_1$ and $X_0 \prec X \prec X_1$.
\vskip1ex}
\noindent Intuitively the reason that this assumption is important is that it ensures that the equivalence classes of reversible accessibility and financial equilibrium ``cut across'' each other and avoid ``degeneracy''.\footnote{A thermometer is such a degenerate case--a system which varies along a single dimension, so that for all $X$, $Y$ at different temperatures $X \not\sim Y$. For discussion of how to deal with this case, see  \cite{LY99}. We will introduce financial thermometers in Section~\ref{sec:thermometer}.} This assumption is crucial below in deriving what we will call the second law of thermal macroeconomics. Here we give a plausibility argument for this assumption in the economic context; \rev{}{but the reader who is happy to accept that temperature-order and accessibility do not collapse into a single notion may wish to move directly to the next section.}

First let $M$ be the amount of money in state $X$, $M'$ be some fraction (e.g.~$\frac12$) of $M$, and let $X_0', X_1'$ be $X \mp M' \odot$. 
Then $X_0' \prec X \prec X_1'$ because, as in Section~\ref{sec:accessibility}, we have assumed that an economy accepts a macroscopic amount of money from the trader but will not give a macroscopic amount of money back to the trader for nothing. Thus, the trader can just give $M'\odot$ to move $X_0'$ to $X$ and $M' \odot$ to move $X$ to $X_1'$, and each change is irreversible (so we get $\prec$).\footnote{We do allow money to flow from one economy to another, as in Section~\ref{sec:fineqm}, but we preclude the possibility that an economy will voluntarily hand over money \emph{to the trader}.}  

Intuitively (though this is not actually required), under financial contact, money will tend to flow from an economy where it has less ``value'', to an economy where it has greater value.  Thus, consider two economies with the same goods, but different quantities of money. Within an economy, the value of any unit of money will be lower where the amount of money is greater (just as money is devalued when a central bank prints money). This means that the value of money will be lower in state $X_1'$ than in $X_0'$. 

So, clone the system into $A_0, A_1$, with initial states $X_0', X_1'$. We wish to find a reversible path from $X_1'$ to a new state $X_1$ and a reversible path from $X_0'$ to a new state $X_0$, such that the resulting states $X_1,X_0$
are in financial equilibrium. This can be done by the trader 
reversibly buying goods from $A_0$.\footnote{Note the implicit assumption that the economy has some type of good other than money; the trivial case of only money has to be excluded, else A14 is impossible.} This increases the amount of money in comparison with the volumes of goods in that economy (and hence reduces the value of money there, to be formalised in Section~\ref{sec:temperature}). At the same time, the trader can reversibly sell those same goods to $A_1$ (increasing the value of money in that economy). The trader ends up in the same state except for amount of money. 
With sufficient trading, it is plausible that the two copies can be brought to  
financial equilibrium.  Let $X_0 \equiv X_1$ be the states achieved, then $X_0 \sim X_0' \prec X \prec X_1' \sim X_1$.

\section{Scaling symmetry:~Extensivity}
\label{sec:extensivity}

Next comes the crucial assumption 
\rev{\vskip1ex
\noindent{\bf Axiom A4}: One can scale an economy $A$ by any positive real number $\lambda$ to obtain a scaled economy $\lambda A$, and  if $X\precsim Y$ for $A$ then $\lambda X \precsim \lambda Y$ for $\lambda A$.
\vskip1ex}
\noindent This assumption seems plausible if we think of an economy as consisting of lots of small units with a network of purely local interactions based on geographic proximity and we ignore the discreteness of the set of agents and possibly of some types of good.  
Then one can imagine a scaled version of the economy and it is plausible that the accessibility relation would inherit scaling symmetry.  We call an economy satisfying A4 {\em extensive}.  

Yet it may be a poor assumption for real economies.  The concept of economies of scale is well-known and already argues against scaling symmetry.  Indeed, an economy might need to be at least a certain size before it could include certain types of industry at all (say aircraft manufacture or investment banking). {Moreover}, real economies have discrete components, in particular the number of agents is an integer.\footnote{But perhaps the formulation by Lieb and Yngvason can be modified to ask only for scaling by (positive) integers, as does Giles \cite{Gi}.} 

In sum, the assumption of scalability is strong, and unlikely to hold in general in many aspects of real economies. Nonetheless, the analogy with thermodynamics in the natural sciences suggests that the assumption may be useful for many purposes. For example, classical entropy, which is based on the same scaling assumptions, has been of central importance in analysing chemical reactions inside biological cells, although cells, like economies, are clearly enormously intricate mechanisms containing specific structures and processes operating at specific scales \rev{and with small numbers of molecules of some important types}. In any case, we proceed with the scalability assumption, to explore whether it may provide a potentially useful idealization in some circumstances. Later we will briefly consider whether and how it may be relaxed.

A further assumption concerning scaling is \rev{that economies can be ``sliced up.''}
\rev{\vskip1ex
\noindent{\bf Axiom A5}:~Any system can be subdivided into two parts in an arbitrary ratio $\lambda:1-\lambda$ by cutting connections, and in particular the state $X$ can be decomposed into a state $(\lambda X, (1-\lambda)X)$ of the pair of systems, with $X\sim (\lambda X, (1-\lambda)X)$. 
\vskip1ex}
\noindent Again, it is not clear that this holds in the economic context:~cutting internal connections might make a substantial change to an economy, and the economy might not be homogeneous so the two parts might not be just scaled versions of the whole.\footnote{Again, there is perhaps an escape by replacing A5 with $(m+n)X \sim (mX, nX)$ for all positive integers $m,n$.}
If we ignore the {question of} discreteness, there is no problem with assuming $(\lambda X, (1-\lambda)X) \preceq X$, because putting the two scaled copies in contact merges their total goods and money, so {that} by A0 we obtain $X$.
But it is not clear that we should expect $X \preceq (\lambda X, (1-\lambda)X)$, because cutting connections in a finite system will typically lead to a division of the goods and money in ratios differing from $\lambda : 1-\lambda$ by order $1/\sqrt{N}$.  Perhaps one could imagine the trader cutting the connections in a slow smooth way by enacting only a slowly diminishing fraction of each exchange, which would thereby reduce fluctuations arbitrarily close to zero at the macroscopic level.
We proceed nonetheless (and leave implications of weaker assumptions for later work).


\rev{Following Lieb and Yngvason, we also need a technical assumption (which can be ignored on first reading)},
\rev{\vskip1ex
\noindent{\bf Axiom A6}:~If $(X,\eps Z_0) \precsim (Y,\eps Z_1)$ for some $Z_0,Z_1$ and a sequence of $\eps \to 0$ then $X \precsim Y$.
\vskip1ex}
\noindent In economic terms, this means that if two economies are reversibly accessible when each is conjoined with an arbitrarily tiny scaled copy of some other economy, then they are also reversibly accessible from each other if that arbitrarily tiny scaled copy is entirely absent.  This is automatic {in the present analysis} because we built into the definition of accessibility that the trader is allowed to end up with a small change in its external system {as long as they can make that change arbitrarily small}.

\section{Convexity assumptions}
\label{sec:convexity}
 
We need a few more technical axioms from \cite{LY}, \rev{which can be skipped over on first reading}.  

Firstly, we assume 
\rev{\vskip1ex
\noindent{\bf Axiom A7}:~For states $X,Y$ of a simple {economy} and $t \in (0,1)$, $(tX,(1-t)Y) \precsim tX+(1-t)Y$.
\vskip1ex}
\noindent This seems straightforwardly true:~the economy formed by merging two economies is accessible from the pair of separate economies.

\rev{\vskip1ex
\noindent{\bf Definition (Forward sector)}: 
The {\em forward sector} of a state $X$ in a state space $\Gamma$ is
$A_X = \{Y \in \Gamma: X \precsim Y\}.$
\vskip1ex}
\rev{\noindent It is the set of states accessible from $X$.
A consequence of A7 is that 
the forward sector 
of $X$} is convex:~{that is, for $t' \in (0,1)$, if $W \in A_X$, and $Z \in A_X$, then $t'W+(1-t')Z \in A_X$}. 

Secondly, we assume 
\rev{\vskip1ex
\noindent{\bf Axiom A9}:~Each forward sector $A_X$ has a unique support plane at $X$.  This tangent plane is furthermore assumed to be a Lipschitz-continuous function of $X$ as defined in \cite{LY}.
\vskip1ex}
\noindent In both the physics and economics contexts, this technical assumption is required to allow the analysis to proceed. Is it justified? Indirectly, we may evaluate this based on whether the analysis that it allows generates useful results.
It is an interesting topic for future work to consider where this assumption may fail in real economies, and what implications this might have.

Thirdly, we assume 
\rev{\vskip1ex
\noindent{\bf Axiom A10}:~The boundary of each forward sector $A_X$ is connected. 
\vskip1ex}
\noindent This technical assumption looks innocuous in an economic context---at least, it is not clear under what plausible economic circumstances it would be violated.

\section{Economic Entropy}
\label{sec:entropy}

With the above assumptions A0-A15 we are now ready to begin to reap the rewards.  The first reward, \rev{an economic notion of entropy}, comes from Lemma 4 and Theorem 1 of \cite{LY}. We identify a simple system with its state space $\Gamma$.
Given a simple system  $\Gamma$, define a {\em multiple scaled copy} to be the product system with state space $\lambda_1 \Gamma \times \ldots \times \lambda_n \Gamma$ for some $n\ge 1$ and positive numbers $\lambda_1,\ldots \lambda_n$. Define its {\em weight} to be $w = \sum_i \lambda_i$. \rev{Recall from Section~\ref{sec:accessibility} that two states $Y,Y' \in \Gamma$ are said to be {\em comparable} if $Y \preceq Y'$ or $Y' \preceq Y$ (including the possibility of both).}
\vskip 1ex
\noindent{\bf Theorem 1}: For any simple {economy} $\Gamma$, there is a real-valued function $S:\Gamma \to \R$ on its state space such that for any states $Y = (\lambda_1 Y_1,\ldots, \lambda_n Y_n)$ and $Y'=(\lambda'_1 Y'_1,\ldots \lambda_{n'}'Y_{n'}')$ of any two multiple scaled copies with equal weights
then $Y$ and $Y'$ are comparable and
$$
    Y \precsim Y' \mbox{ iff } \sum_i \lambda_i S(Y_i) \le \sum_j \lambda'_j S(Y'_j).
$$
Furthermore, the function $S$ is unique up to orientation-preserving affine transformations:~$S \mapsto a S + b$ with $a>0$.  
\vskip 1ex
In keeping with thermodynamics, \rev{and as noted above}, we call $S$ {\em entropy}.  We extend it to scaled copies of $\Gamma$ by $S(\lambda X) = \lambda S(X)$ and to products of {economies} by $S(X,Y) = S(X)+S(Y)$, and thus identify the two sides in the above inequality as the total entropies of the two states.

Note that this is a very strong result:~in economic terms, it means that for any two such states, $Y$ and $Y'$, of any pair of subdivisions of a simple system into scaled copies ({i.e., where} the summed weights {are} both 1), the trader can either move the system from $Y$ to $Y'$, or from $Y'$ to $Y$, or both, by receiving or giving money or goods and arbitrarily small change in its external system.  Furthermore, the direction of the possible change is encoded by the sign of the change in the total entropy.
The restriction to multiple scaled copies having the same weight is natural because the trader can not change the weight of a multiple scaled copy, only \rev{how it is subdivided}.

We illustrate the construction of $S$ from \cite{LY} in Figure~1. 
\begin{figure}[htbp] 
   \centering
\includegraphics[width=2.0in]{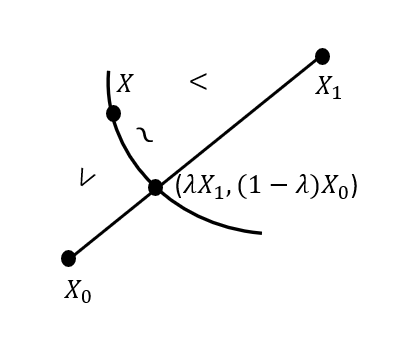} \ 
\caption{Sketch proof of existence of entropy:~given $X_0,X_1$ as in A14, we draw the straight line of convex combinations in the space of product systems and deduce that any $X$ between them in the accessibility order is equivalent to one point on the line; the resulting $\lambda$ is the desired entropy. 
Here we make the natural identifications $X_0 \sim (0X_1,X_0)$, $X_1 \sim (X_1,0X_0)$, $X \sim (\lambda X, (1-\lambda)X)$.
Changing the choice of $X_0\prec X_1$ makes an o-p affine transformation of $S$.}
   \label{fig:entro}
\end{figure}

Theorem 1 tells us that entropy determines accessibility in multiple scaled products of a single simple system, but we can go well beyond that by combining
Theorem 2 and Theorem 4 of \cite{LY} \rev{to obtain our next result.}  

\vskip1ex
\noindent {\bf Theorem 2}: For any collection of simple systems $\Gamma_k$ \rev{for $k$ in some set $K$,} and entropy functions $S_k$ as per Theorem 1, there are positive constants $a_k$ such that defining $S = a_k S_k$ on $\Gamma_k$ and extending it to all product systems formed from scaled copies of the $\Gamma_k$ by $S(\lambda X) = \lambda S(X)$ and $S(X,Y)=S(X)+S(Y)$, then all states $Z,Z'$ in products with  weights $w_k = w'_k$ for each $k$ are comparable and 
$$Z\preceq Z' \mbox{ iff } S(Z) \le S(Z').$$
\vskip1ex

This says that the entropy functions for different simple systems can be calibrated by a scale-change to make total entropy represent the accessibility pre-order in product systems.  It leaves free only an arbitrary overall positive multiplicative constant and the arbitrary additive constants (which play no role because of the restrictions to equal weights). 
\rev{As an aside, note that if we assume} one more axiom \rev{from \cite{LY}} then Theorem 7 of \cite{LY} shows that the latter freedoms can also be reduced to allow incorporation of manufacturing to combine some goods into others 
\rev{(\cite{LY} take this approach in order to model how entropy works in chemical reactions)}, but we leave that for future work.

The calibration procedure is based on their Lemma 5, which says that for any two simple systems $\Gamma_A, \Gamma_B$ there are states $X_0,X_1 \in \Gamma_A$ and $Y_0,Y_1 \in \Gamma_B$ such that $X_0 \prec X_1$, $Y_0 \prec Y_1$ and $(X_0,Y_1) \sim (X_1,Y_0)$.  
Thus the multiplicative constants for $S_A$ relative to $S_B$ can be fixed by scaling one or other to make $S_A(X_1)-S_A(X_0) = S_B(Y_1)-S_B(Y_0)$.

Alternatively, but equivalently, the scale for entropy can be fixed by choosing a scale for temperature (to be defined in Section~\ref{sec:temperature}).

\rev{With these preliminaries, we are now in a position to} deduce the all-important\footnote{We note that this is unrelated to the ``second law of economics'' 
\rev{presented} in \cite{Ku}.}
\vskip 1ex
\noindent{\bf Second law of thermal macroeconomics}:~On putting two or more exchange economies into contact (in any way), the total entropy can not decrease. 
\vskip 1ex
The second law suggests that we can think of entropy as an aggregate utility or value of an economy; \rev{and that putting several economies together can never decrease this aggregate utility}.  But in contrast to standard utility theory in which the utilities of different agents are typically not viewed as being on a common scale \cite{ER}, one can meaningfully compare, and indeed add, the entropies of different parts of an economic system. \rev{Moreover}, it is only the total entropy that can not decrease. Thus, in some interactions between two economies, the entropy of one economy can decrease, as long as this is compensated by an equally large or larger increase in the entropy of the other economy. For example, this happens when money flows out of a \rev{high temperature} economy where it has less value to a \rev{low temperature} economy where money has greater value. 

The second law of \rev{thermal macroeconomics} can be seen as a substantiation of Adam Smith's ``invisible hand'':~at the macro-level, putting economies in contact never decreases the total value of the joint system.  

To prove the Second Law, take systems $A$ and $B$ in states $X_A$, $Y_B$, respectively and put them in contact.  It could be just financial contact or permitting arbitrary flow of goods and money or permitting flows corresponding to a given price.
The joint system goes to a state equivalent to some $(X'_A,Y'_B)$ (dependent on the type of contact).  The trader does not do anything, so we have $(X_A,Y_B) \precsim (X'_A,Y'_B)$.  By Theorem 2, we deduce that $S(X_A)+S(Y_B) \le S(X'_A)+S(Y'_B)$.

Note that \cite{LY} consider the second law to be the existence of entropy with the property of Theorem 2, but we consider it worth stating this particular consequence.  Note also that the terminology ``second law'' is historical:~the first law of thermodynamics is conservation of energy.  It took physicists as much thought and experiment to settle on the first law as on the second one (for one account of this history, see \cite{Sas2}). In our exchange economies, each type of good, including money, is assumed to be conserved, so that the ``first law'' of the conservation of money (and indeed, of goods) is trivial. In principle, there could be further conserved quantities needed to describe a macro-state, but at least in our context of exchange economies satisfying A0, there are none.  

The consequences of entropy become significant for systems consisting of several parts with specifications of what transfers are possible.  
For example, for a system with two parts, having a common money and a common type of good, with transfers only via a trader who wishes to end up with the same state except for money, the state space for the system is given by the amounts $G_j, M_j$ of goods and money in parts $j=1,2$, but the allowed changes conserve $G_1+G_2$.  Within this 3D space, the 2D surface of constant total entropy separates the space into those states that are accessible from a point on this surface from those that are not accessible.   Changes within the surface of constant total entropy are reversible.  The others are not. This will be illustrated by a simple example in the next section.

Alternatively, one could leave out the trader, but allow transfers of goods and money between the two parts: that is, the economies trade directly with each other, rather than each transaction being mediated by the trader.  Then the possible changes conserve both $G_1+G_2$ and $M_1+M_2$.  In this 2D space there are curves of constant total entropy which separate accessible states from inaccessible ones. The picture is analogous to the indifference curves for individuals in standard economics, but now at the level of a whole multi-part  economic system.
Again, this will be illustrated in the next section.

\rev{The existence of entropy with the property of Theorem 2 has remarkably strong implications.} On the one hand, it provides a very restrictive constraint on possible transitions from one equilibrium state of a system or set of systems to another:~it precludes any transition that leads to a decrease in (total) entropy, whatever the trader (or other exogenous agent or agents) does. 
On the other hand, it specifies that there is always an available path by which the trader or other external agency can move the economy in \emph{any} entropy-increasing way in the state space.  Entropy-conserving changes can also be achieved in a limiting sense (in practice, requiring transitions to be carried out arbitrarily slowly). 

If we interpret entropy as corresponding to aggregate utility, then the first observation implies that trade, or financial flow, between any two (exchange) economies does not decrease aggregate utility. That is, there will always be gains of trade, except for the limiting case of zero gains. Indeed, if any number of economies are put into contact using any network of trade or financial connections, trade or financial flows will yield non-negative gains of trade for the compound system. Furthermore, any reduction in restrictions on trade (e.g., removing a barrier to trade for one or more goods, or removing a prohibition on moving money from one economy to another) will lead to non-negative gains of trade. Note, though, that the gains of trade may be unequally distributed---indeed, one or more economy may lose entropy, as long as the entropy of the whole system increases. We analyse trade in detail in Section~\ref{sec:trade}.

One consequence of existence and extensiveness of entropy  \rev{that has interesting implications for standard macroeconomic theory} is 
a rationale for the 
notion of \rev{the} representative agent,\footnote{The notion of a representative agent, is widely and somewhat controversially used in providing a microfoundation for macroeconomic behaviour. \cite{H96} traces the notion back to Alfred Marshall \cite{Mar}, while \cite{K92} finds similar ideas in Francis Edgeworth \cite{E81}. The notion of a representative firm was critiqued by Lionel Robbins \cite{L28}; and critiques of the broader coherence of the notion of a representative agent have continued \cite{K92}, leading to richer modelling approaches \cite{KMV}.} though with a twist. Given entropy function $S(M,G)$ for an economy with $N$ agents, we can define $U_r(m,g) = S(Nm,Ng)/N$ and consider it as the utility for a representative agent to possess amounts $m$ of money and $g$ of goods.  Maximising the sum of the entropies of several economies then corresponds exactly to maximising the average utility of their representative agents, weighted by the numbers of agents in each economy. 
There is no need, however, to scale entropy to the level of individual agents.  Our approach says that each economy has a utility function (its entropy) and a collection of economies in contact act to maximise the sum of their constituent utilities (i.e., maximising the entropy of the collection of economies as a whole). Note that this maximisation of summed utility over the system of economies can arise purely from the interactions of myopic or selfish agents, from their ability to trade and to move money from one economy to another, as long as there is convergence to a unique statistical equilibrium (Section~\ref{sec:eqm}).

The ``twist'' is  that there is no requirement that each economy attempts to ``maximise its own utility.'' \rev{As we noted above,} 
money can flow from one economy to another, lowering the entropy of one and increasing the entropy of the other, as long as the entropy of the whole system does not decrease. We will, though, also consider in Section~\ref{sec:MBtrade} the important special case of what we will call ``mutually beneficial'' trade, which is defined by not decreasing the entropy of any economy (i.e., leading to an analogue of Pareto optimality at the level of economies).

To end this section, we discuss to what extent the entropy function depends on the choice of num\'eraire (money in our presentation so far).\footnote{We are not aware whether Lieb and Yngvason discussed the analogous question in the physical case, but their discussion didn't have any other conserved quantities apart from energy.} 
We restrict attention to num\'eraires for which it is always desirable {at the aggregate level} to have more (required for Sections~\ref{sec:accessibility} and \ref{sec:accfineqm}).
The accessibility concept does not depend on the choice of num\'eraire, but the concept of financial equilibrium does.  Nonetheless, the end result is a characterisation of accessibility in terms of an entropy function $S$ that is unique up to \rev{orientation-preserving} affine transformations, so the entropy functions determined by two different choices of num\'eraire can differ only by \rev{such a} transformation.

\section{A toy illustration with interacting agents}
\label{sec:toy}

One of the distinctive characteristics of {the conception of thermal macroeconomics developed here} is that it proceeds entirely at the level of aggregate behaviour, positing axioms about trade and financial flows and deducing conclusions such as the second law. Thus, just as classical thermodynamics makes no reference to the molecular basis of pressure, heat, entropy and so on, so TM makes no reference to the micro-foundation of economic behaviour, in terms of the actions of individual agents. A great advantage of this strategy, both for physics and economics, is that the micro-foundations of real-world systems are often of great complexity and only partially known, so that providing a detailed micro-level analysis is not realistic---but nonetheless, laws at the aggregate level are still derived.

It is interesting, however, to illustrate the general theory by special cases in which micro-foundations are sufficiently simple that they can be analysed directly. \rev{The results of this analysis will also help build intuitions concerning how the general ``macro'' account works. On first reading, the reader might wish to skim the technical details of the micro-foundational set-up, and focus on the formula for entropy that we derive in (\ref{eq:SCD}) and the results plotted in the figures later in this section}. 

In the physics context, a prime example \rev{of a micro-foundational model in thermodynamics} is the molecular model of  ideal gasses, leading to a foundational explanation of temperature and pressure and the entropy function $S = N \log v u^{d/2}$, where $N$ is the number of molecules, $v$ and $u$ are the volume and energy per molecule and $d$ is the number of degrees of freedom per molecule (e.g.~for a diatomic molecule like $O_2$ or $N_2$, $d$ is the sum of three translational plus two rotational degrees of freedom, at normal temperatures where vibration is suppressed quantum-mechanically).
Note that it is conventional to add $N(d+2)/2$ \rev{to the above formula for $S$} and multiply by Boltzmann's constant, but that is just an affine transformation.

In particular, let us compute the entropy for a toy example of an exchange economy
with a single good in addition to money and $N$ independent identical ``Cobb-Douglas'' \rev{(CD)} agents.\footnote{\rev{Precursors come in \cite{SF, SST}} but they use the term slightly differently from us.} 
By Cobb-Douglas agents we mean 
that they have a ``utility'' function 
\begin{equation}
U(m,g) =  g^{\alpha-1} m^{\eta-1},
\label{eq:U}
\end{equation}
for amounts $g$ of good and $m$ of money, for some $\alpha, \eta > 0$.
It is natural to think of both goods and money as being desirable for each agent, in which case $\alpha, \eta > 1$.\footnote{We adopt the form (\ref{eq:U}) with minus ones in the exponents because it makes future formulae simpler.}
Exponent $\alpha=1$ corresponds to indifference concerning amount of goods, while $\alpha<1$ corresponds to preference for fewer goods, which allows one to take into account cases like pollutants.
Nonetheless, our theory requires only that money is desirable and only at the aggregate level.  Perhaps surprisingly, we will see that this is satisfied for all $\eta>0$.

Importantly, in contrast to standard economic usage, our agents are not utility maximisers; instead, they use their utility functions to bias the probability rate for the outcome of random pairwise encounters towards increased utility for both, but with some ``errors'' (though these can be made arbitrarily small, by choosing high values of $\alpha$ and $\eta$). Thus, our agents can be viewed as ``noisy'' utility maximizers, where the level of noise can be as small as desired but not zero.

We define the stochastic dynamics more precisely now.  Let us consider an encounter between one agent $i$ and another agent $j$. Label the amounts of money and goods for each agent before the encounter by pairs $m = (m_i,m_j),g=(g_i,g_j)$ of positive 2-component vectors, and after the encounter by $m',g'$. We suppose that transitions are restricted to those that conserve both goods and money:~$m'_i+m'_j = m_i+m_j$ and $g'_i+g'_j = g_i + g_j$.
We denote
the probability rate from $m,g$ to $m',g'$ by
$q(m,g,m',g')$.  
Each agent biases the probability rate by making it proportional to their utility of the outcome.  Thus we take
$$q(m,g,m',g') = k_{ij}(g'_i g'_j)^{\alpha-1} (m'_i m'_j)^{\eta-1}\, \delta(g'_i+g'_j-g_i-g_j)\, \delta(m'_i+m'_j-m_i-m_j),$$
where $k_{ij}\ge 0$ is a \rev{(symmetric)} encounter rate for $i$ and $j$ and the $\delta$-functions enforce conservation of goods and money.  We suppose that the outcome is independent of all previous encounters of all pairs, conditional on the current state. The constraints $\alpha, \eta>0$ make the rate integrable and hence the stochastic process well-defined.  
Assuming the graph formed by positive encounter rates between agents is connected, for given total goods $G$ and total money $M$ \rev{we believe} this stochastic process has a unique stationary probability that attracts all initial probability distributions (this property is called ``ergodicity''). If the money and goods were quantised, ergodicity would follow from a standard result for irreducible continuous-time Markov processes on finite state spaces \cite{Sen} (or positively recurrent ones on countable spaces, e.g.~Thm 1.6 of \cite{Ke}), but here the state space is continuous. Nevertheless, 
result\rev{s of this form} can be obtained for continuous spaces by use of more sophisticated techniques.  For one treatment, see \cite{DGS}.  Their models have only money, but extension to an arbitrary finite number of types of good is straightforward.  They give a detailed proof only in the case $\eta=1$ and for equal encounter rates between all pairs of agents, but state that their method generalises.  In another line of work, \cite{CCL} proves ergodicity for the Kac model of a 1D gas, which is equivalent to the case $\eta=\frac12$ under the same assumptions on encounter rates.  
\rev{Recently,} one of us 
has proved ergodicity \rev{for fully connected CD economies} more directly and more generally \rev{\cite{M25}}. 

\rev{A useful feature is that} the process is reversible (in the sense of Markov chains) with respect to the distribution with density $\rho$ proportional to the product of the utilities, i.e.,~
\begin{equation}\rho(g,m) = Z^{-1} \prod_i g_i^{\alpha-1} m_i^{\eta-1} \ \delta\left(\sum_i g_i-G\right)\, \delta\left(\sum_i m_i - M\right),
\label{eq:CDdensity}
\end{equation}
where $g,m$ now represent the vectors $(g_1,\ldots g_N)$ and $(m_1,\ldots m_N)$.  Hence we deduce that this distribution is the stationary one.
It is a product of two Dirichlet distributions (one for the partition of goods, the other for money), and the normalisation constant is
\begin{equation}
Z = \frac{G^{N\alpha-1}\Gamma(\alpha)^N}{\Gamma(N\alpha)} \frac{M^{N\eta-1}\Gamma(\eta)^N}{\Gamma(N\eta)}.
\label{eq:Z}
\end{equation}

The marginal distribution for one agent in this stationary distribution is a product of two independent Beta-distributions (one for goods, one for money).  For example, the Beta-distribution for money has density
$$\rho(m) = \frac{\Gamma(N\eta)}{\Gamma(\eta)\Gamma((N-1)\eta)} \frac{m^{\eta-1} (M-m)^{(N-1)\eta-1}}{M^{N\eta-1}}.$$
It is an illustration of how inequality of wealth can arise (on which there is a large literature, e.g.~\cite{C4}), though note that that for $\eta>1$ and $N$ large, the bulk of the distribution is peaked close to the mean $M/N$, and that each agent samples the whole distribution as time evolves.\footnote{In particular, wealth does not concentrate in a fixed set of agents.} 

To check the axioms are satisfied and to determine the entropy of this toy economy we need to spell out how each agent interacts with an external trader.  We specify this to be consistent with how agents interact with the other agents, i.e.,~biased in the same way by their utility function.  
We suppose the trader offers to buy or sell arbitrary amounts at a price $\mu>0$.  Then the probability rate for each agent $i$ to transition from $(m,g)$ to $(m',g')$ is
$$q(m,g,m',g') = K_i {g'}^{\alpha-1} {m'}^{\eta-1} \, \delta(m'+\mu g' - (m+\mu g)),$$ 
for some $K_i\ge 0$, which is the trader's encounter rate with agent $i$.  There is no contribution to the dependence on $(m',g')$ from the trader because it has unlimited goods and money.  {As the transition is along the ``budget'' line $m'+\mu g'= m+\mu g$ we can eliminate $m'$ in favour of $g'$ and write the rate as a function of just $g'$ (given $m,g,\mu$):~$$q(g') = K_i g'^{\alpha-1}(m+\mu g-\mu g')^{\eta-1}.$$ This is a Beta-distribution again (though unnormalised), and its mean is}
$$\langle g' \rangle = \frac{\alpha}{\alpha+\eta} \left(g+\frac{m}{\mu}\right).$$

Below we specify in more detail 
how the trader can interact, and in particular how it can donate money. But if the reader is happy to accept that the axioms can be satisfied, then we can make a simple derivation of the entropy function just from the above.

First, we give a heuristic {derivation}.  The means over the initial stationary distribution were
$\langle g\rangle = \frac{G}{N}, \langle m\rangle = \frac{M}{N}$.
Thus the change from the mean of $g$ to the mean of $g'$ is
$$\frac{\frac{\alpha}{\mu} M - \eta G}{N(\alpha+\eta)}.$$
Thus in a time $\Delta t$ we obtain a change $$\Delta G = K \left(\frac{\alpha M}{\mu} - \eta G\right)$$
in the quantity of goods, where $K = \frac{\Delta t}{N(\alpha+\eta)} \sum_i K_i$.  $\Delta G$ is positive, zero or negative according as $\mu$ is less than, equal to, or greater than 
\begin{equation}\mu_c = \frac{\alpha M}{\eta G}.
\label{eq:muc}
\end{equation}
The change in the amount of money is $\Delta M = -\mu\, \Delta G$.
So we see that $$\alpha \frac{\Delta G}{G} + \eta \frac{\Delta M}{M} = \left(\frac{\alpha}{G}-\mu\frac{\eta}{M}\right)\Delta G =\frac{K}{GM\mu} (\alpha M - \mu\eta G)^2 \ge 0,$$ with equality iff $\mu = \mu_c$.
This is precisely the condition $\Delta S\ge 0$ for infinitesimal changes in
\begin{equation}
S = N \log\left(\left(\frac{G}{N}\right)^\alpha \left( \frac{M}{N}\right)^\eta\right).
\label{eq:SCD}
\end{equation}
The factors of $N$ have been included to make this function scale correctly with $N$, and by chance, the logarithm makes it additive.
Thus we deduce that (up to an orientation-preserving affine transformation) the entropy of the toy economy is this function $S$.  

Next, we give a more complete version of this derivation, again assuming the axioms are satisfied.  On putting the CD economy with initial $M_0,G_0$ in contact with the trader at a given price $\mu>0$, the system goes to a Dirichlet distribution for the quantities $m_i, \mu g_i$, with exponents $\eta$ for the $m_i$ and $\alpha$ for the $\mu g_i$ and $\sum_i m_i + \mu g_i = M_0+\mu G_0$. This is because the CD+trader process is reversible with respect to this probability, and ergodic (which could be proved along the same lines as for just the CD economy).
The resulting equilibrium has mean money $\bar{M} = \frac{\eta}{\eta+\alpha}(M_0+\mu G_0)$ and mean goods $\bar{G} = \frac{\alpha}{\alpha+\eta}(\frac{M_0}{\mu}+G_0)$.
For large $N$ the distribution of $M$ and $G$ is tightly concentrated around the mean on the macro-scale.  Thus after disconnecting the trader, the CD system has moved to essentially the above mean. In the case that $\mu = \mu_c = \frac{\alpha M_0}{\eta G_0}$ the displacement is zero, but for nearby $\mu$ it is along a curve that has slope $dM/dG = -\mu_c$ at $M_0,G_0$, because $dM/d\mu = \frac{\eta}{\eta+\alpha}G_0$ and $dG/d\mu = -\frac{\alpha}{\eta+\alpha}\frac{M_0}{\mu^2}$, both non-zero and $\frac{d}{d\mu} M+\mu_c G = 0$ there. 
Integrating $dM/dG = -\frac{\alpha M}{\eta G}$ from $(G_0,M_0)$ we obtain that the trader can move the CD economy arbitrarily close to anywhere along the curve $\log M^\eta G^\alpha = \log M_0^\eta G_0^\alpha$ by taking $\mu$ just above or below $\mu_c$ at all times. Then the trader can use an external system to move the system onto the curve with an arbitrarily small change in the external system. {For example, choose an external system with lower temperature and put it in financial contact; then some money will flow out of our system, thereby reducing $\log G^\alpha M^\eta$.}  Furthermore, the trader can move the system backwards along this curve.  So the curve consists of reversibly accessible states.
This establishes that the accessibility relation is encoded by $\log G^\alpha M^\eta$. Including factors of $N$ to make it scale correctly and noting that the result is additive, we deduce (\ref{eq:SCD}) for the entropy, up to orientation-preserving affine transformation.

In Appendix~\ref{app:axioms}, we check the axioms for CD economies, suitably interpreted, and in Appendix~\ref{app:CDentropy}, we give a derivation of the entropy for CD economies using the construction of proof of \cite{LY}, and also show that for CD economies with different exponents, the entropies of (\ref{eq:SCD}) are already calibrated, so {that} Theorem 2 can be applied directly.

For completeness of description of the model to fit with the axioms, however, we must specify a complete list of the ways the trader can interact with the economy.  In particular, we must
also specify how the trader can exchange money with the economy.  We propose that the trader makes available a macroscopic amount $M_T$ of money {in what we call a ``pot''};\footnote{{We do not make available the trader's full (unlimited) money because equilibrium would never be reached.}} agent $i$ interacts with the {trader's pot} at rate $K_i$ and updates its money $m_i$ according to probability density proportional to $u_i \Theta(M+M_T-\sum_j m_j)$, where the Heaviside function $\Theta(x) = 1$ if $x > 0$, $0$ if $x\le 0$, and $M$ is the initial total money in the economy.  This represents the trader being indifferent to how much money is in the pot.  It produces a stationary density proportional to $\prod_i u_i\, \Theta(M+M_T-\sum_j m_j)$, which is just a Dirichlet distribution for the money in the agents and the pot, with exponents $\eta$ for the agents and $1$ for the pot. 
The money in the CD economy is distributed as Beta$(N\eta,1) (M+M_T)$, which has mean $\frac{N\eta}{N\eta+1}(M+M_T)$ and relatively small variance, so for a macroscopic $M_T$, essentially all the money from the pot goes to the economy.

To complete the list of allowed actions of the trader, {we allow that} it can disconnect parts of a system at equilibrium, and it can put parts of a system in various forms of contact, for example, financial contact or complete contact, but also {contact} where one or more constraints {are} imposed, of the form $\sum_i a_i dG_i = 0$, where money is included as $G_0$ and the coefficients $a_i$ may have either sign.  Finally, the trader is allowed to connect an external system, but it must be one that satisfies the axioms, and for a move to be accessible it must be achievable with arbitrarily small nett change in the external system.

In the design of such micro-economic models, it is important to exclude the possibility of the trader acting as a ``Maxwell demon'', e.g.~targeting those agents with more money than goods and offering to sell to them for a price above the market price ($\mu_c$ in the present case), which would probably lead to all states being accessible from all other states, and hence to failure of A8 and A14.

It is important to note that even though this example has homogeneous agents, the utility $U_r(g,m) = \log g^\alpha m^\eta$ for the resulting representative agent is not the utility for a single Cobb-Douglas agent, which recall is $U(m,g) =  g^{\alpha-1} m^{\eta-1}$. {Indeed,} it is not even a function of the utility for a single agent.  This is not too surprising, however, because our agents use their utility function to bias the probability rates, not to maximise utility. Nonetheless, if the exponents $\alpha$ and $\eta$ of the Cobb-Douglas utility function are scaled up sufficiently, the agents will approach maximising behaviour (as will be explained in Section~\ref{sec:temperature}).

\begin{figure}[htbp] 
   \centering
\includegraphics[width=2.5in]{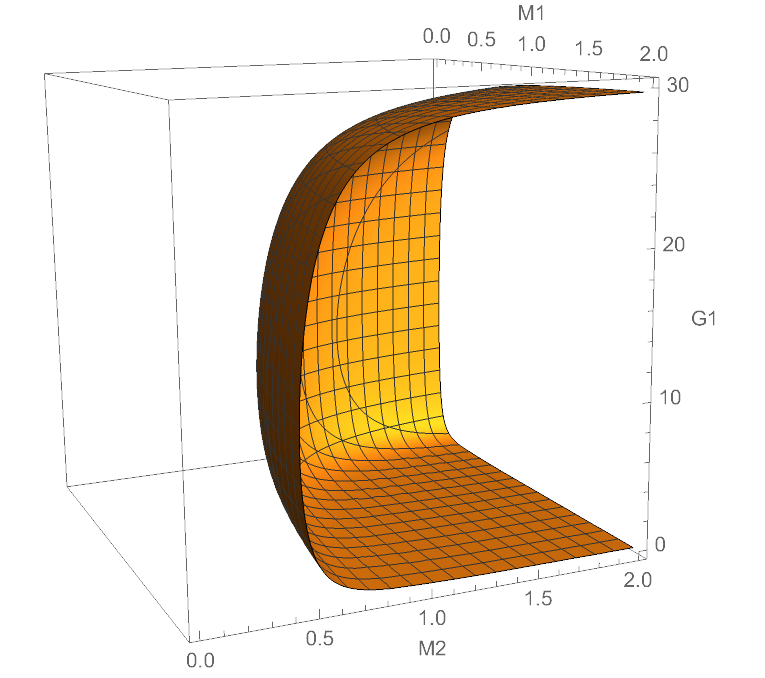}
\caption{Boundary of accessible region for a two-part Cobb-Douglas economy from an initial endowment (anywhere on the displayed surface).  The accessible region is to the right. The economies have numbers of agents in the ratio $N_1:N_2 = 1:2$, and the same exponents $\alpha=1, \eta=\frac52$.  The total goods $G=30$.}
   \label{fig:AccRegion}
\end{figure}

\begin{figure}[htb] 
   \centering
\includegraphics[width=2.5in]{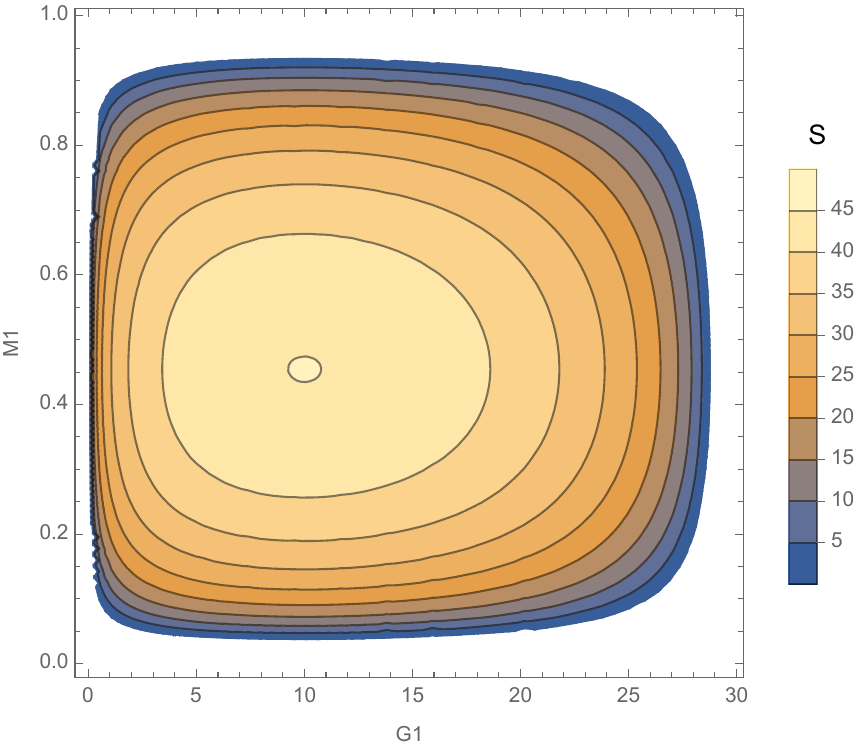}
\caption{Contours of entropy $S$ for a two-part Cobb-Douglas economy.  The economies have numbers of agents $N_1=10,N_2 = 20$, the same exponent $\alpha=1$ for goods, but different exponents $\eta_1=\frac52, \eta_2=\frac32$ for money.  The  total goods $G=30$ and total money $M=1$.}
   \label{fig:Splot}
\end{figure}

To illustrate the consequences of entropy, Figure~\ref{fig:AccRegion} 
shows the boundary of the region accessible from an initial endowment of a system of two Cobb-Douglas economies\footnote{The choice $\alpha=1$ is perhaps unrealistic, because it implies the utility is independent of the amount of goods, but similar figures are obtained for $\alpha>1$.}.  We see that adding money to either economy is allowed, but also adding money to one while removing money from the other is allowed, within limits.  More importantly, goods can be moved from one to the other but this might require adding money to one or both.
Figure~\ref{fig:Splot} shows contours of the entropy for {the }division of goods and money between two Cobb-Douglas economies with no external trader.  The allowed transitions are those that do not decrease the total entropy.

The reader may recognise (\ref{eq:SCD}) as the entropy of an ideal gas (discussed earlier), where $N$ is the number of molecules, $G$ represents volume of the container, $M$ represents internal energy, $\alpha=1$, and $\eta = 1/(\gamma-1)$ for ratio $\gamma$ of specific heats (e.g.~$\gamma=\frac75$ for a diatomic gas), modulo a conventional additional term $N\gamma/(\gamma-1)$ that has no effect if $N$ is constant.  Indeed, that analogy motivated the route to deriving the entropy function in (\ref{eq:SCD}). 
The analogy should not be pushed too far, however.  In particular, in the more general case where there are different types of good, the quantities of the types of good correspond most naturally to the quantities of different types of molecule in the gas, rather than to the volume of the mixture of gases.\footnote{It is interesting to consider the possibility that, instead of amount of goods, the number of agents in an economy may be more closely analogous to volume. Thus, population size could be represented explicitly, and allowed  to change, in the short term through migration, and in the longer term, through births and deaths. Questions of population change and flow are, of course, of considerable relevance to analysis in political economy and this is a direction deserving study. Note, though, that the current analysis allows the possibility that agents own goods and money in more than one economy, so it is not yet clear what it would mean to assign them to a particular economy.}

One could ask why the entropy of the trader was not considered in this analysis, but the assumption that the trader has unlimited goods and money can be regarded as saying that its entropy is independent of how much goods and money it has.
More precisely, we are assuming that the trader has no preferences for amounts of goods and money, so all changes are allowed, which means all its states must have the same entropy.  This conclusion is true \rev{in our example}, at least to a close approximation:~for financial contact with the trader we gave the trader exponent $1$, but that is negligible compared with the $N\eta$ for the CD economy.  Strictly speaking, to talk about entropy of the trader we require  the trader to be an extensive trading economy.

This toy example also has wider significance. We have stressed that thermal macroeconomics, like thermodynamics, is defined purely at the ``macro'' level, and does not depend on specific microfoundational assumptions. This makes macro-analysis possible for systems for which a detailed microfoundational analysis is infeasible---which will typically be true for realistic economic systems, in the light of the heterogeneity and complexity of human behaviour. 
It is nonetheless reassuring to find an explicit illustration of how there can be a bridge between the micro- and macro-accounts, where the micro-foundations are sufficiently simple to be tractable---just as statistical mechanics provides a micro-foundation for thermodynamics of sufficiently simple physical systems (e.g., ideal gases). Note this toy micro-economic model is also the starting point for the extensive tests of TM by computer simulation reported in \cite{LMC}\rev{, which generally show good agreement between theory and simulation.}

A purely macro-analysis is sufficient to derive the second law and a variety of specific economic results\rev{, as we will show} in following sections; and macro-quantities, such as economic entropy and temperature can in principle be measured directly; these issues will be discussed further below. But a micro-foundation, where available, is nonetheless required to explain the specific form of the entropy function $S$, and hence the detailed behaviour of an economic system. 

This raises the interesting question of the impact of different stylized micro-foundational assumptions on aggregate behavior and how this behaviour aligns with the macro-description in a thermal macroeconomic framework, and in particular, the question of how different micro-foundational assumptions modify the entropy function. 
Many such assumptions are possible. Starting from our Cobb-Douglas model, for example, different agents $i$ could have different $\alpha_i$ and $\eta_i$ {(this leads to relatively straightforward extension of the above results, in particular the entropy has the averages of $\alpha_i$ and $\eta_i$ in place of $\alpha$ and $\eta$)}.  Exchanges could be restricted to change of goods and money of opposite signs. Agents could make just a fraction $f_i$ of their holdings available at each interaction.
There can be more than one sort of good (in addition to money). The utility function might incorporate substitution effects or complementarity effects.
Goods and money might come in discrete amounts. Moreover, agents might be subject to behavioural biases, such as a status quo bias in favour of whatever goods they hold at the time \cite{SZ} or tendencies to compare with, or imitate, their neighbours. We anticipate that, unlike this toy example, the preferences of agents in general result in an entropy function that is not just a sum of functions of the quantity of each type of good. Examples of some of these effects are treated in \cite{LMC}.

It is interesting to note that there is, to our knowledge, no parallel conventional macroeconomic theory which captures the behaviour of the CD economy. \rev{Thus, the TM framework appears to provide a powerful explanatory framework for understanding at least some economic systems. It is to be hoped that the additional explanatory power of thermal macroeconomics will also extend to more realistic economic models (and of course to the real economy).} While the Cobb-Douglas set-up explored here yields analytically tractable results, the implications of more complex micro-assumptions will not typically be analytically tractable and will require simulations. This is an important direction for our on-going and future work. In the rest of the present paper, we largely set microfoundations aside, and return to explore the implications of the purely ``macro'' TM theory.

\section{Economic Temperature}
\label{sec:temperature}

The second reward, at which we hinted in Section~\ref{sec:fineqm}, is that there is a positive function $T$ of state of all simple {economies} such that when two simple {economies} $A,B$ are in financial contact there is nett money flow from $A$ to $B$ if and only if $T(X_A)>T(Y_B)$, and $A,B$ are in financial equilibrium if and only if $T(X_A)=T(Y_B)$.  
In keeping with thermodynamics again, we call $T$ {\em temperature}. \rev{We noted above that we can intuitively think of temperature as inversely proportion to the marginal aggregate utility of money; and from this point of view, the flow of money from ``warmer'' to ``cooler'' economies makes sense---money flows from where the marginal unit of currency has the greater ``value.'' This provides a view of the value of money, and derived notions such as inflation, which has a well-defined meaning, and is not dependent on, for example, the specification of sometime arbitrary choices of the ``basket of goods'' that a sum of money can purchase.}

\rev{To make these intuitions rigorous, we begin by noting that it follows from our axioms that} $S$ is differentiable with respect to $M$ (Theorem 5 of \cite{LY}), and a function $T$ satisfying the above is given explicitly by $$\frac{1}{T} = \frac{\partial S}{\partial M},$$ holding amounts of all goods fixed except money $M$.\footnote{We note that it is important in such ``ceteris paribus'' claims to specify precisely what is held constant.  When we come to Section~\ref{sec:cross}, for example, it will be essential to be clear about this.}

Note that temperature $T$ has unit $\odot $ over entropy, which depends on the scales for both money and entropy.  
Nevertheless, ratios of temperatures are uniquely defined, independently of the scales. 
This implies that one can fix a scale for temperature by one reference system in a reference state.\footnote{In physics, the standard reference state is the triple point of water, the temperature and pressure at which water can exist as a solid, liquid or gas).}  Then, as promised, a fixed scale for entropy follows:~it is the scale such that the change in entropy $\Delta S$ is $\Delta M / T$ for an infinitesimal change $\Delta M$ in money with fixed goods.  The unit of temperature is arbitrary (just as the unit of temperature, the Kelvin, is arbitrary in physics), but then entropy has units of money over temperature.  We consider it simplest to measure temperature in units of money.  For example, the reference system could be a Cobb-Douglas economy with $1\odot$ per agent and exponent $\eta>0$ for money, and then, as we will see shortly, it makes sense to define its temperature to be $\eta^{-1}\, \odot$.
This has the consequence that entropy becomes unitless.

Following Skilling (personal communication) for physical thermodynamics, we call $$\beta = \frac{1}{T}$$ the {\em coolness} of the economy.
Its economic interpretation is the marginal aggregate utility of money (analogous to the marginal utility of money for an individual agent, a concept that goes back to at least \cite{H}).

It is useful to refer back to the toy example of Section~\ref{sec:toy}, with Cobb-Douglas agents where the probabilities of trades are proportional to their utilities $U(m,g) =  g^{\alpha-1} m^{\eta-1}$, for amounts $g$ of good and $m$ of money. 
Then the resulting entropy function $S = N \log (M/N)^\eta (G/N)^\alpha$ implies that
$$\beta = \frac{N\eta}{M},\quad T = \frac{M}{N\eta}.$$
Note that the toy economy has the special feature that temperature is proportional to the amount of money per agent and independent of the amount of goods.
To give an interpretation of temperature here,
if one multiplies both exponents $\alpha$ and $\eta$ by a constant $c > 1$ then the probability rates and the stationary distribution, proportional to the utilities, will be more strongly dominated by transitions corresponding to maximising utility. Thus, with increasing $c$, the degree of noise will decrease, and the transitions will be closer to pure utility-maximizing trades. 
If $\alpha$ and $\eta$ are multiplied by $c$, the temperature $T=\frac{M}{N\eta}$, is divided by $c$.  Thus we see for this example that temperature quantifies deviations from utility maximisation by the agents.  So standard economics {with noiseless utility-maximizing agents} might be considered to be the case of zero temperature.\footnote{Note though that the context in which Hotelling \cite{H} introduced marginal utility of money was utility maximisation subject to a given budget; the marginal utility of money entered as a Lagrange multiplier to relate changes in money to changes in utility and there is no reason for it to be zero.} 

We will see that temperature being proportional to amount of money and independent of amounts of goods is a feature of any economy with ``pure money'', to be discussed in Section~\ref{sec:puremoney}. 


When economies $A$ and $B$, with $T(X_A)>T(Y_B)$, are put into financial contact there is nett money flow from $A$ to $B$ before they equilibrate.  If the changes are made quasistatically (that is, via a sequence of equilibrium states) so that the changes are reversible, then the money flow $dM_{AB}$ from $A$ to $B$ in each step is given by $T_B dS_B = dM_{AB} = -T_A dS_A$.  Then we see that $A$ loses entropy, $B$ gains entropy, and in fact $B$ gains more than $A$ loses, because $B$ is at a lower temperature.
\rev{This is true independently of the way in which} 
the changes are made. The cooler economy gains entropy at the partial expense of the hotter one. On the other hand, at the level of individual agents, the benefit flows in the opposite direction:~agents with money in the cool economy, where money initially has more value, find that the value of their money is eroded by an influx of money from the hot economy; conversely, 
agents who move money from the hot to the cool economy find the value of their money goes up, which is of course why they have the incentive to do so.\footnote{The government of an economy might attempt to prevent such flows by imposing capital controls.} Note that the coolness of an economy should not be confused with the buying-power of money, because the latter depends also on what goods are available.

We can simulate financial contact of two of our toy economies, with exponents $\eta_A, \eta_B$ for money not necessarily equal, by having agents $i \in A$ interacting with agents $j \in B$ at some rates $k_{ij}$ (many of these coefficients could be zero; it is enough to have non-empty subsets of $A$ and $B$ in financial contact) and dividing their combined money according to distribution proportional to $m_i^{\eta_A-1}m_j^{\eta_B-1}$ (a Beta-distribution again).
On average, this results in a split of the money between $A$ and $B$ in the ratio $\eta_A : \eta_B$. 


Note that because the ideal trader of our accessibility notion has constant entropy, their temperature is infinite.  This might seem to imply infinite money flow from the trader on putting them in financial contact with an economy, but we do not use the trader in this way.  Rather, the trader makes available only a finite pot of money, or decides at what price to buy or sell.

To end this section we discuss how the concept of temperature depends on the choice of num\'eraire.  Suppose one economy $V$ has a scarcity of apples and surfeit of bananas, and the other $W$ has the opposite. If apples are used as money, then financial contact will mean that apples flow from $W$ to $V$, and hence that $W$ is hotter than $V$. But if bananas are used as money, then financial contact will mean that bananas flow from $V$ to $W$, and hence that $V$ is hotter than $W$. Thus, even the ordinal notion of temperature is dependent on the choice of which good we take to function as money. Note, though, that the choice of money does not affect which state-transitions are possible in the economy (subject to the desirability assumption of Section~\ref{sec:exchange}), thus that \emph{entropy}, and in particular the second law, is independent of the choice of num\'eraire (up to 
\rev{orientation-preserving} affine transformations).

\section{Money capacity and inflation}
\label{sec:moneycap}

Having obtained a concept of temperature for an economic system, we define a subsidiary notion that will be useful in analysing how changes in the amount of money in an economy affect its value.  Following the terminology of heat capacity for a physical system, we define the {\em money capacity} of an economic system to be
$$C = \frac{\partial M}{\partial T}$$
for changes in temperature keeping amounts of all goods fixed except money.\footnote{Note, though, that this quantity is a derivative, and hence might more aptly be called marginal money capacity (and the physical concept would analogously be marginal heat capacity). As the physics terminology is established we stay with ``money capacity'' in the economic context.} 
From the viewpoint that $M$ and $G$ are the independent variables, it is better to define $C$ as the reciprocal of $\frac{\partial T}{\partial M}$ at constant $G$.
Money capacity has a particularly simple form in our toy economy, $$C=\eta N.$$

Note that money capacity is positive, by the following argument.  From \cite{LY} and Section~\ref{sec:convexity}, $S$ is concave, in particular with respect to $M$ at constant goods, so $\partial^2 S/\partial M^2 \le 0$.  Now $\partial S/\partial M = \beta$, thus concavity of $S$ gives $\partial \beta/\partial M \le 0$.  But $\beta = 1/T$, so $\partial T/\partial M \ge 0$ and hence $C$ is positive or infinite.

Money capacity is relevant to various economic questions. For example, money capacity is inversely related to the inflationary impact of adding money to an economy, while keeping the quantities of all other goods the same.  The added money could flow in from a neighbouring economy or from the government printing money. 
Recall that the marginal utility of money, $\beta$, is the inverse of temperature, i.e., $1/T(M)$ (writing $T$ as a function of $M$, as quantities of goods are held constant).  Thus for one unit of money added to the economy (with no change in other goods), the {change $\Delta \beta$ in the marginal utility of money is $\frac{\partial(T^{-1})}{\partial M}$ = $-T^{-2}\frac{\partial T}{\partial M}$ = $-\frac{1}{CT^2}$, so that} the relative change is
$$\frac{\Delta \beta}{\beta} = -\frac{1}{CT}.$$  
Notice that this reduction in the value of money corresponds to inflation:~indeed, the money capacity is a key ingredient deciding the inflationary effect of adding money to an economy, at least in our exchange economies.  We can define the {\em inflation rate} to be $$I =\frac{1}{T}\frac{dT}{dt}$$ (equivalently $-\frac{1}{\beta}\frac{d\beta}{dt}$) with $t$ representing time. Thus for rate of increase $\dot{M}$ in money, $I = \frac{\dot{M}}{CT}$. Note that this definition of inflation is entirely abstract and not dependent on choices of specific price indices and their associated weighted baskets of goods.

Money capacity also provides a useful relation between temperature and entropy, namely 
\begin{equation}
\frac{\partial S}{\partial T}=\frac{C}{T},
\label{eq:STCT}
\end{equation}
where again the derivative is taken with fixed amounts of all goods except money.
The proof is that the change $dM$ in money can be written as $T dS$ from the formula  $T^{-1}=\frac{\partial S}{\partial M}$, and also as $C dT$ from the formula $C^{-1}=\frac{\partial T}{\partial M}$.
From relation (\ref{eq:STCT}) and positivity of $C$ we deduce that $S$ increases with $T$.
In words, the increase in aggregate utility from unit increase in temperature is the money capacity divided by the temperature.

Money capacity also turns out to be important in quantifying fluctuations of the amount of money in a subsystem, as will be discussed in Section~\ref{sec:thermometer}.

Finally, we comment that the above proof of positivity of money capacity $C$ includes the possibility of $C=+\infty$.  This is a situation analogous to physical systems at a phase transition.  For example, on heating a solid at constant rate, its temperature rises until it reaches its melting temperature; the temperature remains constant there until it has all melted and then goes back to rising.  The heat capacity is infinite when part-melted and part not.  The possibility of phase transition in macroeconomics, without breaking our unique equilibrium assumption A0, merits exploration. It is possible, for example, that something akin to a phase transition occurs when inflation is so high that coins become as valuable for their constituent metal as for their face value; or, more extremely, as in Weimar Germany, bank notes are used as cheap paper or fuel.

\section{Values and market prices for goods}
\label{sec:price}

The third reward is that for each type of good there exists a market price, $\mu$, for exchange between the good and money, at which, if offered by an external trader, the economy will make no nett exchange between the good and money with the trader. Thus, we will arrive at an analogue of the celebrated general equilibrium result of Arrow and Debreu, albeit with a simple exchange economy, and with a statistical notion of equilibrium. Notice that this approach makes no assumptions about the rationality or otherwise of individual behaviour. 

To derive this result, we first define a ``value'' for each type of good, actually a marginal value for one more unit of the good.\footnote{It is important to note that this value is measured in units of entropy, not money, just as we defined the value of money in units of entropy.}
For each type of good, with amount $G$, $S$ is differentiable with respect to $G$ (under the axioms of Section~\ref{sec:convexity}). 
For a good other than money, let $$\nu = \frac{\partial S}{\partial G},$$ 
keeping the amounts of all other goods fixed, including money.  We call $\nu$ the {\em value} for the good. This is an important concept in our theory.  It is the analogue for goods of the coolness $\beta$ for money (Section~\ref{sec:temperature}). Our quantifications of value of goods and of money provide a new theoretical basis for discussions of value that go back to Adam Smith's ``Wealth of Nations'' (see, e.g.~\cite{Pag}).

Note that in contrast to money, however, the value of a good might be negative.  This caters for goods that are undesirable or perhaps desirable if other types of good are scarce but not if they are plentiful. For example, a bulky good might be desirable if no alternative is available but undesirable if compact alternatives are cheap and plentiful; or a good might be desirable only when complementary goods are available (side-cars may be desirable if motorcycles are available but undesirable ``junk'' if not).

If we allow changes in the amounts $G$ of good and $M$ of money by connection to an ideal trader then 
\begin{equation}
dS = \beta\ dM + \nu\ dG.
\label{eq:dS}
\end{equation}
{There will be no spontaneous nett exchange of goods and money between the economy and the trader if and only if the exchange between money and goods is reversible, i.e., leads to no change in entropy ($dS=0$). By rearranging (\ref{eq:dS}) with $dS=0$, we can see that a} reversible exchange  between money and the good is possible if and only if the trader\footnote{Note that as mentioned in Section~\ref{sec:toy}, the trader has constant entropy so its state does not affect the allowed changes.} is offering what we call the {\em market price} $$\mu = {-\frac{dM}{dG}} = \frac{\nu}{\beta}.$$

For the example of Section~\ref{sec:toy} where we explored our simple Cobb-Douglas economy, 
$$\nu = \alpha \frac{N}{G}, \quad \mu = \frac{\alpha}{\eta}\frac{M}{G}.$$
This $\mu$ corresponds to the critical price $\mu_c$ (\ref{eq:muc}) we found in Section~\ref{sec:toy}.

In view of the centrality of the second law in thermal macroeconomics, we have focused on changes in economic entropy $S$ in (\ref{eq:dS}). By contrast, it is common in thermodynamics to give priority to energy rather than entropy. In an economic context, the analogue of the first law of the conservation of energy in physics is the much less interesting conservation of money. Indeed, in an exchange economy, money and all goods are conserved by assumption. Nonetheless, if we choose to give priority to money over entropy (as is the case in standard economics) then the analogous relation to (\ref{eq:dS}) can be written
\begin{equation}
dM=T\ dS-\mu\ dG.
\label{eq:dM}
\end{equation}

The relations (\ref{eq:dS}) and (\ref{eq:dM}) extend naturally to a vector $G$ of amounts of goods and corresponding covectors $\nu$ of values and $\mu$ of prices.
Furthermore, the relation (\ref{eq:dS}) extends to exchange between any pair of goods, not necessarily including money. 
The coolness $\beta$ is simply the special case of the value for money.  For exchange of two types of good $G_1, G_2$, then $dS = \nu_1 dG_1 + \nu_2 dG_2$, so reversible exchange happens if and only if the trader offers the exchange ratio $\nu_2/\nu_1$ for amount of $G_1$ per amount of $G_2$.

\section{Pure money}
\label{sec:puremoney}

What differences does it make 
to our theory if money has no intrinsic value (e.g.~paper or digital money), as opposed to some good like gold? Just as did Hume years ago \cite{Mu},
we argue in this case that for fixed vector $G$ of amounts of goods, (i) the price $\mu$ of any type of good, and (ii) the temperature $T$, should be proportional to the amount $M$ of money in the economy, because the value of a unit of money will be determined by how large a fraction it is of the total quantity of money in the economy. 
Thus, declaring each aurum to be $\lambda$ new-aurum should make no difference other than to multiply all prices and the temperature by $\lambda${---indeed, it is scarcely more than a superficially trivial notational change, like deciding to use the word ``dollar'' to refer to cents}.\footnote{The situation is more complex, though, for less trivial changes of currency, such as when individual European countries joined the Euro. In such cases, there are other factors in play alongside a simple rescaling of the currency, and these may have non-trivial economic consequences, including on real prices. For example, in the short term, traders may ``round up'' prices differently in the new currency; and there will be anticipated and real impacts of long-term economic convergence.}

The above two conditions (i) and (ii) imply that the value $\nu = \mu/T$ for each type of good is independent of $M$, and that the coolness $\beta $ is $K(G)/M$ for some function $K$ of  $G$. Using $\nu_j = \frac{\partial S}{\partial G_j}$ and $\beta = \frac{\partial S}{\partial M}$, we deduce from the first condition that $\frac{\partial \beta}{\partial G_j}=0$.  Thus, from the second condition, $K$ is independent of the amounts of goods. As a result, 
\begin{equation}    
S = K \log M + F(G)
\label{eq:Spure}
\end{equation}
for some constant $K>0$ and function $F$ of the amounts of goods. 
An example is the toy economy of Section~\ref{sec:toy} (where also the other good is pure). From (\ref{eq:Spure}) we see that purity of money can be viewed as a macro-analogue of the microeconomic concept of ``quasilinear utility'' \cite{SF}.

The above holds for an economy with pure money and a fixed number $N$ of agents.  To generalise to a variable number of agents, the dependence on $N$ should be made extensive, that is
$$S = N(k \log M/N + f(G/N))$$
for some constant $k$ and function $f$.

Money is likely to be ``pure'' to a good approximation in modern economies, and we discuss the implications at various points below. One implication that we can make already is that for pure money, the money capacity $C=K$ 
is constant.  In particular it is never infinite.  But historically, some currencies have certainly not been pure, for example silver.  The requirement to use silver as currency for trade with China clashed with its increasing use in the UK as silverware for serving food and drink, a problem that notably Adam Smith and Isaac Newton addressed \cite{Fay,Pag}.

\section{Financial thermometers \rev{and other macroeconomic measurements}}
\label{sec:thermometer}

\rev{A crucial question about our theory is how one might hope to measure macroeconomic quantities such as temperature, goods' values and entropy for a real economy. This will require extending our approach beyond exchange economies, to include production, consumption, manufacturing, finance, and so on, a project which is on-going and that we discuss further below. But measurement challenges can nonetheless be addressed in the limited case of exchange economies, in ways that seem likely to extend naturally to more complex cases.
We focus on the measurement of temperature, ending the section with short comments about measuring other quantities.  We have tested the proposed measurement procedures for temperature and value on simulated economies \cite{LMC}, and are in the process of testing the measurement of entropy.}

Having deduced the existence of temperature in an equilibrium economy (and hence a concept of the value of money, and the meaning of inflation, as in Sections~\ref{sec:temperature} and \ref{sec:moneycap}), a natural question is whether and how temperature can be measured.  A difficulty is that in practice an economy is in general not in equilibrium because it has inputs and outputs. Moreover, rather than the pure exchange economies we are treating here,  real economies have production of raw materials, manufacturing, consumption, services, and so on. Nonetheless we address the question 
for equilibrium exchange economies, with the hope of later generalisations.

Here, we focus on the question of how financial temperature can be measured in principle---i.e., how to construct an ``ideal'' financial thermometer. We leave the question of how well such an ideal can be approximated in practice for later work. To measure the temperature of an economy $A$, one could imagine floating a relatively small economy $s$ on a ship, putting it in financial contact with $A$ (with no transfer of goods) and studying the changes to $s$ to infer the temperature of $A$.\footnote{Calling $s$ a ship thermometer is a natural metaphor when measuring the temperature of mainland economies with sea frontage, but of course applies quite generally.} The smallness of the ship thermometer $s$ is needed (just as for a physical thermometer) so that the contact with $s$ does not have a significant effect on the temperature of $A$, the economy whose temperature is being measured. 
    
The intuition is that money will flow into, or out of, the ship thermometer from the mainland, bringing the temperature of $s$ into balance with the temperature of $A$. If $T_s < T_A$, people in $A$ will tend to move their money to $s$ because goods are cheaper there (that is, the aggregate marginal utility of money is greater); this influx of money will in general raise the temperature (by positivity of money capacity, Section~\ref{sec:moneycap}), until the temperature reaches that of the mainland.  If $T_s > T_A$, the opposite will happen. 
Then the amount of money on the ship after equilibration is an increasing function of the temperature, so with calibration it can be used to measure the temperature.

Perhaps the amount of money on the ship is not readily visible, however.  In this case, one could instead measure the price of some good on the ship, which is more likely to be visible (after averaging), and convert it to temperature.
One can make a plausible story that if $T_s < T_A$ then the influx of money can
be used to buy goods in $s$, thereby driving up their price until $T_s$ reaches $T_A$, though we have not proved that the price has to increase.
If $T_s > T_A$, the opposite can be expected to happen.
Thus the market price of each good on the ship will settle down to a function of the temperature of the mainland $A$, so one can read off the temperature from the market price after calibration. 

This calibration is very simple in the special case where the currency is ``pure'', with no intrinsic value except as a medium of exchange (as in the previous Section). The amount of pure money in an economy is proportional to the temperature of that economy. Thus, if the amount of money in the ship thermometer is easily measurable, it provides a direct measure of the temperature of $A$. 
Indeed, from the previous section on pure money, $T=M/K$, so the temperature on the ship is proportional to the amount of money.  Furthermore, for an economy with pure money, the price of a good is $\mu = T \frac{\partial F}{\partial G}$.  The amounts of goods on the ship do not change, only money, and $F$ is independent of $M$.  So we deduce that the price of any good is proportional to the temperature.\footnote{The ship thermometer with pure money is therefore analogous to an ideal gas thermometer in physics, for which volume is proportional to temperature at constant pressure.
Note that if the currency is not pure, but a good with some intrinsic value such as gold, then the picture is not so simple. Then an influx of gold into the ship may change the relative prices of other goods on the ship, which might be complements to, or substitutes for, gold. For example, plentiful gold might drag down the value of silver (if silver and gold are substitutes) as compared with, say, butter. These same interactions with other goods mean that the value of gold cannot be read off merely by measuring the ratio change in the quantity of gold on the ship.} 

Note that this measure of temperature agrees with a conventional price-index for comparing value of money in two economies, say $A_1$ and $A_2$, if the goods in $A_1$ and $A_2$ are the same but all prices differ by a fixed factor. But our approach is much more general. Indeed, there need not be any overlap whatever between the goods in $A_1$, $A_2$, or indeed $s$.

As an example of a ship thermometer, we can use the toy economy of Section~\ref{sec:toy}.
Combining previous results, $\mu = \frac{\alpha N}{G} T$, and noting that $N$ and $G$ remain fixed, we see that $\mu$ gives a result proportional to $T$.  So the market price of the good on the ship can be used to measure the temperature of the mainland economy.  
Alternatively, if one can measure the amount $M$ of money on the ship then one can use the more direct formula $T = \frac{1}{N\eta} M$, proportional to $M$, but almost any quantifier on any type of ship thermometer would respond in some nontrivial way to the temperature of $A$.

As mentioned above, the scale for temperature is arbitrary (at least from our construction).  So what can really be measured is ratios of temperatures. Our ship thermometer allows us to do this.

If we choose to use the amount $M$ of money on the ship to infer the temperature, it is worth noting that if the ship thermometer is too small then fluctuations in $M$ will make the temperature difficult to measure (and indeed, prices on the ship will be difficult to measure also).  Indeed, the analogue of Einstein's formula for fluctuations in energy of a small subsystem, derived from the canonical ensemble in statistical mechanics, is that the variance $$\Var\, M = C T^2,$$ 
where $M$ is the amount of money on the ship, $C$ is the money capacity of the ship ({more precisely}, at constant $\nu$ rather than $G$, but for a CD economy there is no difference), and $T$ is the (common) temperature of the ship and $A$.  Using the toy economy of Section~\ref{sec:toy} for the ship, the measured temperature $T_m = \frac{M}{\eta N}$ and $C=\eta N$, so $$\Var\, T_m = \frac{\Var\, M}{\eta^2N^2} = \frac{T^2}{\eta N}.$$  
It is natural to compare $\Var\, T_m$ to $T^2$, hence to divide both sides of the above equation by $T^2$.
Thus the conclusion is that the temperature will be measured more precisely if the number $N$ of agents on the ship is larger and if their exponent $\eta$ for money preference is larger.  
The results generalise to arbitrary ship econom{ies}:~since $\frac{\partial T}{\partial M} = 1/C$, $\Var\ T \sim T^2/C$ for large $C$ as long as it is still much smaller than the money capacity of the economy being measured.
As with the design of physical thermometers, however, there are other considerations, such as how much the financial contact with the ship disturbs $A$ and how fast the ship temperature equilibrates with that of $A$.  We have explored these, and other, issues in a variety of simulated exchange economies reported in \cite{LMC}.

Note that the above analysis of fluctuations in money for a subsystem depends on an identification of entropy as representing the logarithm of available volume.  That is true for the toy economies, where the entropy turns out to be the logarithm of the normalisation factor for the stationary distribution (section~\ref{sec:liberty}), and a general proof under suitable conditions will be presented elsewhere \rev{\cite{M25b}}.  It is also not a
consequence of standard thermodynamics (though see \cite{Mi} for a proposed extension).  But assuming the {identification of entropy with logarithm of available volume}, the discussion of fluctuations
extends to other quantities, such as price:  
$$\Var\, \mu = -T\frac{\partial \mu}{\partial G}_{|_S},$$ 
{along standard lines for statistical mechanics, e.g.~Ch.15 of \cite{PB}.}
The partial derivative will be related to other quantities in Section~\ref{sec:cross}, equation (\ref{eq:diagflex}) (and shown to be non-positive and to contain another factor of $T$).
{The above equation assumes that price is quantified from amounts of goods and money in the subsystem via $\mu = \nu/\beta = \frac{\partial S}{\partial G}/\frac{\partial S}{\partial M}$.}\footnote{See \cite{HS} for a discussion of pitfalls in talking about fluctuations of intensive rather than extensive quantities.}
There are even fluctuations in the entropy of a subsystem, and there are computable covariances between fluctuations (but the relative sizes of all of these scale like $O(1/N)$ as one approaches the thermodynamic limit, as was made explicit for $T_m$ above).

{Note also that the above formulae for variances both give results proportional to $T^2$.  This shows that, under the assumption that entropy represents logarithm of available volume, temperature is not just the inverse marginal utility of money but is also connected fundamentally with the level of activity at equilibrium.}

{A natural question is whether one obtains the same temperature wherever one measures it in an economy.  One might imagine that if the agents in one port have different preferences from those in another port then one would find different temperatures in the two ports.
But under our assumptions that the economy is in equilibrium and there are no internal barriers to money flow in an economy, the equilibration under financial contact with the ship thermometer produces the same state on the ship in both ports.}

We reiterate the contrast between our approach to measuring economic temperature (remember this is the inverse of the marginal utility of money) {and conventional measures of value of money within countries (e.g., the Consumer Price Index) or} cost of living ratios between countries (e.g.~Penn World Tables, World Bank's International Comparison Program), which depend on rather arbitrary choices of baskets of goods. \rev{The approach also avoids somewhat arbitrary decisions in practical inflation indices, concerning whether to weight prices changes for the items in the basket of goods by their initial quantities sold in a time period (Laspeyres Index), their final quantities (Paasche Index), or a blend of the two (e.g., the widely used Fisher Index). Moreover, the approach outlined here also provides an alternative conceptual standard to the ``ideal'' Cost of Living Index of modern economics which is defined at the ``micro'' level, aiming to capture the change in the amount of money required for a household to achieve a certain level of utility \cite{SN10}. Estimating this quantity even for a single household is challenging; and aggregating such estimates across the economy is even more difficult. Switching to a purely ``macro'' approach based on entropy (interpreted as aggregate utility, rather than utility of individual households) provides an interesting alternative strategy, which abstracts away from individual households and comparable baskets of goods.} Indeed,  there is no need for two economies to have any good in common (except money) in order for us to take their temperature ratio.  For realistic applications, we need to extend to the case where different economies use different currencies.  We make some remarks on this in Section~\ref{sec:currencies}. \rev{If practicable macro measures of inflation can be developed, it will be interesting to compare the results of these measurements with those obtained from existing methods, to establish whether they might could be developed to provide useful complement to array of traditional approaches to measuring inflation.}

{To close this section, we discuss {some other} measurement questions. {Firstly,} how one could measure the market price of a good in the ship thermometer? This might be a more practical method for measuring temperature of a real ship economy than measuring the amount of money on the ship, and also 
has independent interest.  One way is to have an external trader propose a price and to monitor the {time-averaged} nett flow of goods and money.  If there is nett flow of money to the trader, the trader increases the price; if in the other direction, the trader decreases the price.  By this process the trader should settle on the market price.\footnote{This is similar to the use of a Wheatstone bridge in physics to measure resistance.}
Again, the price should not depend on the particular agents to which the trader is connected.}
\rev{In the context of simulations, if the ship economy is CD then one can simply measure $\mu = \frac{C_G M}{CG}$, where $C_G$ is the goods-capacity of the ship; this is the method used in \cite{LMC}.} 

One can propose meters for measuring other aspects of an economy than temperature.  For example, the value of a type of good in the economy can be measured by putting the economy in contact for that type of good with a small economy, waiting for equilibration, and measuring the value of the good in the small economy, using some surrogate for value after calibration of the small economy.  Or one could measure temperature and price simultaneously by putting the economy in contact for both money and the good and measuring them in the small economy, again after calibration.

\rev{Establishing methods for measuring temperature, value and price makes it possible to measure the change in entropy when the economy moves from one state to another. This therefore allows us to measure} the entropy function (up to an additive constant), \rev{by analogy with calorimetry in physics.}  The idea is simply that for quasistatic changes, $dS = \beta dM + \nu dG$.  So the change in $S$ along a quasistatic path is $\Delta S = \int \beta dM + \nu dG$, which can be estimated by measuring how $\beta$ and $\nu$ change with $M$ and $G$ along the path.  Recall that $\beta = 1/T$ and $\nu = \mu/T$.  \rev{This line of reasoning provides a critical test for when entropy is well-defined in an economic system:~that the estimated entropy change in moving from one economic state to another should be \emph{path independent}.} \rev{Just as in Physics and Chemistry, measuring entropy is also crucial in applying the theory to economic systems where micro-foundations are not known.  As noted above, tests of the approach outlined here are currently underway on simulated economies. Extending this approach to work with available economic data for real economies is an important challenge for future research.}


\section{Making money out of price differences}
\label{sec:Price differences}

If two economies have some good in common, as well as money, and the price of the good is different in the two economies, then it is clear that an external trader can make money from the price difference.  The trader buys goods in economy  $A$ at price $\mu_A$ where they are cheaper and sells them in the economy $B$ at price $\mu_B$ where they are more expensive (i.e., $\mu_A<\mu_B$).  For infinitesimal changes (so that the prices do not change),  one aurum purchases $1/\mu_A$ goods in economy $A$, which are then sold for $\mu_B/\mu_A$ aurum in economy $B$.  Thus the trader makes $\mu_B/\mu_A-1$ aurum.

In general, however, as trading continues, the prices will change because the amounts of goods and money in the two economies change.  How much money can the trader make in this way? 

In the simplest context with just one good besides money, the state of the joint system follows the path of increasing $G_B$ with $dG_A= -dG_B$, $dM_A = \mu_A dG_B$, $dM_B = -\mu_B dG_B$, $dM_T = (\mu_B-\mu_A)dG_B$ (using subscript $T$ for the trader) until $\mu_B=\mu_A$. The path is contained in the intersection of the surfaces of constant entropy for the two economies, because trade at market prices is reversible and so does not change entropy.  Thus one way to describe the final point is that it minimises $M_A+M_B$ subject to conservation of $S_A$, $S_B$ and $G_A+G_B$,  and the amount of money made by the trader is the difference between the initial and final amounts of money in the economies.

\begin{figure}[h]
    \centering
\includegraphics[height=2.5in]{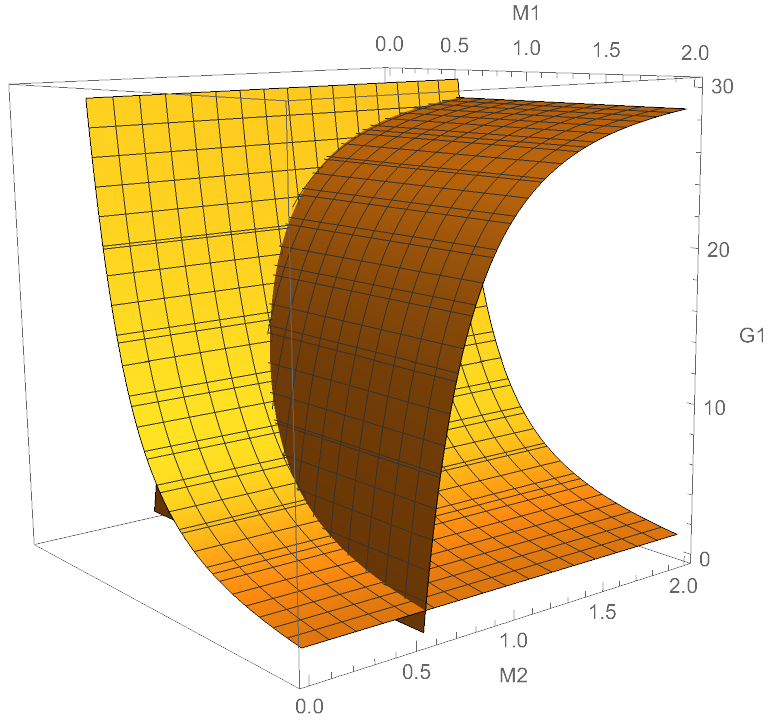}
\caption{The intersection of surfaces of constant entropy for the two economies. This intersection lies in the surface of constant total entropy shown in Figure~\ref{fig:AccRegion}.}
    \label{fig:cstS}
\end{figure}

Let us compute the maximum profit that a trader can make from the two Cobb-Douglas economies of Figure~\ref{fig:AccRegion} without decreasing the entropy of either economy.  Maximum profit is made by restricting to no change in either entropy $S_1, S_2$.  
Then this is a curve on the boundary of the accessible region.  See Figure~\ref{fig:cstS}.
The maximum profit is given by the tangency of this curve to a plane $M_1+M_2=$ cst. A computation with Lagrange multipliers (allowing different exponents $\alpha_i$ and $\eta_i$) yields the tangency at the state with $$G_1 = \frac{\alpha_1 M_1/\eta_1}{\alpha_1 M_1/\eta_1 + \alpha_2 M_2/\eta_2} G, \quad
G_1^{\alpha_1}M_1^{\eta_1} = e^{S_1/N_1}, \mbox{ and } (G-G_1)^{\alpha_1}M_2^{\eta_2} = e^{S_2/N_2}.$$  This does not have an explicit solution $(M_1,M_2)$, but the trader's profit is given by the difference between the initial and final $M_1+M_2$.

This description extends to the case of more than one tradeable good, where the total amounts of each good are conserved.  The first-order conditions for a minimum enforce equality of prices in the two economies for each good at the final point.  The existence of the final point and its connectedness to any initial condition with the given entropies requires some more analysis, of course.  In particular, it is proved in Section~\ref{sec:cross} that {the price of a good decreases as the quantity of that good increases by trade at market price} (i.e.~taking into account compensating change in money), and vice versa.  

We have seen that the maximum amount of money, $M_T$, the trader can extract through exchanging goods and money is achieved by minimizing $M_A+M_B$ subject to conservation of $S_A$, $S_B$ and $G_A+G_B$. But can the trader do better? From the second law of thermal macroeconomics, one might ask whether the trader can find a way of minimising the total amount of money left in the economies $A$ and $B$ subject to conserving goods and \emph{total} entropy, $S_A+S_B$, rather than the more restrictive condition of conserving the individual entropies $S_A$ and $S_B$.  This is indeed achievable by the trader, but not in general by the above strategy.  It requires a more general approach, to be explained in Section~\ref{sec:gains}.  To get there we first explain how one can make money out of temperature differences with no transfer of goods.

\section{Making money out of temperature differences}
\label{sec:Carnot}

The industrial revolution was powered by the observation that one can extract useful work from physical temperature differences. Moreover, in physics, one important consequence of the thermodynamic concept of temperature\footnote{as opposed to purely empirical measures of temperature.} is that it gives a formula for how much work can be extracted from temperature differences (more precisely from temperature ratios). Indeed, the efficiency of an ideal heat engine (called a Carnot cycle) is one way to introduce the thermodynamic temperature scale. So the process of extracting work from a temperature difference turns out to be both practically and theoretically very significant.

It is therefore natural to ask whether there is an interesting analogy in economics. 
As spelt out in the previous section, a trader can make money out of price differences by buying goods in an economy where they are cheaper and selling where they are more expensive.  

But the question here is narrower:~how much money can an external trader extract from a temperature ratio between two economies, without any nett change in amounts of goods in either economy. 
That is, how can the trader profit purely from the flow of \emph{money} from the hot economy to the cold economy (just as a heat engine extracts work purely from the flow of heat between hot and cold bodies)?  

To do this, we propose an economic Carnot cycle.  The trader will use a ``boat economy'' $B$ (essentially our ship thermometer of the previous section, but independent of its use to measure temperature), which can shuttle costlessly between two mainland economies $H$ (hot) and $C$ (cold) (perhaps across a narrow strait), with temperatures $T_H>T_C$, and put into financial contact with either. The trader will make reversible trades with $B$, but not with either $H$ or $C$;
there will, though, be flows of money from $H$ to $B$ and from $B$ to $C$.  

\begin{figure}[htb] 
   \centering
   \includegraphics[height=2.5in]{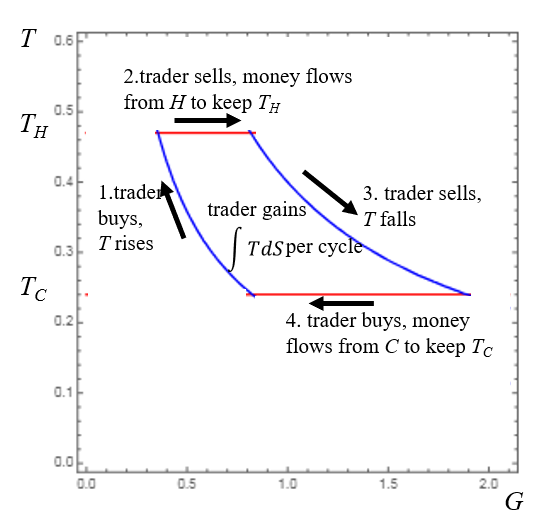}
   \caption{An economic Carnot cycle in the plane of amount $G$ of goods and temperature $T$ for an island economy with $\alpha=2, \eta=5/2$ operating between two mainland economies at temperatures $T_H=0.47$ and $T_C=0.24$.}
   \label{fig:Carnot}
\end{figure}

The trader can extract money from the economic temperature difference between $H$ and $C$ by applying the following four steps. First, the trader
moves $B$ near-reversibly to the temperature of $H$, by continually buying goods slightly above the current market price in $B$, thus gradually, and hence reversibly, pushing up prices on the boat until its temperature is $T_H$. Second, now that $B$ and $H$ are at the same economic temperature, the trader puts them into financial contact and begins selling goods to $B$ slightly under the market price. Now there is less money and more goods in $B$, marginally increasing the buying-power of money, and hence drawing in money from $H$, while keeping the temperature of $B$ at $T_H$. The third step is for the trader to disconnect the financial contact between $B$ and $H$, and to continue reversibly selling goods to $B$. Now the buying power of money on $B$ increases, as the boat contains many goods and little currency with which to buy them---i.e., the temperature of $B$ reduces to $T_C$. The fourth step is for the trader to put $B$ and $C$ in financial contact and to begin buying goods from $B$, thus marginally lowering the buying power of money in $B$ and thus leading to money to flow from $B$ to 
\rev{$C$}. The trader can then disconnect $B$ %
\str{to $L$} 
\rev{from $C$} and begin the cycle again. This process is illustrated in Figure~\ref{fig:Carnot}.

The nett effect of each cycle is to take some money from $H$ and distribute it between $C$ and the trader; the trader obtains money by feeding off the temperature difference.
If carried out reversibly, the trader takes a fraction $(1-T_C/T_H)$ of the money leaving $H$.  Namely, the money leaving $H$ is $-T_H dS_H$ and the money entering $C$ is $T_C dS_C$, and for reversible changes the total change in entropy is zero, so $dS_H = -dS_C$ ($B$ returns to the same state after a cycle so there is no nett change in $S_B$); hence the ratio of the money leaving $H$ to the money entering $C$ is $T_H/T_C$.  This is the same formula as for the efficiency of a physical Carnot cycle, because it relies on the same mathematics. In \cite{LMC}, we report on demonstration of a Carnot cycle in computer simulations of exchange economies, and explore its efficiency.

\begin{figure}[htb] 
   \centering \includegraphics[height=2.1in]{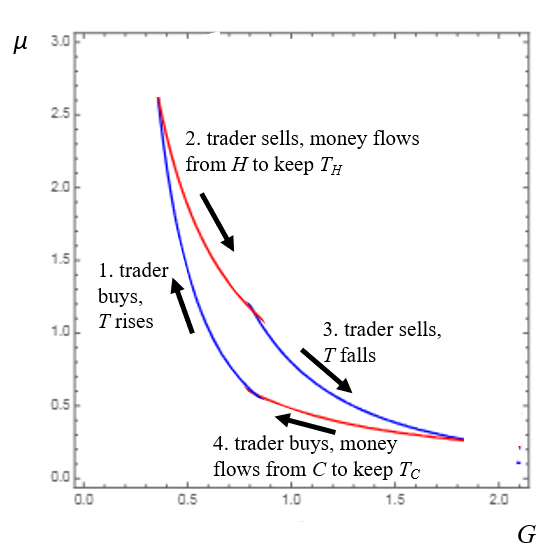} \quad \includegraphics[height=2.1in]{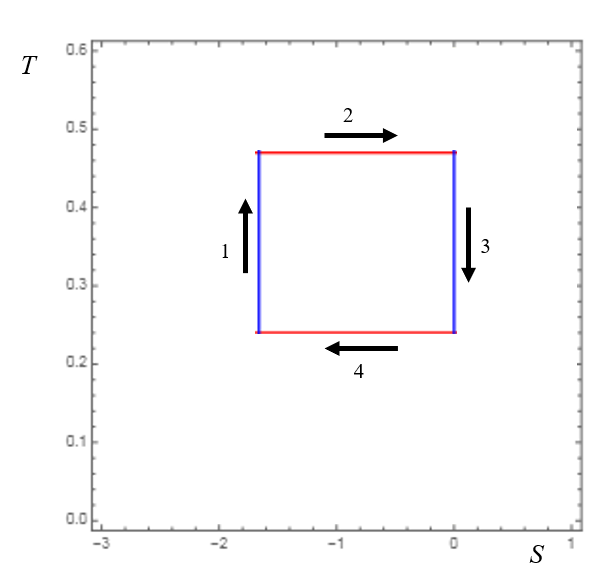}
   \caption{Carnot cycle replotted in the planes of (a) amount $G$ of goods and price $\mu$, (b) entropy $S$ and temperature $T$.}
   \label{fig:Carnotreplots}
\end{figure}

In physics, it is most common to plot a Carnot cycle in the plane of pressure and volume. That would correspond here to price and amount of goods.  But for some purposes the best is the plane of temperature and entropy.
So we replot the figure in these two ways in Figure~\ref{fig:Carnotreplots}. 

More generally, in a cyclic process, the money a trader can extract per cycle with reversible moves is the area of the loop in the $(S,T)$-plane, as in Figure~\ref{fig:Carnotreplots}(b).  To see this, denote the money and goods on the boat by subscript $B$ and for the trader by subscript $T$. For the boat economy, $dM_B = T\, dS - \mu\, dG_B$, and reversible moves imply that the change in the trader's money $dM_T= \mu\, dG_B$.   So we deduce that $dM_T = T\, dS- dM_B$.  But in a cycle, $M_B$ returns to its initial value, so $\int dM_T = \int T\, dS$.

Note that the above discussion in terms of a cycle assumes that the hot and cold economies are sufficiently large that their temperatures are not affected by the flow of money in one cycle.  In reality, the temperatures of the two economies would change slightly and the cycle would have decreasing amplitude in the temperature.  \rev{Both cases are treated in \cite{LMC}.}

A Carnot cycle can also be used in reverse, just as in physics (e.g.~refrigerator or heat pump), to create economic temperature differences by putting in money, or more generally to drive money flow against a temperature difference.
The trader puts in money $M_T$, the cold economy puts in $M_C$ and the hot economy receives $M_H= M_T+M_C$, with the ratio $M_H/M_C$ anything less than $T_H/T_C$.  In particular, the hot economy can receive anything up to $M_T/(1-T_C/T_H)$, a considerable enhancement on just receiving $M_T$ directly from the trader, especially if $T_C$ is only a little less than $T_H$. We call the ratio $M_H/M_T$ the {\em coefficient of performance}, by analogy with heat pumps.\footnote{Note that for refrigerators the coefficient of performance is conventionally measured instead as $M_C/M_T$, which is $M_H/M_T - 1$.} 

{These observations have potentially interesting practical implications. Clearly a large and rich trader can intervene directly to change the buying power of money in a small state (say an isolated island). For example, it can make a small state ``hotter'' (i.e, its money having less value) simply by sending money to the smaller state. But it can generate the same effect more efficiently if it has access to another island, and a ``boat'' which shuttles between the islands (or some equivalent mechanism). The trader can now more cost-effectively cause inflation in one state by simultaneously causing deflation in the other (even if the second small island is initially cooler, so that this flow of money is from cooler to hotter). The efficiency gain will be a factor up to a theoretical maximum of $1/(1-T_C/T_H)$ = $T_H/(T_H-T_C)$. If the value of money is similar in each economy (i.e., $T_H-T_C$ is small), the maximum efficiency gain can be arbitrarily large. It is interesting to ask whether approximations to such phenomena can occur in real financial markets through government or private spectulator interventions. A fuller analysis of such cases is likely to require an extension of the present account to allow that different countries have different currencies, a point we touch on briefly in Section~\ref{sec:currencies}.}

\section{Trade} 
\label{sec:trade}

A key question that has motivated us from the outset is how to determine which trades are viable between two economies. {In the previous two sections, we have considered special cases through which a trader can extract money from a pair of economies by exploiting differences in prices (Section~\ref{sec:Price differences}) and differences in temperature (Section~\ref{sec:Carnot}). Now we are in a position to develop a general account of trade, whether mediated by a trader or involving direct exchange between economies.}

According to the present analysis, nett exchange of goods and money between two economies can occur if and only if it does not decrease total entropy. Thus we obtain an analogue of the traditional notion of `gains of trade'. 
Yet the gains might not benefit each economy individually. 
We distinguish trade that increases the entropy of all parties by the term `mutually beneficial trade', or {\em MB trade} for short. Note that the mutual benefit here is at the level of economies, not individuals. So, for example, when economies at different temperatures are put into financial contact, money will flow from the hotter to the cooler. This may be beneficial at an individual level (because money has more value when transferred from the hotter to the cooler economy), but the outflow of money will transfer entropy (aggregate utility) from the hotter economy to the cooler. Of course, if an economy has a government, that government might try to maximize the aggregate utility of its own economy, and hence might attempt to put restrictions on such ``capital flight''.{\footnote{There are also many other possibilities. For, example, an external trader could operate a Carnot cycle to exploit a temperature difference and then use the proceeds to make unprofitable trades. Or it could operate a Carnot cycle to make or enhance temperature differences by putting in money made from profitable trades.  We amplify on this below, but first talk about MB trade.}}
A goal of our theory is to provide a framework for quantitative analysis of the benefits of trade, including finding which trades are mutually beneficial. 

\subsection{Mutually beneficial trade}
\label{sec:MBtrade}

For a single good and money, and infinitesimal trade maintaining equilibrium, there is a simple (and very natural) criterion for the entropy of both economies to increase:~the goods move from the economy with the lower market price, say $\mu_A$, to that with the higher market price $\mu_B$, and the price for the trade is anything in between the two.  This results from positivity of temperature and equation (\ref{eq:dM}), which we rewrite here as $T\, dS = dM+\mu\, dG$ for reversible changes.  Note that in a sequence of trades the market prices might well change, so this condition has to be satisfied at all stages of the trade.

Trades might involve more than one good and not necessarily include money.  Our MB condition is again that the entropy of both economies increase.  Writing $dS=\sum_i \nu_i dG_i$ for reversible changes (where one of the goods could be money, in which case its value $\nu_i$ is $\beta$), the condition for the package of transfers $dG_i$  (which can have either sign, in particular that for money is likely to have opposite sign to most of those for goods) from $A$ to $B$ to be MB is that $\sum_i \nu_i^A dG_i<0$ and $\sum_i \nu_i^B dG_i > 0$.  This does not have as simple an interpretation as in the case of one good and money, but allows combinations that might not be MB pairwise.
Trade can (and often does) involve goods that do not exist originally in the receiving economy.  It is natural to assign value $\nu=+\infty$ for an unavailable good, but the value will become finite as soon as some of the good is imported.

Reversible trades must take place slowly---in particular, they are quasi-static, meaning that they occur slowly enough for both economies to be at equilibrium throughout the trading process. But trades need not be quasistatic. All that is required for MB is that the entropy of neither economy decreases. 

\begin{figure}[htb] 
\centering
\includegraphics[width=2.5in]{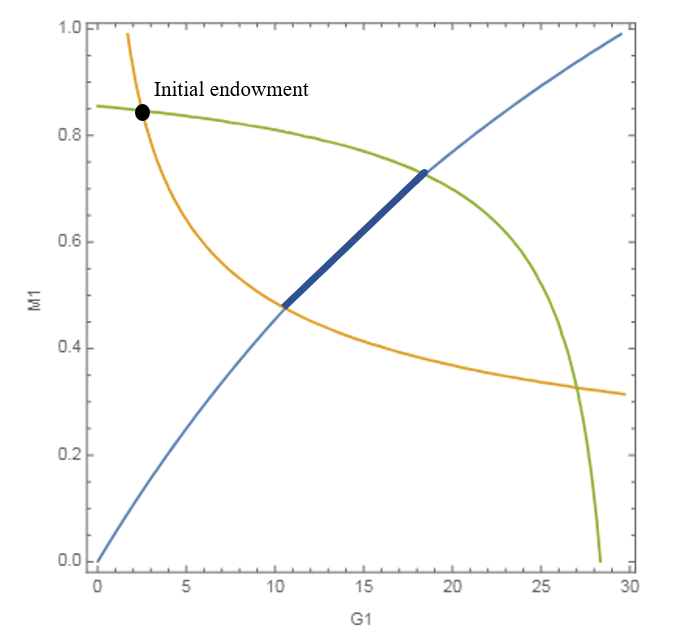}\quad
\includegraphics[width=2.5in]{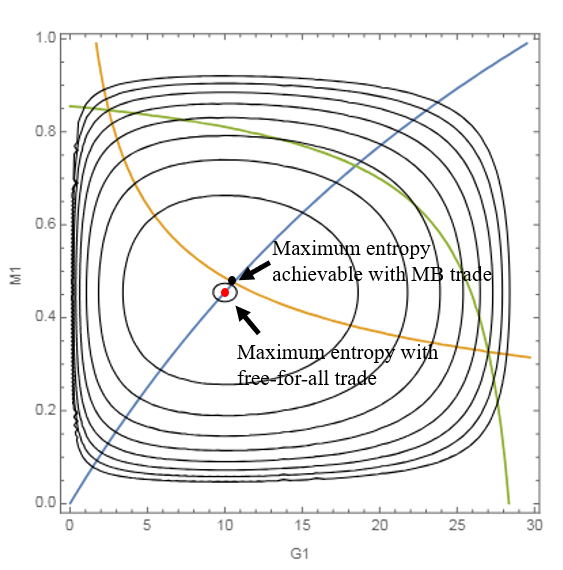}
\caption{(a) Edgeworth box plot for the two Cobb-Douglas economies of Figure~\ref{fig:Splot}.  The orange and green curves are 
\emph{isentropes} for the first and second economy, respectively (i.e., entropy is constant along the curves).  The initial endowment is at their upper left intersection.  The blue curve is the Pareto frontier of equal prices in the two economies, i.e., the set of points where the isentropes for the two economies touch; it is thickened for the part achievable by MB trade from the initial endowment. (b) The same plot with contours of total entropy superimposed, from Figure~\ref{fig:Splot}; the maximum entropy point is inside the smallest contour of total entropy.}
   \label{fig:EdgeBox}
\end{figure}

MB trade is very similar to exchange between rational agents with preferences given by total preorders, often illustrated by the ``Edgeworth box'' \cite{St} (or Edgeworth-Bowley box in \cite{SF}).
Here we obtain the same picture at the macroeconomic level, with the preferences being given by the entropies of the two economies. It is illustrated in Figure~\ref{fig:EdgeBox}(a) for the same pair of economies as in Figure~\ref{fig:Splot}.
The families of curves of constant entropy (isentropes) for the two economies (of which only one is shown for each here) are tangent to each other along a curve called the Pareto set (the set of states from which no economy can gain without some other economy losing).
The tangency condition of the isentropes is that the market prices are the same in both economies.  This is the condition for nett trade to stop (though equal and opposite trades may continue, at the common market price).
It is important to note that there is not a unique final state from given initial state:~any point on the part of the Pareto curve of tangencies of constant entropy curves between the initial constant entropy curves is a possible final state.  Total entropy maximisation (subject to the total goods and money) would select a particular point of the Pareto curve, but it might not be in the allowed interval.  Indeed for the chosen initial endowment of Figure~\ref{fig:EdgeBox}, the maximum entropy point is not in the allowed interval, as can be seen in Figure~\ref{fig:EdgeBox}(b). So a purely macro analysis gives no determinate answer to how prices will settle---presumably, factors such as the relative patience and skill of the negotiators on each side will decide the outcome.\footnote{Or, according to the theory of the Edgeworth box, the final point ought to be the one on the Pareto curve from which the common tangent to the isentropes goes through the initial endowment point, meaning that the whole sequence of trades is conducted at the price corresponding to the slope of that straight line.} Notice, in particular, that MB trade (without financial contact and hence flow of money) does not require that the temperatures of the two economies become equal (recall from Section~\ref{sec:temperature} that for Cobb-Douglas economies, $T= \frac{M}{N\eta}$, so equal temperature corresponds to the ratio $M_1:M_2 = N_1\eta_1:N_2\eta_2$, hence for $N_1=10, N_2=20$, $\eta_1=\frac52, \eta_2 = \frac32$ and $M_1+M_2=1$, we get $M_1=\frac{5}{11}, M_2 = \frac{6}{11}$, which includes the maximum entropy state but nothing else on the Pareto curve). 

\subsection{General trade}
Let us continue some more analysis of Figure~\ref{fig:EdgeBox}(b), but dropping the requirement that trade must benefit both economies. Now our only restriction is that trade must not decrease the entropy of the system overall. Thus, in the ``contour map'' of overall entropy, trades can move only ``up hill'' or, as a limiting case, around the contours. 
In particular, this allows trades to lead to final states in the region below the orange curve but inside the constant total entropy curve through the initial state, where economy 1 loses entropy but economy 2 gains more.  Notably, in this example, that includes the maximum entropy state itself.

\begin{figure}[htb]
    \centering
\includegraphics[height=2.5in]{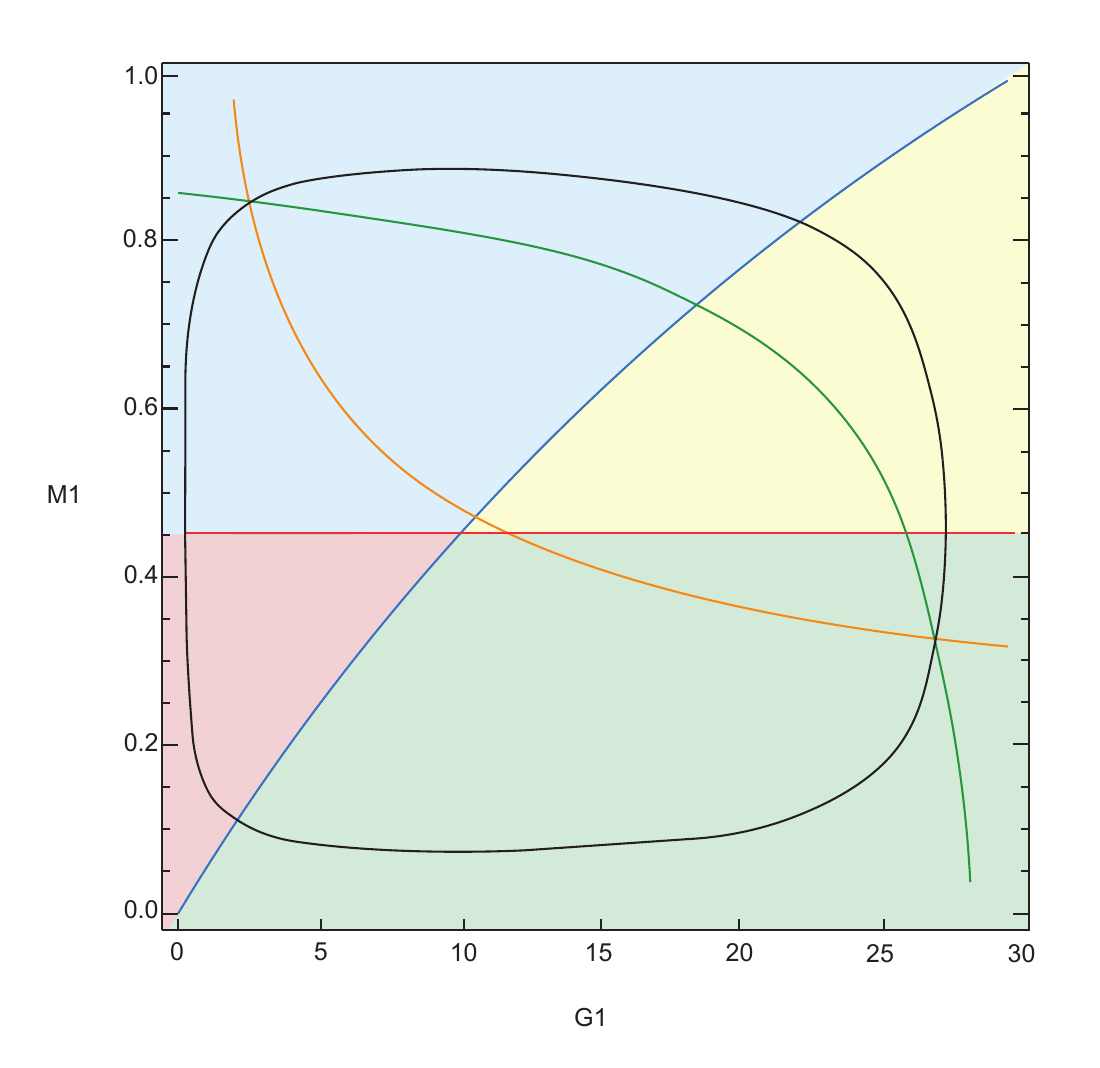}
\caption{Figure~\ref{fig:EdgeBox}(a) with the addition of the line of equal temperatures (red) and just one total isentrope (mauve). The (red) line of equal temperatures and the (blue) curve of equal prices divide the space into four quadrants.  The top left quadrant (shaded in blue) is where $T_1>T_2$ and $\mu_1>\mu_2$. The acute-angled quadrant (yellow) to the upper right is where $T_1>T_2$ and $\mu_1 < \mu_2$.  The lower right quadrant (green) is where $T_1<T_2$ and $\mu_1<\mu_2$.  The lower left quadrant (red) is where $T_1<T_2$ and $\mu_1>\mu_2$.}
    \label{fig:Plot3}
\end{figure}

It is useful to classify the possible moves of the pair of economies in this simple case with money and one good.  We divide the state space into ``quadrants'' according to the signs of the temperature difference $T_2-T_1$ and price difference $\mu_2-\mu_1$. 
See Figure~\ref{fig:Plot3}.
Recall from the end of the previous subsection that in our toy economies, $T = \frac{M}{N\eta}$, so the curve of equal temperatures is a line of constant $M_1$, namely $M_1 = 5/11$ for the chosen parameter values. Recall also that the curve of equal prices is the Pareto curve.
From a starting point other than that of maximum total entropy, the previously considered case of MB trade takes place in the cone of directions where $\mu_2 dG_1 < -dM_1 < \mu_2 dG_1$.  Following the second law, this always points to higher entropy.  Investment takes place in the direction with $(T_2-T_1)dM_1 > 0$ and $dG_1 = 0$.  This, too, always points to higher entropy.  Taking convex combinations of the two we obtain a larger cone of possible directions, depending on the quadrant. But the space of all directions that don't decrease total entropy is a half-space, which raises the question of how to
achieve the remaining entropy-increasing directions. 

One way to achieve them is by allowing connection to a third economy 3 under the control of an external trader T. T uses 3 to create a Carnot cycle to extract money from the temperature difference between 1 and 2, or in reverse to move money against a temperature difference. 
To see how this works in more detail, let us pick a particular set-up, for example, where
$T_1>T_2$ and $\mu_1>\mu_2$ (this is the top left quadrant in Figure~\ref{fig:Plot3}; the other quadrants can be dealt with similarly).  Economy 1 can invest money $m$ in 3, which gets divided into $em$ for 2 and an initial profit $(1-e)m$ for our trader, T. The value $e \in (T_2/T_1,1)$, which is the complement of the efficiency (i.e.~the efficiency is $1-e$), can be arbitrarily close to $T_2/T_1$ as we approach an ideally efficient Carnot cycle, but cannot exceed this value.  T then uses the profit, $(1-e)m$, to buy $g$ goods from 1 at a price  $\tilde{\mu}_1 > \mu_1$ and sell them to 2 at a price $\tilde{\mu}_2 <\mu_2$. Of course, the trader, T, will lose money in doing so, specifically $(\tilde{\mu}_1-\tilde{\mu}_2)g$. The trader may wish to break even, thus balancing the initial profit $(1-e)m$ against this loss $(\tilde{\mu}_1-\tilde{\mu}_2)g$, i.e., by choosing $m$ and $g$ so that $\frac{m}{g} = \frac{\tilde{\mu}_1-\tilde{\mu}_2}{1-e}$.
In bilateral trade, of course, money and goods flow in opposite directions. But by the trader using a third economy, it has managed to arrange a nett flow of money and goods in the same direction.
Here there are both a nett money flow $\tilde{m} = \frac{e\tilde{\mu}_1 - \tilde{\mu}_2}{1-e} g$ from 1 to 2 and also a nett goods flow $g$ from 1 to 2.  By choice of $e$, $\tilde{\mu}_1, \tilde{\mu}_2$ in the indicated ranges, the ratio $\tilde{m}/g$ can be made anything greater than $\frac{T_2\mu_1-T_1\mu_2}{T_1-T_2}$.  The latter can be written as $\frac{\nu_1-\nu_2}{\beta_2-\beta_1}$, which is the slope of the curve of constant total entropy.  So the trader can achieve $dG_1<0$, i.e., moving goods from 1 to 2, with any slope steeper than the isentrope.

Similarly, still in the case $T_1>T_2$, $\mu_1 > \mu_2$, changes with $dG_1>0$ and $dM_1>-\mu_2 dG_1$ can be made, up to the slope of the isentrope.  T buys goods $g$ from 2 at price $\tilde{\mu}_2$ above $\mu_2$
and sells them to 1 at price $\tilde{\mu}_1$ below $\mu_1$, but with $\tilde{\mu}_1>\tilde{\mu}_2$.  The profit $m = (\tilde{\mu}_1 - \tilde{\mu}_2)g$ is used to drive money flow from 2 to 1 against the temperature difference, causing money to flow into 1 of up to $m/(1-T_2/T_1)$.  The nett money flow into 1 is then given by subtracting $\tilde{\mu}_1 g$ that 1 paid the trader, T, for the goods.  The result of this process is to increase the temperature difference between 1 and 2---analogous to refrigeration in the context of physics, where the temperature difference between the inside and outside of the fridge is increased by expending energy. Here, the expenditure of money is required to increase the economic temperature difference between two economies.
Choice of the efficiency of this process of ``refrigeration'' and the prices $\tilde{\mu}_j$ yields $dM_1/dG_1$ anywhere between $-\mu_2$ and the slope of the isentrope.

\rev{Nonetheless,} the use of an external system by the trader is not necessary to achieve these changes.  They can {instead} be achieved by the trader enabling exchange at a sequence of prices that are not necessarily positive.  To see this, any state $X_1$ with higher entropy than the initial state $X_0$ can be connected by a smooth curve along which entropy has positive derivative.  If the trader enables exchange at the price corresponding to the tangent to this curve at each moment then the state will move along it.  This argument extends to arbitrary numbers of goods.

Let us consider entropy as a measure of the aggregate welfare of an economy, and consider a ``benevolent social planner'' deciding whether trade should be allowed between two economies. Of course, if the trade were to reduce the summed welfare between two economies, the social planner would disallow it---but we know that, in the present set-up, this cannot happen:~trade cannot decrease overall entropy, by the second law. The social planner might, instead, allow only trades which benefit both economies (where the benefit for one may be 0)---this is, of course, MB trade. But this may seem too restrictive. There may be trades that lead to large welfare benefits for one economy, 
and only small losses for the other economy.
How can we make this precise? The TM framework has a simple answer:~the social planner may allow trades for which the curve of constant total entropy through the final state intersects the region where both economies gain entropy. 
This is because such a final state could subsequently be moved along the isentrope (arbitrarily closely) to end up in the MB region.
Such trading outcomes are what we might call potentially-MB trades.\footnote{Potentially-MB trades may remind the reader of the Kaldor-Hicks criterion in welfare economics. Here, though, compensation is assessed in terms of transfers of utility---corresponding to reversible trades along an isentrope---rather than transfers of money.} An equivalent way to put the condition {is that a trade is potentially-MB if and only if} the final entropy is not larger than the maximum entropy over the MB region.

Note that such compensation is not always possible:~i.e., not all trades that increase overall entropy are potentially-MB. Figure~\ref{fig:EdgeBox}(b) provides an illustration. Suppose that the initial state of the two economies (before trading is allowed) at the top left intersection of the orange and green curves moves to the maximum entropy point or close to it, then no entropy-preserving change can move it to the other side of the orange curve. Thus the entropy of economy 1 remains below its initial value as we move along the isentrope, and hence this trade is not potentially-MB.




Does MB trade have an analogue in physics?  It is usual in thermodynamics to consider increase of total entropy as the sole principle.
Yet consider a cylinder with an insulated movable partition between two gases (the famous ``piston problem'' \cite{Wr}). If the pressures $P_j$ on the two sides are different then the the partition will move, thereby transferring volume from one side to the other, and also the pressure difference will do work ($(P_1-P_2)\, dV$ in the quasistatic case).  But there is a continuum of choices for where the work goes.  We could suppose the partition is attached to a dry friction device (that produces resistive force independent of the speed of movement) located in the left-hand side, whose friction force is controlled to be just less than the force on the piston resulting from the pressure difference; then the work goes to heat the left-hand side. Or a dry friction device on the right-hand side, or a combination of both. These situations correspond to MB trade.

In contrast, one could just let the partition fly and after some oscillations it would settle down to some division of the volume and the energy between the two parts.  In this case there is no reason that the individual entropies should both increase.  
There is, for example, the maximum entropy state where in addition to the pressures equalising, the temperatures also equalise.  Even with a perfectly insulated piston, this will be achieved over a long timescale via fluctuations in the position of the partition due to finite numbers of molecules on each side (this effect is outside the domain of {classical} thermodynamics, but can be expected to apply to economics too).  In the case of a small or zero pressure difference between the two sides, the entropy of the initially hotter side decreases. 

The economic analogue of the latter case is a free-for-all flow of goods and money with no restriction to exchange at prices between the market prices and allowing money, or goods, to flow {simply because} they are more valuable in another economy. 
The only constraint is that the total entropy can not decrease.  For infinitesimal transfers $dM_{AB}, dG_{AB}$ this constraint reads as 
\begin{equation}
(\beta_B - \beta_A)\,dM_{AB} + (\nu_B - \nu_A)\,dG_{AB} \ge 0,
\label{eq:entropycond}
\end{equation}
from (\ref{eq:dS}), but it is possible that the entropy of one economy would decrease in the process.

In particular, we note that financial contact between economies does not lead to gains in entropy for both economies, only for the one receiving money, as we saw in Section~\ref{sec:temperature}. To take a concrete example, 
suppose two closed economies are identical, except regarding the amount of money available in each (say, one has ten times the quantity of $\odot$ than the other). This means that prices for all other goods will be higher in the economy with more $\odot$. Indeed, if we suppose that $\odot$ have no intrinsic use, aside from as a means of trade (unlike gold, which can be crafted in jewelry, for example), then it is reasonable to assume that the prices in one economy will be ten times higher than in the other, following Section~\ref{sec:puremoney}). Then, if unlimited financial contact (but no trade) is allowed between the economies, money will rapidly (i.e., not reversibly) flow from the economy with high prices (the `low temperature' economy) to the economy with low prices (the `high temperature' economy), and the prices of goods will fall in first economy and rise in the second. This corresponds to an increase in entropy in the first economy, and a decrease in entropy in the second.

\subsection{Gains of trade}
\label{sec:gains}

One of the most fundamental questions in economics concerns the maximum potential gains from trade that can be achieved by allowing trade between economies that were not previously in contact. 

The present approach allows maximum gains from trade to be studied in a particularly simple way. We propose that rather than trading directly, all trade is carried out via an intermediary---the now-familiar trader---who can exchange goods for money with either economy.  Trade of a bundle of goods $(dG_i)_{i \in I}$ from $A$ to $B$ and suitable amounts of money can be facilitated by the trader if $\sum_{i\in I} (\mu_i^B-\mu_i^A)dG_i > 0$, this being the maximum amount of money that the trader siphons off, because the minimum money that $A$ is willing to receive is $\sum_i \mu_i^A dG_i$ and the maximum that $B$ is willing to pay is $\sum_i \mu_i^B dG_i$. 

The trader can then simply donate the money that it has siphoned off to the two economies (in any proportions), thus increasing the economic entropy of both to some extent. In this way, the gains of trade are split between the two economies. This will also allow us to analyze the maximum increase in economic entropy that can be achieved --- which will be reached by donating the gains of trade to whichever economy is economically cooler, because by definition the marginal aggregate utility of money will be greater in that economy.\footnote{Donating money to an economy of finite size will raise its economic temperature, because we proved that money capacity is positive.} This will continue until the two economies are the same temperature, after which the gains should be distributed between them to keep them at equal temperatures.

We can simplify the problem of finding the maximum profit that the trader can extract by noting that this must be via purely reversible trades, i.e., which will not change the economic entropy of the compound system comprising the pair of economies (although it may shift economic entropy reversibly between one economy and the other, as we note below). This follows because reversible trades will minimize the economic entropy of the compound system. Thus any irreversible trades will lead to the compound system having a higher economic entropy. Thus the trader can guide the system to any such state by first engaging in irreversible trades and then raising the economic entropy of the compound system by donating some or all of its gains of trade. But in doing so, the profit retained by the trader is thereby reduced. So we conclude that, in order to retain the gains of trade, the trader should use only reversible trades.

Note that this analysis extends to the case of many economies. If we have three economies, $A$, $B$ and $C$, for example, the maximum amount of money that can extracted by the trader will be determined in part by the maximum amount that can be extracted by trading reversibly between $A$ and $B$. This will yield a compound system in which the prices and economic temperature are the same for $A$ and $B$---if this were not so, then further trading (or voluntary flows of money from the higher temperature economy to the lower temperature economy) could yield more money for the trader---but by assumption the trader has extracted all money possible from the pairwise trade. This compound system can then be conjoined with $C$ and the trader can engage in further reversible trades until $A$, $B$ and $C$ all have the same prices and temperature. 

Interestingly, the maximum amount of money that the trader can extract by bringing $A$, $B$ and $C$ into equilibrium is independent of the order or manner in which the trades are carried out. For examples, the trader might directly make reversible trades between $A$, $B$ and $C$. Or the trader might initially allow trade only between $B$ and $C$, and then trade with $A$; or the trader might interleave trades between $A$ and $B$ with trades between $A$ and $C$; or trade in any other pattern, so long as the trader ultimately fully exploits any trading opportunities. That is, while the trader can bring the economies into equilibrium with different equilibrium prices and temperature, only one set of equilibrium prices and temperature will allow the trader to extract the maximum amount of money. For suppose the contrary, that there are two distinct equilibria that the systems could fall into, which would both yield this maximum. Then we could split the entire economy of $A$, $B$ and $C$ in half; extract this maximum amount of money by shifting to the distinct equilibria in each half (thus extracting, supposedly, the maximum that the trader can obtain). But then the trader could make reversible trades exploiting the price and/or temperature differences between the two equilibria, thus extracting more money---contradicting the hypothesis that the equilibria had allowed the trader to extract the greatest possible amount of money. Thus, the maximum amount of money that can be extracted from a system of economies (once all reversible trades have been exploited) corresponds to a single equilibrium of prices and temperatures for the entire system. 

Moreover, the path of trades by which the prior states of $A$, $B$ and $C$ are shifted to that equilibrium by reversible trades does not affect the amount of money extracted by the trader. Again we can prove this by contradiction. Suppose that the trader could obtain the maximum $M$ money by following one path of reversible trades, but only $M'<M$ by some other path of reversible trades. Then there would be a ``loop'' of reversible trades which would take the initial state of the system to the equilibrium, and back to its original state, as follows: follow the path yielding $M$ to be trader and then follow the other path in reverse, yielding $-M'$ to the trader. Thus, by following this loop, the system ends where it started, and the trader has gained $M-M'>0$. But this is impossible, because money has been created and given to the trader ``from thin air.''\footnote{Comparing with physics, this is analogous to finding a way of generating energy ``for free,'' thus violating the conservation of energy.}

Thus, we can define ``gains of trade'' between any set of economies as the maximum amount of money that can extracted by the trader by effecting trade between these economies. This gain of trade also measures the ``benefit'' that can be obtained through direct trading between economies, without the trader siphoning off any of the surplus (noting, as above, that any entropy-increasing set of trades can be achieved by the trader by first extracting a profit by making reversible trades, and then distributing that profit to the various economies). Where, for example, there is free-for-all direct trading between economies, the gain of trade generates changes in entropy in the constituent economies (which, in sum, will be non-negative, although one or more economies may lose entropy).

Summarising the above discussion, various natural economic results follow, albeit only in the context of exchange economies. On the ``positive'' side for the benefits of trade:

\begin{enumerate}
\item First, gains of trade, as defined above, are always non-negative for the union of two or more countries, and will almost always be positive. These gains are independent of the order in which trades are carried out.
\item Second, the gains of trade for many countries trading together is at least as large as the gains of trade for any partition of those countries into ``trading blocks.'' 
\end{enumerate}

But on the negative side, none of these benefits need arise from spontaneously free-for-all trade between economies. 

\begin{enumerate}
     \item Free-for-all trade may lead to disbenefits (i.e., loss of entropy) for some economies (even though the set of economies as a whole will gain). 
     \item Such free-for-all trade can sometimes, but not always, be compensated by a benevolent social planner who can engage in reversible trades (which therefore don't affect the entropy of the system as a whole), such that trade becomes mutually beneficial.
 \end{enumerate}

The notion of gains of trade can be viewed as analogous to a generalisation of the notion of ``available energy'' in physics or ``exergy'' in engineering:~i.e.,
the amount of work that a physical system  can do by bringing it into equilibrium with a large reference environment (see \cite{M20} for a pedagogical introduction).
By extracting money from one or more economies, the trader is, analogously, extracting value in a usable form, i.e., which could potentially be used to buy goods from some other party. 
A difference 
\rev{from} the physical notion of exergy is that we do not make the simplification of treating one system as a large reference environment but consider the interaction of two or more economies of roughly comparable size. Of course, the same can be done in physics and engineering; it is simply more natural in many physical contexts to consider the environment as large.

In engineering, exergy is often useful for making calculations (e.g.~of the capacity of an energy store), and it is likely that this will be true also in the economic contexts. Note, though, that no new principle is being invoked here, over and above the operation of the second law. 

Another concept in physics is ``free energy''.  This applies to systems that are maintained at a given temperature (Helmholtz free energy) or temperature and pressure (Gibbs free energy).  It quantifies the amount of work that can be done by the system, this time because some other aspects, like chemical reactions, are not yet in equilibrium.  This is likely to be relevant in economics too, when we reach the stage of considering manufacturing.  We briefly discuss this further below in Section~\ref{sec:freeenergy}.

A key issue in this idealized world of exchange economies (and one relevant to real-world economic policy) is the impact of trade barriers on gains of trade. Suppose that the barriers are sufficiently sparse that, for all goods, $G_i$ and all pairs of economies, $E_j$, $E_k$, there is possible sequence of allowable bilateral trades (possibly via intermediate economies) by which  a $G_i$ in $E_j$ can be exchanged for money in the $E_k$ and vice versa. We suppose also that money can flow from any economy to any other (possibly via intermediate economies).  This will be sufficient to ensure that the system of economies reaches equilibrium prices and temperature.  This will be the same equilibrium as if there were no barriers in place; and the gains of trade are unaffected. We consider a weaker form of trade barrier, in the form of tariffs, below.


In the case that economy $A$ is large compared to $B$, the maximum profit (not respecting non-decrease of entropy in either economy) that can be made is, as noted above, given by the `exergy'
of $B$ relative to the temperature and value for $A$:
$$W = M_B-M_B^e - T_A (S_B-S_B^e) + \mu_A (G_B - G_B^e),$$
where superscript $^e$ denote the values if $T_B=T_A$ and $\mu_B = \mu_A$.  
But again, the entropy of one of $A$ and $B$ will decrease, because the total entropy remains constant for maximum extraction of money. 

For two economies of general sizes, the trader makes maximum profit by conserving total entropy while extracting money to bring the pair to a state of equal temperatures and prices, where they can be merged.  
As an example, for two Cobb-Douglas economies with the same exponents $\alpha$ and $\eta$, the maximum money that can be extracted is given by minimising $M_1+M_2$ over the surface $S(M_1,M_2,G)=S^0$ shown in Figure~\ref{fig:AccRegion}, where $S^0$ is the initial entropy.  This is given by the point of contact of the tangent plane $M_1+M_2=$ cst with the surface, equivalently by the conditions $\beta_1=\beta_2$ and $\nu_1=\nu_2$.  We can solve this even if the two economies have different $\alpha_i$, $\eta_i$, as follows. $G_i = p_i G$, where $p_i = \frac{\alpha_i N_i}{\alpha_1 N_1 + \alpha_2 N_2}$. $M_i = \eta_i N_i T$, where $T$ denotes the final temperature, to be determined by making $S=S^0$.  Denoting $q_i = \eta_i N_i$ and initial values by superscript $^0$, we obtain
$$(q_1+q_2)\log T = \sum_{i=1,2} \alpha_iN_i\log\frac{G_i^0}{p_i G} + q_i \log\frac{M_i^0}{q_i}.$$  Thus
$$T = \left[\prod_{i=1,2} \left(\frac{G_i^0}{p_i G}\right)^{\alpha_iN_i} \left(\frac{M_i^0}{q_i}\right)^{q_i}\right]^{\frac{1}{q_1+q_2}}.$$
The final amount of money in the system is
$$M_f = (q_1+q_2)T = \left[\prod_{i=1,2} \left(\frac{G_i^0}{p_i G}\right)^{\alpha_iN_i} \left(\frac{M_i^0}{q_i \bar{M}}\right)^{q_i}\right]^{\frac{1}{q_1+q_2}} M^0
,$$
where $M^0=M_1^0+M_2^0$ and $\bar{M} = M^0/(q_1+q_2)$.
This is at most $M^0$ and the difference is the trader's profit. 

There are many paths by which {the trader can move the two economies to a state where they can be merged reversibly}.  Some {paths} involve the trader using a Carnot cycle in an external system.  But {the merged state} can be approached by a sequence of moves where the trader {simply puts} the two parts in contact with a specified ``exchange rate'', i.e.~so that exchanges conserve $aM+bG$ in each economy for some $a,b$ not both zero.  Importantly, however, the exchange rate must be allowed to be negative.  As in (\ref{eq:entropycond}), the two-part system will then move along the line in the direction of increasing total entropy.

\subsection{Tariffs}
\label{sec:tariffs}
The possibility that trade need not be mutually beneficial, and can reduce the economic entropy of a participating economy, raises the question of how governments might seek to control such trade, e.g., using tariffs or subsidies.\footnote{Of course, the government might also imposes tariffs or quotas in response to special interests or simply to raise revenue.} 
We consider how the thermal macroeconomic account addresses such questions.

Suppose $B$ has to pay a tariff $\odot \tau_i$ per unit of good of type $i$ exported to $A$.\footnote{We could just as well assume that $A$ has to pay the tariff to import the good, as is more typical in practice. Just as in tax incidence theory in standard economics \cite{KS} the equilibrium outcome is the same whether the importer or exporter bear the cost of the tariff.}  We suppose first that the resulting revenue {goes to some external location (e.g.~to our trader)}. 
Denote the amount of money raised from the tariffs by $P$, so  $dP = \sum_i \tau_i dG_{iA}$, which we denote by $\tau \cdot dG_A$.  Then $dM_B = -dM_A-dP$ and $dG_{iB} = -dG_{iA}$, so under quasistatic changes, 
$$dS = \beta_A dM_A + \nu_A dG_A - \beta_B(dM_A+\tau \cdot dG_A) - \nu_B \cdot dG_A = (\beta_A-\beta_B)dM_A + (\nu_A - \nu_B - \beta_B \tau)\cdot dG_A.$$
We see, not surprisingly, that the condition $dS\ge 0$ is the same as without the tariff {except that} $\nu_B$ {is} replaced by $\nu'_B = \nu_B + \beta_B\tau$, or equivalently, the price $\mu_B$ {is} replaced by $\mu_B+\tau$, as long as we are considering changes with $dG_A\ge 0$ for each type of good.  
If $dG_A<0$ for some type of good then the tariff does not apply. 

To simplify the discussion, suppose there is just one type of good, the tariff $\tau>0$, and $A$ and $B$ are always in financial equilibrium.  If the price-difference $\mu_A-\mu_B$ is larger than the tariff $ \tau$, then goods flow from $B$ into $A$, with concomitant changes in prices, until we get $\mu_A-\mu_B = \tau$.  If $\mu_A-\mu_B < 0 $, goods flow from $A$ to $B$ until we get $\mu_A = \mu_B$.  If $0\le \mu_A - \mu_B \le \tau$ then there is no direction that increases the entropy, so there is an interval of possible equilibria. {Interestingly, we see that conventional irreversible tariffs contradict our axiom A0; exploring the implications of such a violation is an interesting direction for future analysis and simulations.}

Similar considerations hold if instead $B$ gets a subsidy $\odot \sigma$ per unit for export of goods.  {As with tariffs, we again}, for the moment consider the subsidy to come from some external source.  Then for $\mu_A-\mu_B >-\sigma$, entropy can be increased by goods flow from $B$ to $A$.  For $\mu_A-\mu_B <0$, it can be increased by goods flow from $A$ to $B$.  In this case there is an interval of the distribution of goods between $A$ and $B$ in which entropy can be increased indefinitely by oscillations in the regime $-\sigma<\mu_A-\mu_B < 0$.  This entropy increase is powered by the unlimited subsidies.  So there is no equilibrium (again contradicting A0).
Combinations of tariffs and subsidies in each direction can be analysed similarly.

The violations of A0 {disappear} in the special {though purely theoretical} case of what we call ``reversible tariffs'', defined as follows.  If an economy $A$ imposes a tariff $\tau_i^A$ per unit of good $G_i$ arriving in $A$, it also gives a rebate of $\tau_i^A$ per unit of good $i$ leaving $A$. This means that the amount of revenue collected depends only on the {nett} change of distribution of goods before and after trade occurs. The tariffs may be negative, i.e.~subsidies for imports.
For a reversible tariff with the money going to and from an external place, the effect is just to shift the condition for equilibrium from $\mu_A=\mu_B$ to $\mu_A = \mu_B + \tau$ (and $\beta_A=\beta_B)$.

{So far we have assumed that tariff revenues (and subsidies) leave the economies under consideration, e.g., to an external trader.} More realistically, and returning to general tariffs, we could suppose that the revenue from tariffs imposed by $A$ goes {back} into
$A$. {We allow the possibility, however, that the money from tariffs might go to a money-component of $A$ other than its distinguished money-component.}  Similarly, subsidies on exports from $B$ might come from a separate money-component of $B$.  We could even have subsidies offered by $A$ on imports into $A$, coming from the separate compartment in $A$ {and so on}.
In such cases, we need entropy functions that take into account the amounts of money in the extra compartments.  Let's say we have $S_A(M_A,G_A,P_A)$, with $P_A$ being the money in the extra compartment, and {with analogous notation} for {economy }$B$.  Then there are coolnesses for the extra compartments, $\tilde{\beta}_A = \frac{\partial S}{\partial P_A}$ and {$\tilde{\beta}_B = \frac{\partial S}{\partial P_B}$}.  To illustrate what happens, consider the case where $A$ collects a tariff $\tau$ on imports of a single good into $A$.  Then for $dG_A>0$, we have 
$$dS = \beta_A dM_A + \nu_A dG_A +\tilde{\beta}_A \tau dG_A -\beta_B(dM_A+ \tau dG_A) - \nu_B dG_A.$$
We see that for $\beta_A=\beta_B$ 
\rev{the} entropy increases iff $\nu_A-\nu_B+(\tilde{\beta}_A-\beta_B)\tau >0$.  So this is equivalent to a modified price $\mu'_B = \mu_B+(1-\tilde{\beta}_A/\beta_B)\tau$.  In particular, if $\tilde{\beta}_A = \beta_B$ then the tariff has no effect on movements of goods in that direction.  But the tariff has no effects in the other direction either, because it does not apply then.  So if for example, $A$ were to put the revenue from tariffs directly back into the distinguished money component of $A$ then the tariffs would have no effect.  This result might sound counter-intuitive, but we believe it is a question of timescales.  If it takes a long time for distribution of tariff revenue to ``trickle down'' into the economy then tariffs do have an initial effect.



What is the effect of a tariff or subsidy on the total welfare, as quantified by entropy?  Denote by $S_{\max}(\tau)$ the maximum entropy with tariff $\tau$ for import of goods into $A$, over the possible distributions of goods and money.  Then $\frac{dS_{\max}}{d\tau} = (\tilde{\beta}_A -\beta_B)(G_A-G_{A0})$, where $G_{A0}$ is the initial amount of goods in $A$ at equilibrium.  So, since the tariff applies only for $G_A>G_{A0}$, we deduce that the effect of a tariff is entropy-increasing only if $\tilde{\beta}_A > \beta_B$.

The revenue raised (or subsidy paid, if negative) by each government depends on the volumes $\Delta G_i$ of goods of type $i$ that have moved from economy $A$ to $B$, and the imposed tariffs/subsidies.
As a simplification to start future discussion, we consider this in the context of ``reversible'' tariffs. 
Then, explicitly, the revenues $R_A, R_B$ raised by the governments of $A, B$, respectively, are
$$R_A = -\sum_i \Delta G_i \tau_i^A, \quad R_B = \sum_i \Delta G_i \tau_i^B.$$

One could ask what is the optimal set of tariffs/subsidies for a government. One objective is for the government to maximise the entropy of its own economy, assuming that the tariffs imposed by other governments are either absent or have some fixed value. Such tariffs may, of course, be to the detriment of other economies with which trade is being conducted. It is also interesting to consider the strategic setting in which multiple governments each set tariffs to maximize the entropy of their own economies, in the light of the tariffs of the others---which is likely to require a game-theoretic analysis. A further interesting variation is where the government aims to maximise the entropy of its economy, while raising some required amount of revenue. The latter question relates to the problem of optimal taxation, where a government wants to raise some amount tax while minimising the reduction in entropy of its economy.\footnote{In our exchange economy, the most natural tax is applied to the stock of goods, and we can ask how to determine the optimal tax rates across the different types of good.}   Such questions have been subject to extensive analysis using standard economics (e.g., \cite{Ram}), and it will be interesting to explore them in the present framework. We leave these topics for future study.


\subsection{Trade frictions}
\label{sec:friction}
In reality, trade involves irreversible costs, for example, for transport of goods or money-transfer fees, and, as we have discussed, irreversible tariffs. The corresponding money goes to agents in one or both economies or to an external trader.  This implies that MB trade requires a sufficient gap between the prices to cover the costs.  It also implies a reduction in the amount of money that an external trader can make from facilitating trade.

Irreversible effects also occur when large trades are made.  For example, the prices an external trader proposes to two economies for a large quantity of goods have to be different from the market prices, which are for infinitesimal amounts of good.  By convexity, the trader makes at most as much money as by slow continuous trading, and typically strictly less money.
\rev{We leave more thorough study of trade frictions for future work.}

\section{Relations between partial derivatives:~analogues of principles of Le Chatelier, Hotelling, Slutsky and more}
\label{sec:cross}

A central theme of Samuelson's Nobel Lecture \cite{S72} is the power of explanation based on maximisation in economics. Contrasting his viewpoint with what he considers vague appeals to optimisation or rationality in other social sciences, he highlights that it has been possible to find testable, quantifiable relationships from the assumption that economic agents are maximisers. His examples focus on individual economic agents as maximisers. 

In the light of the second law of thermal macroeconomics, the present analysis allows all of those relations (and more) to be derived at the level of whole economies, where what is maximised is entropy/aggregate utility. We outline these results here, with a variety of natural extensions. As we noted in Section~\ref{sec:precursors}, Samuelson was sceptical of any analogue of entropy to be maximised by the economy as a whole; but it turns out that the maximisation analysis that Samuelson provided at the level of individual agents can be extended in this way. Specifically, here we explore the consequences of existence and concavity of aggregate utility $S$ on partial derivatives. In economics the partial derivatives most commonly studied are ``elasticities'':~change in demand for a change in price, but we  consider it more appropriate to treat the inverse matrix, change in price for a change in amounts, known as ``flexibilities'' (and are not alone in this \cite{Hou}). 

First, we show that the concavity of the entropy (aggregate utility) function leads to non-positivity of many partial derivatives, and signs for other suitable combinations of them.  These are analogous to Le Chatelier's principle in physics, that the response to an external change is to oppose the change.  They are also analogous to results obtained by Samuelson for the effects of prices on demands, but our results are for an economy in aggregate, rather than a single agent. 

Second, the symmetry of second partial derivatives leads to relations between various first partial derivatives.  In thermodynamics, these are known as Maxwell relations.  In economics, similar symmetries of cross-elasticities follow from existence of a utility function, which Samuelson called the Hotelling conditions
\cite{H} (though they appear to go back to Slutsky \cite{Sl}). Again, however, the Hotelling conditions apply to individual agents (or firms), whereas our relations apply at the level of the macro-economy.\footnote{This parallel between Maxwell and Slutsky's results has been noted in the context of individual agents, e.g., \cite{S72}.}

Third, we derive relations between partial derivatives holding different variables constant.  These are analogous to Slutsky's equation relating compensated to uncompensated demand, and to Samuelson's extension of Le Chatelier's principle.

In the following subsections, we treat these three aspects, and make some additional remarks about symplectic structure and areas.

\subsection{Non-positivity results}

We begin with the dependence of values of goods on their quantities. Recall that the values $\nu_i$ for goods of type $i$ are defined by $\nu_i = \frac{\partial S}{\partial G_i}$ (including that for money, which we denote by $\beta$). The value for a good quantifies the marginal utility \rev{(quantified by entropy)} for one unit of the good.  It follows from concavity of $S$ that 
\begin{equation}
\frac{\partial \nu_i}{\partial G_i} \le 0.
\label{eq:derivneg}
\end{equation}
It says that the value for a type of good decreases (strictly speaking, does not increase, but for simplicity we will use the shorter expression) if the amount of the good is increased.
For money, this gives $\frac{\partial \beta}{\partial M}\le 0$, which says the marginal utility of money decreases if the amount of money increases.
We have already seen in Section~\ref{sec:moneycap} that this can be rewritten as non-negativity of money capacity, $C=1/\frac{\partial T}{\partial M} > 0$, because the economic temperature $T$ is the reciprocal of $\beta$.

These results extend to any principal minor of the second derivative of the entropy with respect to amounts of goods and money (namely, minors of odd order are less than or equal to zero, those of even order are greater than or equal to zero).  Thus, for example, for a single type of good, 
\begin{equation}\frac{\partial \beta}{\partial M} \frac{\partial \nu}{\partial G} \ge \left(\frac{\partial \beta}{\partial G}\right)^2.\label{eq:minor}
\end{equation}

The results of the last two paragraphs can be classed as examples of Le Chatelier's principle. Originally formulated for physical systems, it says that in response to an external change, a system adjusts to counter-act the change.  Samuelson derived such relations for a profit-making monopolist \cite{S72}, considering prices as external and the amount of goods bought as response, and then went on to derive refinements, which we will mention below.  Our relations are for macro-economies and we consider amounts of goods as the external variables and their values as responses {(which seems reasonable given that at a macro-level, prices can respond rapidly but volumes of goods change only slowly)}, but if the entropy is strongly concave (i.e.~its second derivative is negative-definite) then one can invert the second derivative and the inverse is also negative-definite. 

We next extend this result to prices instead of values.
To interpret (\ref{eq:derivneg}) for a good other than money, recall that the values $\nu_i$ determine prices $\mu_i$ by $\mu_i = \nu_i/\beta$.  
So for a good other than money the inequality can be written as 
$$\mu_i \frac{\partial \beta}{\partial G_i} + \beta \frac{\partial \mu_i}{\partial G_i} \le  0.$$ 
It is conventional in economics to think about ``demand'' as a function of prices, i.e., in our context of exchange economies, to ask how the amount of goods possessed by an agent
depends on prices.  As we are considering a whole economy, where prices are determined endogenously, here it is more natural to consider how prices depend on amounts of goods.  Nevertheless, if invertible, one can invert matrices of partial derivatives to deduce how amounts of goods depend on prices.

Extending the consequences of concavity a further, we consider infinitesimal variation of any combination of goods and money. 
Since $\nu_i = \frac{\partial S}{\partial G_i}$ (
including money \rev{as $i=0$}), the infinitesimal change $\delta \nu_i$ is given by $\delta \nu_i = \sum_j \frac{\partial ^2 S}{\partial G_i \partial G_j} \delta G_j$.
From concavity of $S$ we deduce that \begin{equation}\sum_i \delta\nu_i\, \delta G_i \le 0.
\label{eq:deltadelta}
\end{equation}
This equation has the units of entropy and it includes a term $\delta\beta\, \delta M$ {(because, for these purposes, money is just one more good, 
with $i=0$}). This is analogous to the formula Samuelson obtained thanks to the lectures by Wilson, mentioned in Section~\ref{sec:precursors} \cite{S72}, but again for a whole economy rather than a single agent, and for prices rather than values (he was treating a revenue function rather than our entropy).  It has the advantage that it expresses the relation in a way that treats values and amounts of goods symmetrically.  

Converting (\ref{eq:deltadelta}) from values to prices for goods other than money by $\nu_i = \beta \mu_i$ and hence $\delta \nu_i = \beta\, \delta \mu_i + \mu_i \delta \beta$, we obtain
$$\delta \beta (\delta M + \sum_{i\ne 0} \mu_i \delta G_i) + \beta \sum_{i\ne 0} \delta \mu_i\, \delta G_i \le 0.$$
This can be rewritten as
$$\sum_{i\ne 0} \delta \mu_i\, \delta G_i \le \frac{\delta T}{T} (\delta M + \sum_{i\ne 0} \mu_i \delta G_i).$$
The expression in parentheses is 
the monetary equivalent of the infinitesimal change.  If the changes are made at market prices then it is zero, so in this case we deduce that \begin{equation}\sum_{i\ne 0} \delta\mu_i\, \delta G_i \le 0,
\label{eq:dmudG}
\end{equation} just as Samuelson's formula.

We discuss possible significance of equation (\ref{eq:dmudG}).  It is relevant to the question of motion on isentropes of a system, at market prices.  In this case, the inequality becomes an equality, because the lefthand side is the second variation of the entropy.  In the language of symplectic geometry,  we deduce that the isentrope in the space of $(\mu_i, G_i)$ is a Lagrangian submanifold:~$\sum_{i\ne 0} d\mu_i \wedge dG_i = 0$ on tangents to the isentrope.  One way to say what this means is that the amount of money one can extract by moving along it is independent of path {(which we touched on earlier in our discussion of gains of trade in Section \ref{sec:gains})}.  This is because $dM = -\sum_{i\ne 0} \mu_i dG_i$.  This view extends Samuelson's ideas in micro-economics (developed by Russell and Cooper \cite{Ru,CR01,CR11}) to macro-economics.  It may be particularly relevant for a multi-part economy, where moving along an isentrope could correspond to reversible trade or the (infinitesimally) slow flow of money from one economy to another.

\subsection{Symmetry results}
\label{sec:symm}
Let us now consider the consequences of the symmetry of second derivatives of twice continuously differential functions.
For example, 
$
\beta = \frac{\partial S}{\partial M}_{|_G}
$
and (considering just a single other good, for notational simplicity) $\nu = \frac{\partial S}{\partial G}_{|_M}$, so \begin{equation}\frac{\partial \beta}{\partial G}_{|_M} = \frac{\partial \nu}{\partial M}_{|_G}.
\label{eq:symm}
\end{equation}
Using $\beta=\frac{1}{T}$, $\mu = \frac{\nu}{\beta}$ and $C = \frac{\partial M}{\partial T}_{|_G}$, this can be written as $$\frac{\partial T}{\partial G}_{|_M} =  \frac{\mu}{C}-T\frac{\partial \mu}{\partial M}_{|_G},$$
which is non-trivial and testable. It relates {inflation} (the increase in the temperature $T$), on introducing more of a good, to the market price, the money capacity and the change in the market price on introducing more money. 
An interesting question is whether one can say something about its sign.  {If a good is desirable ($\mu > 0$) one might expect its price to go up if more money is added to the economy; but this would still not settle the sign of $\frac{\partial T}{\partial G}_{|_M}$.}  It is plausible that if a good is desirable, then when more is added to an economy, its price will go down---which will drag down all prices to stay in equilibrium, so money becomes more valuable, i.e., $T$ goes down: $\frac{\partial T}{\partial G}_{|_M} \le 0$; but it is not clear that this is {always} right.
We think one could make examples with either sign.  We already have examples where it is zero (our toy economy or indeed any economy with pure money).  The only constraint we see is (\ref{eq:minor}).

Similarly, one can consider the effect on the temperature of changing the amount of a good by purchase at the market price.  We denote such a derivative by 
\begin{equation}
\frac{\eth}{\eth G} = \frac{\partial}{\partial G} - \mu \frac{\partial}{\partial M}
\label{eq:compensated}
\end{equation}
(called a ``compensated'' derivative in economics, and derivative at constant entropy in physics). 
From $\frac{\eth \beta}{\eth G} =\frac{\partial \beta}{\partial G} - \mu \frac{\partial \beta}{\partial M}$, $\mu = \nu/\beta$, $\beta = 1/T$ and (\ref{eq:symm}),
we obtain
$$\frac{\eth T}{\eth G} = - T \frac{\partial \mu}{\partial M}.$$
Intuitively again for a desirable good, the righthand side looks negative {(if we add more money to an economy but keep the amounts of goods fixed, we might expect prices to increase)}. We can prove this if money is pure:~using $\mu = \nu/\beta$, we deduce $$\frac{\partial \mu}{\partial M} = T\left(\frac{\partial^2 S}{\partial M \partial G} - \mu \frac{\partial \beta}{\partial M}\right).$$  For pure money the mixed second derivative is zero. Concavity of entropy implies that $\frac{\partial \beta}{\partial M}\le 0$.  So we would deduce that purchase of new goods is deflationary.  On the other hand, it might be that the good is substitutable for money but people prefer money, in which case the price of the good might go down if money increases.  Thus we can not in general deduce a sign for $\frac{\eth T}{\eth G}$. Rather fancifully, we can imagine a prison economy in which the supply of currency, $Au$, is drastically restricted, and people trade primarily with, say, matchsticks. If aurums are allowed to flow into the prison, then matchsticks may no longer be used for trading at all, and their value may collapse to be close to zero. In such circumstances, it may be possible that the \emph{price} of matchsticks, measured in $Au$, may actually fall when more money is added to the economy.



As another example of consequences of symmetry of second derivatives, we obtain the identity $$\frac{\partial \mu}{\partial T}_{|_G} = \frac{\partial S}{\partial G}_{|_T},$$
where the amount of money is allowed to change to accommodate the indicated variations.
The proof is to write $d(TS-M) = S\ dT + \mu\ dG$, and use symmetry of the second derivatives of $TS-M$.\footnote{{Spelling this out in more detail, we know that in general
$d(TS-M) = S\ dT + T\ dS + \mu\ dG$. But when trading at market prices, entropy will be unchanged, so that $dS=0$. Trading at market prices also implies that the change in money $dM$ is precisely balanced by the change in the quantity of the good, $dG$ multiplied by its price $\mu$. This gives $d(TS-M) = S\ dT + \mu\ dG$. The first partial derivatives are therefore: $\frac{\partial (TS-M))}{\partial G}_{|_T} = \mu$ and $\frac{\partial (TS-M)}{\partial T}_{|_G} = S$. Using the symmetry of second derivatives ($\frac{\partial^2 (TS-M)}{\partial T \partial G} = \frac{\partial^2 (TS-M)}{\partial G \partial T}$), gives the required result that: $\frac{\partial \mu}{\partial T}_{|_G} = \frac{\partial S}{\partial G}_{|_T}$}.}
This identity tells us that the marginal change in price of a good, given a change in financial ``temperature'' (fixing the amount of the good), is the same as the marginal change in the aggregate utility, given a change in the amount of that good (fixing the value of money). 
One can extend this to multiple types of good, obtaining symmetry of the derivatives $\frac{\partial \mu_i}{\partial G_j}_{|_T}$.

{Next we consider the changes in prices resulting from purchases of goods, possibly of a different type.  Suppose an additional quantity of goods of type $j$ is bought by the economy from elsewhere at market price.  How much does the market price of type $i$ of good change?  This means we want to know the ``flexibilities''
\begin{equation}
\calM_{ij} = \frac{\eth \mu_i}{\eth G_j} 
= \left(\frac{\partial}{\partial G_j}-\mu_j \frac{\partial}{\partial M}\right) \mu_i,
\label{eq:flexibilities}
\end{equation}
using the above compensated derivatives. 
This is a variant of the Slutsky equation, which normally describes the effect on demand of a change in prices (``elasticities''), taking into account the concomitant change in money, but here we are considering the effect on prices of a change in amounts of goods.
Using $\mu_i = \nu_i/\beta$, $\nu_i = \frac{\partial S}{\partial G_i}$, $\beta = \frac{\partial S}{\partial M}$, after some manipulations we obtain
\begin{equation}\beta \calM_{ij} =  \frac{\partial^2 S}{\partial G_j \partial G_i} - \frac{\partial}{\partial M}\left(\frac{\nu_i \nu_j}{\beta}\right).
\label{eq:Mij}
\end{equation}
We see that this is symmetric in $i$ and $j$.  

One may also ask about the sign of the diagonal components of $\calM$.  They are non-positive, because for $i=j$ the right hand side of (\ref{eq:Mij}) can be written (suppressing the indices) as
\begin{equation}\left[\begin{array}{c c} 1 &-\mu \end{array} \right] \left[\begin{array}{cc} \frac{\partial^2 S}{\partial G^2} & \frac{\partial^2 S}{\partial G \partial M} \\ \frac{\partial^2 S}{\partial M \partial G} & \frac{\partial^2 S}{\partial M^2} \end{array}\right] \left[\begin{array}{c} 1 \\ -\mu \end{array}\right] \le 0,
\label{eq:diagflex}
\end{equation}
by concavity of $S$.  Thus $\frac{\partial\mu}{\partial G}_{|_{S}}$ is $T$ times this and so is non-positive.

Furthermore, we deduce that the flexibility matrix $\calM$ is negative semi-definite, because for any vector $x$ with coordinates corresponding to amounts of types of goods, 
\begin{equation}\beta\, x^T \calM x = \left[\begin{array}{cc} x^T &-x\cdot \mu\end{array}\right] \left[ D^2S \right] \left[\begin{array}{c} x \\-x\cdot \mu \end{array}\right] \le 0,
\label{eq:negdef}
\end{equation}
where $D^2S$ is the matrix of second derivatives including with respect to money as the last coordinate. 

These observations are reminiscent of some standard results in economics. Firstly, for a profit-maximiser who chooses prices as a function of demands for goods, the flexibilities are shown to form a symmetric negative semi-definite matrix, just as we have obtained here for how prices respond to amounts of goods in a macro-economy.
Secondly,
for a utility-maximising agent, where prices are considered as independent variables and demands as responses, the elasticities (derivatives of demands with respect to prices) are shown to form a symmetric negative semi-definite matrix.  We will derive the analogue of the second result for a macro-economy next under a non-degeneracy condition.

If $S$ is strongly concave ($D^2S$ invertible) then for $x\ne 0$ we obtain strict inequality in (\ref{eq:negdef}).  Furthermore, we obtain that $\calM$ is invertible and its inverse $\calM^{-1}$ is negative-definite.  $\calM^{-1}$ describes how amounts of goods in an economy depend on prices.  In particular, the cross-elasticities are symmetric and the diagonal components of $\calM^{-1}$ are negative.}

The use of the above assumption of strong concavity raises the question of whether there is an economic analogue of the phase transitions that are so important in physics, where a continuous change in parameters of a system leads to an abrupt change in its properties (e.g., the cooling of water leading it to turn into ice).  Phase transitions will occur where $S$ is not strictly concave.
If the quantity of goods is in an interval of non-strict concavity then the population may partition into one fraction with average goods per agent at one end of the interval and another fraction with average goods per agent at the other end ({by analogy with liquid/vapour coexistence in chemistry). Exploring this type of phenomenon in an economic context is an interesting direction for future work.} Note that the other apparent {way in which phase transitions might arise, through} a jump in derivative of the entropy, is ruled out in the current theory by the proof of differentiability of entropy.



\subsection{Partial derivatives under different constraints}
Returning to the discussion of results on partial derivatives, we {derive two more} results analogous to a result of Samuelson's that he called the Le Chatelier-Samuelson principle.  

The first result compares the dependence of value for a good on the amount of the good under two conditions:~fixed money or fixed temperature.  The result is that if $\frac{\partial \beta}{\partial M} \ne 0$ (else it might not be possible to hold temperature fixed) then
$$\frac{\partial \nu}{\partial G}_{|_M} \le \frac{\partial \nu}{\partial G}_{|_T} \le 0.$$
The proof is that fixed temperature requires $0=d\beta = \frac{\partial\beta}{\partial G} dG + \frac{\partial\beta}{\partial M} dM$ (where unless indicated, partial derivatives are at constant $M$ or $G$ respectively), so  $dM = -\frac{\partial\beta}{\partial G}/ \frac{\partial \beta}{\partial M}\, dG.$
Then 
$$d\nu = \frac{\partial \nu}{\partial G} dG + \frac{\partial \nu}{\partial M} dM = \left(\frac{\partial \nu}{\partial G} - \frac{\partial\nu}{\partial M}\frac{\partial\beta}{\partial G}/\frac{\partial\beta}{\partial M}\right) dG.$$
So $$\frac{\partial \nu}{\partial G}_{|_T} = \frac{\partial \nu}{\partial G} - \frac{\partial\nu}{\partial M}\frac{\partial\beta}{\partial G}/\frac{\partial\beta}{\partial M}.$$
The factors in the numerator of the second term on the right are equal by symmetry and the denominator is negative.  Hence $\frac{\partial \nu}{\partial G}_{|_T} \ge \frac{\partial \nu}{\partial G}_{|_M}$.  But also the righthand side can be recognised as the quotient of a 2-dimensional minor of $D^2S$ (non-negative) by $\frac{\partial\beta}{\partial M} $ (negative), hence $\frac{\partial\nu}{\partial G}_{|_T} \le 0$.

For the second result, we consider two or more types of good and take into account changes in money required to purchase or sell goods.
As before, we consider the amounts of goods as the independent variables and the prices as dependent.
The result is that 
\begin{equation}\frac{\eth \mu_1}{\eth G_1}_{|{G_2}} \le \frac{\eth \mu_1}{\eth G_1}_{|{\mu_2}} \le 0.
\label{eq:LeC}
\end{equation}
The partial derivatives here are those that take into account compensation by the change in money, as in (\ref{eq:flexibilities}), but holding $\mu_2$ fixed in the second instead of $G_2$. 
For notational simplicity, we will indicate which variables are held fixed only when they are not just the other goods.  We assume that $\frac{\eth \mu_2}{\eth G_2} \ne 0$ (hence negative, by (\ref{eq:diagflex})), else it might not be possible to hold $\mu_2$ fixed.  The proof is similar to the first result.
We use
$d\mu_i = \sum_j \frac{\eth \mu_i}{\eth G_j}dG_j$, to show that if $d\mu_2=0$ then $dG_2 = -\frac{\eth \mu_2}{\eth G_1}/ \frac{\eth \mu_2}{\eth G_2}\, dG_1$ and then 
$$\frac{\eth \mu_1}{\eth G_1}_{|{\mu_2}} = \frac{\eth \mu_1}{\eth G_1} - \frac{\eth \mu_1}{\eth G_2} \frac{\eth \mu_2}{\eth G_1}/\frac{\eth \mu_2}{\eth G_2}.$$
By symmetry of the partial derivatives, the numerator in the second term is non-negative, and by negativity of the denominator we deduce the first inequality in (\ref{eq:LeC}).  The second inequality comes by recognising the righthand side as the quotient of a 2-dimensional minor of $\calM$ by $\frac{\eth\mu_2}{\eth G_2}$.

{Samuelson (and commentators such as \cite{MR}) frame discussion of his relations in terms of short- and long-term, but they seem to us to depend simply on whether goods or prices are held fixed.}

Note that many of these cross-relations simplify if money is pure (in the sense of Section~\ref{sec:puremoney}).  In particular, all the cross-derivatives between money and goods are zero. 
{In more detail, if money is pure, one observation is that $T=M/K$ and so the money capacity $C=K$.  Another is that the compensated derivative $\frac{\eth T}{\eth G} = -C\mu$.  A third is that in the Le Chatelier-Samuelson principle, $\frac{\partial \nu}{\partial G}_{|_T} = \frac{\partial \nu}{\partial G}_{|_M}$.}

There is a further economic consequence that follows nicely in the present thermal macroeconomic account---perhaps not surprisingly, as Samuelson derived it by analogy with thermodynamic considerations. In his Nobel lecture, he noted ``While reading Clerk Maxwell’s charming introduction to thermodynamics..., I found that his explanation of the existence of the same absolute temperature scale in every body could be true only if on the $p-v$ diagram that I earlier referred to in connection with LeChatelier’s Principle, the two families of curves---steep and light or less-steep and heavy---formed parallelograms like $a$, $b$, $c$, $d$ in Figure 2 which everywhere have the property that $\textrm{Area } a/\textrm{Area }b = \textrm{Area }c/\textrm{Area }d$. And so it is with the two different economic curves. It is a consequence of the Hotelling integrability conditions...'' Samuelson then draws out the economic parallel:~that a price change for some good has less influence on the quantities of good held when the quantities of other goods can change (but with fixed prices), than when the quantity of those other goods is fixed. 

\begin{figure}[h]
   \centering
   \includegraphics[width=2.7in]{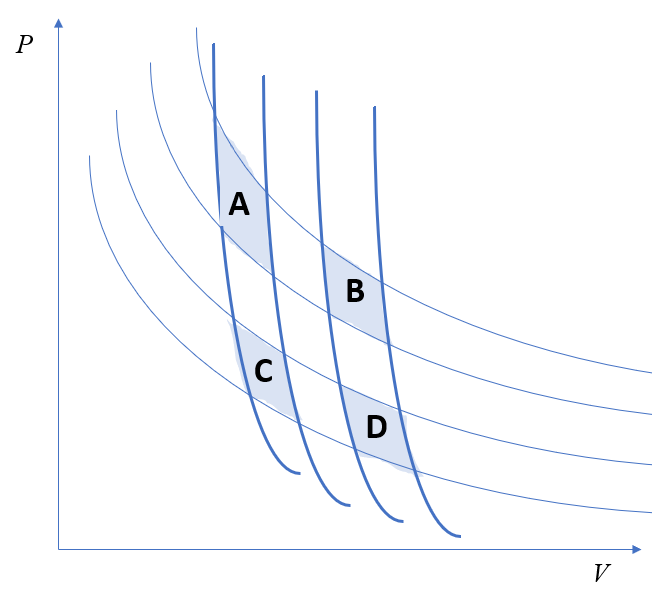} 
\includegraphics[width=2.7in]{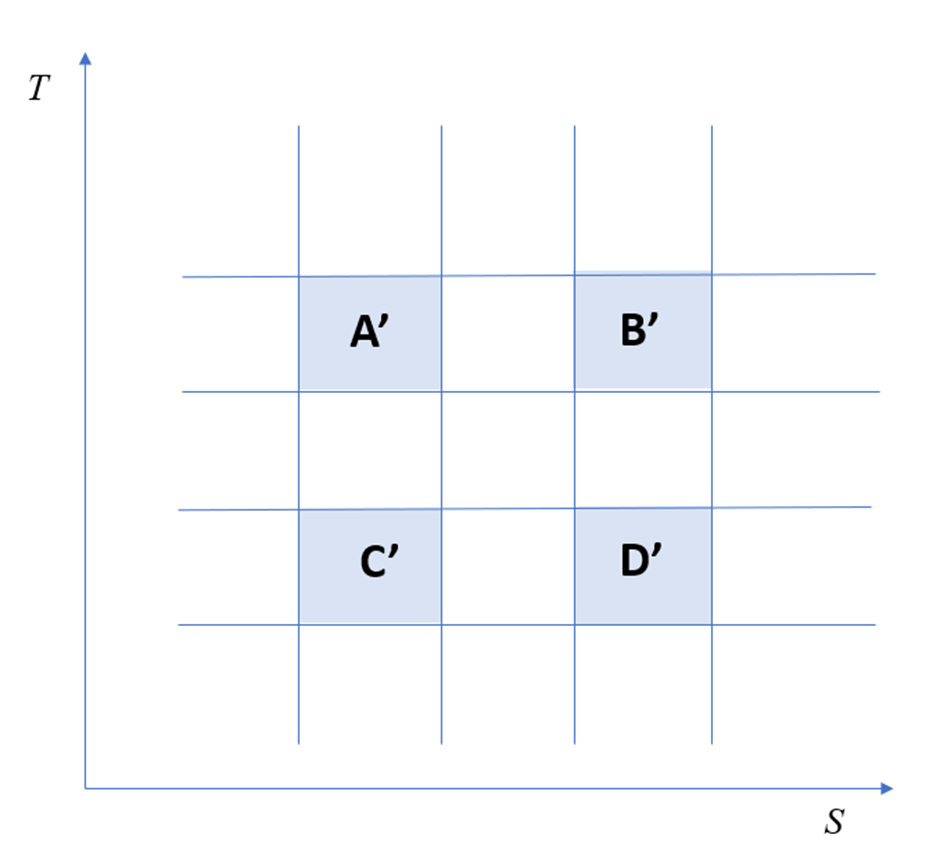}
   \caption{Relationship between volume (quantity of goods) and price, with changes at a constant temperature (light lines) and entropy (heavy lines). The same graph replotted in temperature-entropy coordinates, so that the changes move the system along the lines of a grid. The amount of money that the trader can extract by reversibly running a Carnot cycle around these areas is $\int PdV = \int TdS$---thus the areas $A$, $B$, $C$ and $D$ are the same as the corresponding areas $A'$, $B'$, $C'$ and $D'$. The equal area rule clearly applies to the areas on the grid in TS coordinates, and hence in the original PV coordinates in Samuelson's discussion.}
   \label{fig:EqualArea}
\end{figure}

Samuelson applied his analysis to an individual agent:~specifically a ``profit maximizing monopolist'' who adjusts the inputs required in production. In our account, as usual, we instead  focus on the economy as a whole. In particular, we consider Carnot cycles running around the periphery of the areas $A$, $B$, $C$ and $D$ that Samuelson identifies. The amount of money that can be extracted by the trader in one turn of the Carnot cycle is represented by the area $\int PdV$ of the region whose boundary is enclosed by the path. Replotting in $TS$ coordinates, 
the Carnot cycle now corresponds to running round the sides of rectangles of a grid. The areas are preserved in the re-plotting. That is, in $TS$ coordinates the amount of money extracted by the trader by going round the Carnot cycle is represented by the area $\int TdS$: i.e., $Area_A=Area_A'$, $Area_B=Area_B'$, etc. In the $TS$ plot, it is immediately clear that $Area_A'/Area_C'=Area_B'/Area_D'$, hence giving Samuelson's result for  $PV$ coordinates: $Area_A/Area_C=Area_B/Area_D$. See Figure~\ref{fig:EqualArea}. Note, again, that while Samuelson's result applies only for an individual maximising agent, the present result holds for the economy as a whole.

\section{Further considerations for exchange economies}
\label{sec:further}

Here we raise some further aspects of exchange economies that seem natural topics for future work extending the thermal macroeconomic framework.

\subsection{More than one currency}
\label{sec:currencies}
We have so far restricted ourselves to considering a single currency, the Aurum $Au$. How can the TM approach be extended to deal with multiple currencies, either within a single economy or between economies? 

There are various cases in which multiple currencies are used in a single economy, e.g.~in the long and controversial history of ``bimetallism'' in Europe, the U.S. and elsewhere, where two forms of money, typically silver and gold, are used in parallel (see \cite{Fay}), and in the recent proliferation of cryptocurrencies, though these are only rarely used for trading.\footnote{Although bitcoin has been made legal tender in El Salvador:~https://www.bbc.co.uk/news/world-latin-america-57398274} 
In the TM framework, we can simply view the currencies as different forms of good.  Then as with any other goods there will be a market exchange rate, generalising the notion of market price, as in Section~\ref{sec:price}.  This is the exchange rate that if posted by an external trader would lead to no nett flow of either currency to or from the trader.\footnote{In practice, of course, bimetallism typically involved the government setting a fixed exchange rate between gold and silver.} 

The market exchange rate between two currencies within an economy will change depending on the quantities available---for example, the discovery of large new silver deposits would tend to lower the value of silver currency in relation to gold. The case in which an economy has two forms of pure money is particularly easy to analyze. 
Then $S = k_1 \log M_1 + k_2 \log M_2 + F(G)$, for some constants $k_1,k_2$ and function $F$ of amounts $G$ of all other goods.  It follows that the price $\mu_i$ of a good using currency $i$ is $\frac{\nu M_i}{k_i}$.  So if the amounts of the two currencies are fixed, the ratio $\mu_1/\mu_2$ of prices for a given good is constant ($\frac{M_1k_2}{M_2 k_1}$).

In the second context, where two economies use different currencies but currency-flow between economies is allowed, then currency-exchange is a special case of trade, in which economy $A$ can only buy goods in economy $B$ in two steps:~first exchanging currency $A$ currency for currency $B$, and then using currency $B$ to purchase the good. Thus, the currencies play a key intermediary role  in trade. When two or more economies with different currencies are in equilibrium, the prices of goods will be the same in all economies (taking account of any tariffs, as above) up to a scale-factor depending on the exchange rates between those currencies. The market exchange-rate between economies will be such that there is no nett flow of currency between economies. These conclusions follow directly from the analysis of trade outlined above:~we discussed trade in the context of a single common currency, but that was not really necessary (indeed, the analysis of trade still applies where one allows exchanges of bundles of goods that do not involve any currency). We leave the extension of this type of analysis to more complex cases, including the impact of fixed exchange, currency controls, and so on, for future work. 

A further direction to explore systematically is the impact of the choice of the goods or goods that are treated as currencies. We noted in Section~\ref{sec:entropy} that entropy does not depend on the choice of num\'eraire, but other economically relevant quantities, such as temperature and of course prices, will do so. Moreover, if we take ``frictions'' in exchange seriously (which will make perfectly reversible trades impossible) then it is interesting to ask how (and to what extent) the existence of money reduces these frictions, which may in turn throw light on which num\'eraires are ``good'' choices to serve as currency. Another direction for future work is to consider the economic significance of different forms of money, from commodity money such as gold or silver, to paper money and purely digital currencies; and the impact of IOUs, cheques, the existence of credit, and so on. These and related questions appear tractable within the TM framework, but the details remain to be worked out.




\subsection{Spatial inhomogeneity}
\label{sec:loceqm}
An important feature of real economic systems is that they are geographically and sociologically inhomogeneous. 
How does this affect temperature and prices?
According to our theory, if in equilibrium then temperature and prices should not depend on position in the economy.
Let's compare with physics.  For an atmosphere in a gravitational field, if it is in equilibrium (in particular, with no input or output fluxes of radiation) then indeed the temperature becomes constant.  But the pressure $p$ does not.  Instead it is $p+g\rho h$ that becomes constant (where $\rho$ is the density, $g$ the gravitational field and $h$ the height).  So do prices require some analogous correction?

\subsection{Heat and work}
In thermodynamics the distinction between heat and work is usually important.  Is there an analogue in economics?  It seems that money transfer for goods at market prices should qualify as work and any other money transfer as heat.
Concretely, $dM = T\, dS + \mu\, dG$ for quasi-static changes.  So sale of goods at market price can be regarded as work $\mu\, dG$ done on the system and all other money flow $T\, dS$ as heat flow into the system.

Heat flow and work have different consequences. Work is that component of the trade or other interaction that is reversible: it moves the system on an isentrope, and thus leaves the aggregate utility of the system unchanged. Moving the system isentropically will, though, typically involve the system losing or gaining money, which may be captured, or provided, by some external system, such as our trader. By contrast, ``heat flow'' into a system is not reversible: the system moves to a state with higher entropy, i.e., moving to a new isentrope, with higher aggregate utility. 

Thus, in the context of trade between two economies, we can see the gains of trade being split, in some proportions, between the trader (or, more generally, any intermediary or system of intermediaries as a real economy with importers, wholesalers, retailers and so on) (work) and the agents in the economies (heat flow). At one extreme, the trader trades slowly at market prices and captures all the gains of trade; at the other extreme, there is a free-for-all of direct exchange, leading to new equilibrium prices, and the entire gains of trade are assimilated as ``heat flow'' by the agents in the economy itself (where the amount gained depends purely on the final equilibrium, and not the path by which it is reached).

\rev{One direction we highlighted in Section~\ref{sec:friction} for further work is the incorporation of trade frictions, which intuitively generate ``heat''.}

\subsection{Free energy}
\label{sec:freeenergy}

As touched on in Section~\ref{sec:gains}, it is common in physical thermodynamics to consider systems in thermal equilibrium with a bath at a temperature $T$, and then to replace the principle of non-decrease of entropy $S$ by that of non-increase of free energy $F=E-TS$.
If a network of economies is in financial equilibrium, as one could presume is the case for those with no currency controls, then our theory would give the same temperature (fluxes of inputs and outputs are neglected here).  It then becomes natural to consider the global economy to be at a given temperature and individual economies are then more simply described by $F=M-TS$, considered as a function of $M$ and $G$.  The principle of maximising entropy is replaced by minimising $F$.  Similarly, if goods were freely traded with a large reservoir economy, one could expect a fixed common price $\mu$, and it becomes simpler to minimise $M-TS+\mu G$ as a function of $M$ and $G$. From this perspective, an economy whose free energy can be reduced is one that can effect useful changes.

While notions of free energy, exergy and their generalisations may be useful for analyzing special cases in economics, it is likely to be more useful in general just to look at potential increases in entropy.  For a single economy at equilibrium, of course, it is already at maximum entropy and no increases are possible, but for a composite economy, one can look at the possible increases, as discussed in Section~\ref{sec:trade}.

\subsection{Entropy as liberty}
\label{sec:liberty}
John Stuart Mill's great works of political philosophy, \emph{On Liberty} \cite{MillL} and \emph{Utilitarianism} \cite{MillU}, attempt to promote the idea that a free society, where liberty is maximised, will also maximise welfare.\footnote{The welfare theorems of conventional economics take a step in this direction, if we assume that competitive equilibria are the result of maximally free choice in trading of goods and a Pareto optimum is at least a local optimum in utility for individual agents, given the choices of the others (though competitive equilibria will not in general maximize the summed utilities over agents). But the link between maximizing liberty and welfare has been controversial ever since. 
The thermal macroeconomic framework provides a potentially interesting perspective on the relation between ideas of utility and liberty, by drawing a parallel with the ``macro'' interpretation of entropy of classical thermodynamics and the ``micro'' interpretation of entropy provided by statistical mechanics.} In statistical mechanics, entropy gains a concrete interpretation as the logarithm of the available volume corresponding to given macro-state.  Volume is measured using Liouville measure in phase space \cite{Se} or its quantum analogue (commonly done by counting eigenstates, but more generally by using Fubini-Study metric \cite{BH}).  Is there an interpretation of economic entropy as the logarithm of the available volume of a macro-state?  This would connect to ideas of maximising liberty:~the volume at the ``micro-level'' associated with a macro-state corresponds, intuitively, to the number of different micro-states compatible with that macro-state. Thus, higher entropy macro-states impose fewer constraints at the micro-level, leaving economic agents greater ``freedom of choice.''

For the toy economy of Section~\ref{sec:toy}, the natural notion of volume is the normalisation constant $Z$ of (\ref{eq:Z}) (called ``partition function'' in statistical physics).  {To leading order in $N$, the logarithm of $Z$ depends on $G$ and $M$ as\footnote{The constant $C = \log\Gamma(\alpha)-\alpha\log\alpha - \alpha + \log\Gamma(\eta)-\eta\log\eta-\eta$.} 
$N \left(\log (\frac{G}{N})^{\alpha} (\frac{M}{N})^{\eta} + C\right)$, so indeed the entropy corresponds to logarithm of available volume (we will show elsewhere \rev{\cite{M25b}} that this is true in greater generality)}.
If a scale for entropy can be chosen so that it is logarithm of available volume then the fluctuation formulae from statistical mechanics carry over (mentioned in Section~\ref{sec:thermometer}). In the simple context of an exchange economy, maximising liberty is maximising the number of ways of in which goods and money can be distributed across agents. For example, if we have two economies, $A$ with only apples and money, and $B$ with bananas and money, and allow them to trade, then the volume of possible stocks of goods increases---each agent in both $A$ and $B$ is now ``free'' to have non-zero stocks of apples, bananas and money. Thus, allowing trade can be viewed as increasing the liberty of agents, and this remains true independent of the details of the preference or decision-making mechanisms of those agents. Indeed, increasing liberty might be viewed as a virtue of trade, and indeed markets more generally, irrespective of whether agents have coherent preferences (and hence well-defined utility functions at the level of individual agents).

There is a potentially interesting parallel with Sugden's \cite{Su04} opportunity-criterion defence of the market. He argues that the virtue of markets in an exchange economy such as we have described here is that they maximise the number of possible alternative bundles of goods that an agent can hold in exchange for their current bundle. This is very close to the present perspective. Sugden is considering maximising the number of potential exchanges for each agent, using deterministic agents in a competitive equilibrium. By contrast, the statistical mechanics viewpoint sketched here considers the volume of states generated by actual exchanges in statistical equilibrium considering the economy as a whole. We leave exploring potential links for future work, but note that the present approach potentially yields an ``opportunity criterion'' justification not just for individual markets but for any entropy-increasing transition. That is, certain types of economic processes (e.g., trade, financial contact) may be justified on the grounds that they lead to increased ``opportunity sets'' for consumers, in the terminology of \cite{Su04}. This analysis may also help clarify the alignments, and divergences, between the utilitarian and liberal political objectives that trace back to Mill.\footnote{Connections can also be made with ``future state maximisation'' models of control process underlying animal collective behaviour, as outlined in \cite{CT}, according to which each animal acts to maximize the variety of its possible future sensory states, and with Friston's free energy principle, e.g.~\cite{PPF}.} 

A question for the future is to explain how rules and conventions arise. On the face of it, these seem to reduce liberty by restricting individual behaviour. But good rules and conventions (like agreeing to drive on one side of the road or the conventions of language) actually increase liberty by enlarging the space of actions people can take by coordinating their behaviour \cite{Bi}.

\subsection{Migration}
We have considered each economy as consisting of a fixed number of agents, but suppose we allow them to move between economies. An immediate question is how to define a reversible change, which would thus leave entropy unchanged. A natural approach is for our trader to mediate, as before, but now charging people entry when moving from one country to another (the charge could be negative, incentivising people to move to a country). The trader would the extract money from a system of several countries by facilitating ``migratory contact'' between economies, and extracting the maximum possible amount of money by slowly reducing the charge until there is no nett flow of people. The process could then be reversed by the trader slowly levying the same charges, but with opposite sign. The ``gain of trade'' extracted by the trader will behave as gains of trade when goods are exchanged, i.e., depending purely on the initial and final state. That is, the amount of profit the trader can extract will be independent of the order in which migrations occur. Moreover, when there is free movement of people, and thus no charge from the trader, the entropy of the system will increase (not necessarily to the mutual benefit of each economy). There will, of course, be interesting interactions between migrations of people and flows of money and exchanges of goods. So, for example, suppose that the distribution of preferences might differ across economies, such that in economy 1 people mostly prefer apples to bananas but the opposite in the otherwise identical economy 2. Then we should expect a mutually beneficial exchange of apples and bananas between the economies (which could be used to harness a gain of trade, by our trader). But alternatively, if migration is allowed, some people might leave economy 1 for economy 2 (attracted by the availability of bananas), while people might move from economy 2 to economy 1 (attracted by the availability of apples). Here trade and migration are, to an extent, substitutes. With sufficient migration before trading begins, the difference in tastes between the economies could be eliminated entirely (perhaps generating a large profit for the trader in the process); but if so there will be no further nett trade (and no gains of trade for the trader to exploit). Conversely, sufficient trade will equalise the prevalence and prices of apples and bananas in each economy, reducing or eliminating the incentive to migrate. Clearly any mix of trade and migration is possible---and the gains of trade (and, relatedly, the entropy gain for the pair of economies as a whole) will depend only on the final state (the stock of people, money and goods in the economies), and will be independent of the path by which this state is achieved (e.g., migration first, then trade; the opposite; or any other variation).

We leave further exploration of migration to future work, including the crucial questions of whether agents take their goods and money with them when migrating from one economy to another and taking account of the ``frictions'' involved (which might be caused by the logistical and social costs of moving and adjusting to a new economy). Note, though, that it is possible to make progress (e.g., establishing the aggregate utility increase arising from the movement of people, in this idealized setting) without needing to make microfoundational assumptions about the distribution of preferences, or decision-making procedures, of individual agents.

\subsection{Non-extensivity}
We made explicit in Section~\ref{sec:extensivity} that the present analysis assumes scaling symmetry---i.e., that economies can be divided into identical copies with the same properties apart from scale. It seems unlikely that this is a good approximation. For example, as we noted earlier, economies contain organizations at many scales, and some industries are not viable at a small scale. Nonetheless, there are robust \rev{empirical} power-law scalings
throughout many aspects of the economy \cite{Gab09}, so it is possible that a more general notion of entropy, such as Tsallis entropies, can be applied  \cite{Ts}. Recent work in non-extensive statistical mechanics (e.g.~the special issue \cite{B+} and the paper \cite{LY14}) may provide a useful starting point.

\subsection{Externalities}
A particularly important application of the TM framework will be to consider externalities, where (in our 
\rev{context of} exchange economies) the goods held by an agent affect not only that agent's welfare, but indirectly affect the welfare of others (either positively or negatively). So, for example, we might imagine an economy of leisure activities around a small lake, where people can buy and sell fishing rods among other things. But buying (and, presumably using) a fishing rod will reduce the utility of fishing for others, by reducing the number of fish for them to catch. The trader can slowly and hence reversibly sell fishing rods to the lakeside economy until the price of a fishing rod is near zero\footnote{Reversibility assumes, perhaps implausibly, that fish stocks would fully recover as the number of fishing rods reduces---though the assumption is perhaps less implausible if we recall that reversible trades are required to be arbitrarily slow}. 
At this point there are almost no fish and almost no utility to be gained from owning a fishing rod. 

Thus, when fishing rods are few, a subset of people enjoy fishing (so that the summed utility for fishers may be large); but when the number of fishing rods is high, the summed utility for fishers may be much lower. So the existence of externalities breaks the relationship between entropy and ``utility'' in any sense relevant to the welfare of individual agents. This is analogous to the breakdown of the welfare theorems in conventional microeconomics. The connection between the equilibrium outcomes of a competitive market and Pareto optimality requires that there are no externalities, and does not hold otherwise. 


Following conventional economic analysis, we can address the problem of externalities in our lakeside economy by, for example, having a market for a fixed number of fishing permits (a ``cap and trade'' scheme), or simply fixing the number of fishing rods, so that voluntary exchanges involving fishing rods no longer harms other agents. Notice that the present approach still provides a potential route to analyse whether one set of rules in an economy is ``better'' than another, without having to measure and sum the utilities of individual agents. Specifically, we can make clones of our lakeside economy, each governed by different rules; and then put these in contact and ask about the direction of the flow of money when the clones are put in financial contact\footnote{or indeed migration of people, if we include this in our analysis.}. A natural assumption is that money will flow to the clone with the ``best'' set of rules---which imply that an economy with such a set of rules would have a lower temperature (and thus that the marginal utility of money will be greater). We leave this and related cases (e.g., dealing with positive externalities) for future work. 

We see the fact that the TM framework applies straightforwardly in the context of externalities as a significant strength:~economies will be governed by the second law whether there are externalities or not. All that changes is the \emph{interpretation} of entropy as corresponding to aggregate welfare. Indeed, by operating at the macro-level, the TM approach, and the second law in particular, applies irrespective of the micro-foundations concerning agents and their interactions, so long as the macro-level axioms hold. So, for example, there are no requirements that agents make decisions based on consulting a utility function of any form, that their preferences are transitive or complete, and so on. Moreover, an agent's utility (if it has one) and behaviour may depend in complex ways on the utilities and behaviours of other agents \cite{FF}\footnote{Indeed, we can view the wide variety of social preferences, whether positive or negative, as particular forms of externality, and as generating interdependencies between agents, which will not typically impact the ``macro'' assumptions that underpin the present approach.} or on comparisons with the agent's own prior or current experience \cite{Vl}. Different microfoundations will, of course, change the properties of the entropy function, just as in physics where different microscopic properties of a gas (for example, whether it approximates an ideal gas in which molecules do not interact or whether the gas molecules are subject to van der Waals forces) determine the entropy function. Crucially, though, the macro-level analysis can proceed and macroscopic quantities such as prices, temperature and entropy can be measured, even when the microfoundations are not known or are too complex to model.

\subsection{Varieties of good and agents}
Can one extend the TM theory beyond discrete categories of good? One possibility is a continuum of types of goods of some type.  For example, what one might think of as a single type of good, could actually vary continuously in quality. Relatedly, for pieces of fruit, loaves of bread, art works, and the like, each instance of an object is to some degree unique---this raises the question of whether goods need to be modelled as splitting up into discrete categories.

We see no in-principle reason why TM does not extend to deal with continuous variation between goods or each good being unique. To take a particularly simple example, consider two economies $X$ and $Y$ that are identical except regarding a single good, bread, of which there are distinctive varieties $B_x$ and $B_y$. Suppose that trade in bread is suddenly allowed between $X$ and $Y$. According to a `micro' interpretation of entropy in terms of sizes of state space (as used in statistical physics), this appears to lead to an increase in entropy, because the volume of possible locations for each loaf of bread of type $B_x$ has now expanded beyond $X$ to include the union of $X$ and $Y$, and similarly the locations for each loaf of bread of type $B_y$ has now expanded beyond $Y$ to include the union of $X$ and $Y$. At a micro-level, this increase in entropy seems, at first glance, to be independent of the degree of difference between $B_x$ and $B_y$---all that matters is that they are distinguishable. This creates a puzzling discontinuity as the two types of bread become arbitrarily similar---because at the point where they are identical, there should abruptly be no entropy change when trade is allowed (because the number of microstates now depends on the volume of ways in which a single good can be distributed across the union of $X$ and $Y$, which is unchanged.\footnote{Aside from the small effect of fluctuations in the total numbers of the good in each economy, which can vary when trade is allowed.}  This type of issue is problematic even in statistical physics, where for example for some purposes differences in nuclear spin are irrelevant, but for others (like crystallisation of isotopes of molecular hydrogen under pressure) not.  If one distinguishes between all hydrogen molecules one runs into Gibbs paradox, that entropy is not extensive.

In TM, no such discontinuity occurs at the macro-level, because increasingly similar goods will simply be ever-better substitutes, and the ``gains of trade'' (i.e., entropy increase) obtained from allowing trade in $B_x$ and $B_y$ will smoothly reduce until it is zero where $B_x$ and $B_y$ are perfect substitutes. Whether or not $B_x$ and $B_y$ are distinguishable is irrelevant---what matters is whether they are treated as substitutes by the agents in the economy. The gain of trade (and hence entropy increase) can in principle be measured macroscopically by the amount of money that the trader can extract from the two economies when slowly mediating trade in $B_x$ and $B_y$. Similarly, this quantity will capture the entropy gain if each loaf of bread in each economy is unique, and where gains of trade may arise through aggregate differences in the characteristics in the populations of loaves in $X$ and $Y$ (e.g., due to differences in ingredients, baking traditions and so on). The gain of trade that the trader can extract from such population differences will, of course, provide a measure of the differences in those populations, from the point of view of the agents in the population themselves---i.e., if the agents are indifferent between bread differing only in the types of flour used in $X$ and $Y$, then there will be no gain of trade to be exploited between the two economies. An interesting direction for future work is to consider these cases where entropy is interpreted as volume of microstates. Here a useful starting point may be to consider that what counts as a distinct microstate is determined from \emph{the point of view of the choices of the agents}. This is reminiscent of the ``subjective'' view of micro-state entropy advocated in statistical mechanics by Edwin Jaynes \cite{Ja}. But in an economic, rather than a physical, context we can see the subjective viewpoint as being fixed by the preferences of the agents within the system.

Relatedly, consider cases where the populations of agents in economies $X$ and $Y$ have different preferences. For example, the agents in $X$ may be indifferent between some aspect of a type of good that the population of agents in $Y$ see as very important (e.g., $X$ may be indifferent between physically identical lab-grown or mined diamonds; $Y$ might prize one type much more highly). TM will capture an increase in aggregate entropy (and associated gains of trade) when the two economies are allowed to trade diamonds, and will predict price changes both for diamonds and indirectly for the prices of other goods and the value of money (and indeed, opening up the possibility of trade in diamonds may induce flows of money or other goods already traded between the economies). TM  makes no assumptions about the heterogeneity of preferences of agents within and between economies---it focuses purely on relationships between macroscopic quantities. 

To consider the most extreme case of heterogeneity of preferences and goods, consider the art market. Here, suppose that each `work' is unique; and each agent has idiosyncratic preferences. In a conventional economic analysis, it is difficult to make sense of the question of whether art is cheaper in $X$ or $Y$---there is no ``basket of goods'' in each economy that can be compared. However, we can ask in which direction money flows, if trade in art is opened up between $X$ and $Y$. If money flows towards $X$, this can be interpreted as implying that art is ``cheaper'' in $X$ than in $Y$ (it is tempting to suggest that ``art works'' should therefore flow in the opposite direction---but of course, where each work is unique, there is no meaningful notion of quantity of art works).

\section{Beyond exchange economies}
\label{sec:future}

We have developed our theory in the context of exchange economies, but this leaves out many features of real economic systems.  Here we sketch some possible future developments. We are optimistic that prior applications of thermodynamics, which has been extended to deal with many complex processes ranging across physics, chemistry, atmospheric sciences, physiology, and ecology, may provide powerful tools and analogues for such extensions.

A key feature of real economies is that some goods are produced and some are consumed. So, for example, crops grow, and food is consumed. Also, money is printed by central banks. Thus it will be important to extend thermal macroeconomics to allow fluxes of goods and money into and out of a system.  This appears to pose no in-principle difficulties---indeed, it is a standard situation in physical thermodynamics, called non-equilibrium thermodynamics.  For small deviations from equilibrium it still produces a steady state, which to leading order minimises the rate of entropy production, but for larger deviations it can produce time-dependent states, such as cycles.  

Another key feature of real economies is that goods are processed and combined to make other goods.
To include manufacture is a relatively simple extension from exchange economies.  A combination of goods is put together in some way to produce a new sort of good. 
Manufacturing bears strong resemblance to chemical reactions, so we might expect to be able to carry over many results from chemical thermodynamics, which studies, among other things, which reactions will spontaneously occur, and which cannot, because they would lead to decreasing entropy. Chemical thermodynamics can treat cascade reactions, where the products of one reaction then drive another; and competition for inputs between different reactions.
These all seem to have natural analogues when considering networks of manufacturing processes which may compete with, and feed into, one another.

Chemical reactions are typically assumed to be reversible, at least in principle. For example, the energy released when chemical bonds are formed is assumed precisely to balance the amount of energy required to break those bonds---so that both matter and energy are perfectly conserved. This might appear to contrast with the economic case, where it ``costs'' money both to assemble a car from its component parts \emph{and} to disassemble it back into its component parts. But the contrast may be more real than apparent:~energy is required (in the form of heat) to bake a cake from flour, eggs, butter and sugar, but surely a great deal more energy is needed to go through the extremely complex process of attempting to reconstitute the flour, eggs, butter and sugar from the cake. Un-baking a cake no more releases energy than dismantling a car releases money.



We see a close analogy between businesses and the technologies they deploy in economics and enzymes in biochemistry. By introducing innovative processes for converting (less valuable) inputs to (more valuable) outputs, businesses lower {the cost of surmounting the barrier} to making this transformation; this {seems} analogous to the operation of enzymes that lower the energetic barrier (``activation energy'') required for a chemical reaction. Biological systems involve cascades of enzyme-driven reactions of staggering complexity; and the enzymes are, of course, themselves constructed through similarly complex processes. The emergence of complexity in biochemical and economic systems may have interesting parallel aspects, which may potentially shed light on both.
Indeed, a macro-approach to biology inspired by thermodynamics has been developed in \cite{KF}.

Another direction to develop is theory for fluxes as a result of value differences.
As we have seen, on putting economies into contact, differences in values (including coolness) lead in general to flows between them.
The flows will depend on the nature of the contact.  For given type of contact and small differences in value one can postulate a linear relation between the value differences $\delta \nu$ and the flows $\delta F$.  In general this is given by a matrix $L$: $$\delta F = L\, \delta \nu.$$  
We have confirmed this for some examples of pairs of simulated micro-economies with substitutes or complements effects \cite{LMC}.
The matrix $L$ can have off-diagonal terms, e.g.~a difference in coolness could drive a flow of a good different from money.  This produces entropy at rate $$\sigma = \sum_j \delta F_j \delta \nu_j = \delta \nu ^T L\, \delta \nu.$$ 
In the physical context, an important observation by Onsager is that because the total entropy can not decrease under the flows, the matrix $L$ is positive semi-definite.  The same applies in thermal macroeconomics. {The significance of this result for economics needs exploring.  For example, suppose two types of good are more useful together than not, and one type can move between economies, the other not.  We suppose money can move between them also. 
Then if all moves that increase total entropy are permitted, the mobile good will move to make the values for it equal, and money will move to make the temperatures equal.  Thus the prices for the mobile good will equalise.  
\rev{Is there} an economic analogue of osmotic pressure?}

The above relations between fluxes and value differences predict dynamics for equilibration, in particular for price formation, generally regarded as a mystery in standard economics.  For finite economies, we expect to see a stochastic version, like a multivariate Ornstein-Uhlenbeck process (compare \cite{M21}).  Alternatively, the system of economies could be held out of equilibrium by constant fluxes of inputs and outputs to and from various of its subsystems, and a non-equilibrium steady state would result.  Again, finiteness would make fluctuations visible, analogous to those analysed in \cite{GL}.
One can also consider a continuum of economies, each in local thermodynamic equilibrium. 
Then small gradients in values will produce fluxes.

Onsager went further in the physics context by showing that if the microscopic dynamics is reversible (meaning invariant under reflecting the direction of time and associated momenta) then $L$ is symmetric.  So for example, the heat flow produced by passing unit electric current through a thermopile (Peltier coefficient), when divided by the temperature, equals the voltage produced by unit temperature difference across it (Seebeck coefficient).  It is natural to ask whether this has an analogue in economics?  On the one hand, the same logic (consideration of correlations between components of fluctuations) should apply; but on the other, there is less reason to expect reversible microdynamics in economics.

Two key aspects of real economies that will require building on the thinking in terms of chemical reactions and fluxes are
labour and finance.  Labour is, of course, fundamentally irreversible:~it takes labour to make changes and also to un-make them.  Thus, a self-sustaining economic system with labour requires low entropy influxes and high entropy outfluxes of materials, analogous to how the earth-system feeds off turning low entropy sunlight into high entropy infrared radiation. Labour is, of course, also closely connected with the size of the population; a more complete model will need to capture births, deaths and flows of population, as well as the processes by which people enter or exit the labour market.

Finally, we note that questions of capital, investment, interest rates, central banks and indeed the entire financial system are absent from the exchange economy outlined here. Here again a perspective based on fluxes is likely to be required. 





\section{Conclusion}
\label{sec:conc}

The fundamental result of our paper is that extensive exchange economies have an entropy function that governs the allowed changes in state on interaction between economies:~they are the changes for which the total entropy does not decrease.  
This is in distinction to the Pareto criterion in standard microeconomics, where the allowed changes are those in which the utilities of no agents decrease.  \rev{In standard economics, u}tility functions specify preferences but can not in general be added meaningfully.  In contrast, entrop\rev{ies of subsystems} can be added, thus allowing meaning to our principle that total entropy can not decrease. 
One might ask how the entropy of one economy could decrease in an interaction. \rev{One way this can happen} is that agents can move money to the other economy where it is of more value to them, which is, of course, a feature of the real world.

Entropy leads to a variety of consequent quantities for an economy, like temperature, money capacity, market prices, and relations between them. In particular, we introduce a notion of temperature for an economy.  From the viewpoint of fluctuations, traditional economics can be viewed as focusing on economies at temperature zero.  Thus, if real economies have positive temperature then TM opens up an important new direction in economics.  Some consequences are the nett direction of money flow between economies is from hotter to cooler, the possibility of money pumps that can exploit temperature differences or enhance them, and a definition of inflation as \rev{relative rate of} increase in temperature. \rev{The theory provides a new perspective, without reference to microfoundations, on the existence of market prices, the value of money, the meaning of inflation, and a new justification for some core principles of mathematical economics, including the symmetry and negative-definiteness of a (macro-version of) the Slutsky matrix, the Hotelling conditions and the Le Chatelier-Samuelson principle.}

For a real economy, even an exchange economy, the entropy function will in general be unknown \rev{and it will not generally be possible to calculate it from micro-foundational assumptions}.  But just as in physics, \rev{it may be possible} to measure it.  For example, steam turbines operate on steam, which is far from an ideal gas (\rev{for which entropy can be calculated from micro-foundational assumptions)}; but engineers use empirically-compiled steam tables to compute optimal operation \rev{of a steam turbine}.  By analogy, we might expect that a new branch of econometrics \rev{analogous to calorimetry} can be developed to measure temperature, money capacity, entropy and other relevant quantities. \rev{We have begun to test this in simulated economies (\cite{LMC} and more to be reported elsewhere).}

If permitted to exchange some types of good and/or money, nett trade can occur in any way that increases the total entropy.  This includes mutually beneficial trade, in which the entropies of all partners increase, but also other trades where one gains and another loses entropy.  Entropy allows one to compute the maximum amount of money that a trader could make.

One perspective on our work is that it provides a justification for the  ``representative agent'' of traditional economics. In conventional economic analysis, while the objective is to understand aggregate behaviour of the economy, the concept of utility applies primarily to individual agents. This creates a gap between macro and micro analysis that is often fixed by postulating the existence of representative agents, with the assumption that the behaviour of such agents should match the behaviour of the economy as a whole. In the thermal macroeconomic framework, this rather convoluted line of reasoning is not needed:~the whole \rev{system of economies} 
behaves as if it were maximising a utility function \rev{(entropy)} and hence can be viewed as a rational utility-maximising agent. 
On the other hand, in interactions with other economies, the set of economies maximises their total utility, rather than each attempting to maximise just its own, as mentioned above.  
The benefits of moving money to another economy are not represented by the entropy of the originating economy.

This paper makes only a start on the thermodynamics of economic systems.  Much more can be done.  In particular, one should include production and consumption, manufacturing and recycling, enterprise, labour, migration, births and deaths, interest rates, multiple currencies, non-extensivity, and consider the possibilities of phase transitions and dynamical behaviour.  Many of these directions reflect the fact that real economic systems are not in equilibrium.  Even if they might still be steady-state at the macro-level, they are subject to fluxes of inputs and outputs.  The question of extending the concept of entropy to non-equilibrium states is addressed in \cite{LY13}, but it was already standard to apply thermodynamics to contexts with fluxes such as biological cells, chemical engineering and the climate system.  In the physical context, thermodynamics places limits on the rates at which processes can occur, given the input and output fluxes (e.g.~see \cite{Kl} for the Earth system).  The same can be expected to apply to economics. Recent developments in non-equilibrium thermodynamics of physical systems (e.g., \cite{FE,Ma,MN}) are also likely to be relevant to modelling economic phenomena.

The theoretical and practical significance of thermodynamic ideas, originally developed for understanding the efficiency of steam engines, has had a lasting impact across physics, chemistry, biology and engineering, the ramifications of which are still being explored. This raises the possibility that establishing the relevance of a parallel mathematical structure in an economic domain may, when more substantially worked out, have similarly far-reaching implications. 

We hope that once fully developed and further tested by simulation (as in \cite{LMC}), this theory may provide powerful methods for thinking about and quantifying real-world economic systems at the macroscopic level\rev{, complementing existing approaches in economics}. \rev{We hope too that, just as classical thermodynamics has had far-reaching practical implications in domains such as engineering, thermal macroeconomics may in the long term have relevance} 
for practical matters such as trade negotiations, inflation, investment, regulation, and the management of an economy. 

\section*{Acknowledgements}

We are grateful to Daniel Sprague for taking on an initial exploration of the ideas for his MSc project in 2011, to Eric Smith for a discussion in 2013, to Elliott Lieb and Jakob Yngvason for a discussion in 2014, to our colleagues Peter Hammond, Marcus Miller and Herakles Polemarchakis in the Economics department at Warwick for their comments, and to Shafi Sardar for testing some of our ideas, notably on the Carnot cycle, by simulations in the academic year 2022/3.  We gratefully acknowledge the support of the Economic and Social Research Council (ESRC), via the Rebuilding Macroeconomics Network (Grant Ref:~ES/R00787X/1, subproject:~Why are economies stable?); and through the ESRC Network for Integrated Behavioural Science (grant number ES/P008976/1).  
We thank Angus Armstrong for putting a first version of this paper on the Rebuilding Macroeconomics website for discussion.  We are grateful to the numerous readers of that version for their comments, notably Jean-Philippe Bouchaud, Mike Cates, Russell Golman, Alex Imas, Andreas Joseph, Kunihiko Kaneko, Vassili Kolokoltsov, Michael Kumhof, Wayne Saslow, and Robert Sugden. \rev{Finally, we thank Rebekah Yore for professionalising the shading of Figure~\ref{fig:Plot3} for us.}

\begin{appendix}

\section{Checking {that the axioms of thermal macroeconomics apply to Cobb-Douglas} economies}
\label{app:axioms}

We treat the axioms in the order of appearance in our text, rather than following the order of the numbering given by \cite{LY}.
\begin{itemize}
\item[A0] Convergence to equilibrium:~We refer to \cite{DGS} for a strategy. Although they treat only the case of one good (money), 
\rev{equal encounter rates for all pairs,} and $\eta=1$, it is clear that their approach generalises.  But also it can be tidied up.  One of us has given a general streamlined proof for arbitrary \rev{fully} connected encounter graph and arbitrary exponents 
\rev{\cite{M25}}.
\item[A2]  $\preceq$ is transitive:~trivial.
\item[A1] $\sim$ is reflexive:~trivial.
\item[A8] $\forall X\  \exists Y$ such that $X \prec Y$:~This requires us to be clear about what the trader is allowed to do.  We have shown in section~\ref{sec:toy} that the trader can donate money to any system with no other changes, so for $M>0$, $X \preceq X+M$, but it remains to show that there is no action by the trader that can remove money without any other change. Offering to buy or sell goods at a specified price can reduce money but only by increasing goods.  Separating a system into parts does not change the total money; nor does putting parts in contact.  The trader's external system could probably be used to decrease money without other change but unless that system violates the axioms then if it is to return arbitrarily close to its initial state then the decrease of money in the CD system must be zero.  So we deduce that $X \prec X+M$.
\item[A3] $X_A \preceq X'_A, Y_B \preceq Y'_B$ implies $(X_A,Y_B) \preceq (X'_A,Y'_B)$:~trivial.
\item[A11] $(X_A,Y_B) \preceq \theta(X_A,Y_B)$:~trivial.
\item[A12] $\forall X_A,Y_B\ \exists X'_A,Y'_B$ such that $\theta(X_A,Y_B) \sim (X'_A,Y'_B)$:~trivial.
\item[A13'] $\equiv$ is reflexive:~trivial.
\item[A13] $\equiv$ is transitive:~This requires an analysis of the effect of financial contact between two CD economies $A$ and $B$. We allow them to have different exponents $\eta$ for money.  If the initial amounts of money are $M_A,M_B$, then on financial contact (where the agents continue to use their utility functions) the process goes to a Dirichlet distribution (as far as money is concerned) with exponents $\eta_A$ for agents in $A$, $\eta_B$ for agents in $B$, and total money $M_A+M_B$.  The marginal of this for the money in $A$ is distributed as Beta$(N_A\eta_A,N_B\eta_B)(M_A+M_B)$, which is highly peaked around its mean of $\frac{N_A\eta_A}{N_A\eta_A+N_B\eta_B}$, and the rest is in $B$.  Thus if they are disconnected, they have virtually the same value of $T 
= \frac{M}{N\eta}$ (this will be recognised as the CD temperature, but in checking the axioms we are not allowed to use that interpretation; but we are just about to derive it).  We deduce that no nett money flows between two CD economies on financial contact iff they have the same value of $T$.
Having the same value of something is transitive.
\item[A15] $\forall X_A,Y_B\ \exists M \ge 0$ such that $X_A \equiv Y_B+M$ or $Y_B \equiv X_A + M$:~This follows from the formula $T=\frac{M}{N\eta}$ and the characterisation of $\equiv$ as equal $T$, derived under A13 above.  If $T_A\ge T_B$ then choose $M= \frac{N_B\eta_B}{N_A \eta_A} M_A - M_B\ge 0$ to make $X_A \equiv Y_B+M$.  Reverse the roles of $A$ and $B$ for the opposite case.
\item[A14] $\forall X\ \exists X_0, X_1$ such that $X_0 \equiv X_1$ and $X_0 \prec X \prec X_1$:~again we can construct $X_0, X_1$ explicitly. Choose any $G'$ strictly between $0$ and $G(X)$.  Let $X_0 = X-G', X_1 = X+G'$, then they are in financial equilibrium because they have the same money (the $N\eta$ are the same for both) and they have $X_0 \prec X \prec X_1$ because $G^\alpha$ is strictly increasing in $G$ and we showed accessibility in section~\ref{sec:toy} for increasing $M^\eta G^\alpha$. To show non-accessibility in the other direction, we argue as under A8 that there is no action by the trader that can decrease $G^\alpha M^\eta$. The argument in section~\ref{sec:toy} showed that $G^\alpha M^\eta$ can not be decreased by trading with the trader at any price.  It can not be decreased by the trader donating money, because $\eta>0$.  Separating a system into parts does not change the total amount of $N\log(G/N)^\alpha (M/N)^\eta$, which we call ``entropy'' for short, although we can't yet assume its properties.  Putting parts into any form of contact does not decrease this; to see this requires another calculation.  Let's do it first for financial contact.  Given initial amounts $M^0_A,M^0_B$ of money adding up to $M^0$ in parts $A$ and $B$ of sizes $N_A,N_B$, financial contact produces a new equilibrium with money $M_A$ distributed as Beta$(C_A,C_B)M^0$, where $C_A=N_A\eta, C_B=N_B\eta$ are the money capacities of the two parts.
This has mean $\frac{C_A M^0}{C_A+C_B}$.  Thus the contribution of money to the entropy of the financial join is $N\eta\log (M^0/N)$, which is at least $N_A\eta\log(M^0_A/N_A) + N_B\eta\log(M^0_B/N_B)$ by concavity of $\log$.  A similar calculation shows non-decrease of the entropy on allowing just goods to flow or allowing both to flow.  To treat the case of contact with specified ``exchange rate'', i.e.~$a\, dM + b\, dG=0$ for each part for constants $a,b \ne 0$ (not necessarily both positive), is slightly more complicated.  The resulting distribution, reduced to marginals for the amounts of money and goods in each, is the product of Beta$(C_A,C_B)M^0$ for money and Beta$(D_A,D_B)G^0$ for goods (with ``goods capacities $D_A=N_A\alpha,D_B=N_B\alpha$), conditioned on $aM_A+bG_A = K^0 = aM^0_A+bG^0_B$.  We did not obtain an explicit formula for the mean $M_A$ of this, but for large $N_A,N_B$, the distribution is tightly peaked around a point near the solution of \begin{equation}
b(\beta_A+\beta_B) = a(\nu_A+\nu_B), 
\label{eq:tgt}
\end{equation}
where $\beta_A = N_A\eta/M_A, \nu_A = N_A\alpha/G_A$ etc.  Again by concavity of $\log$, this point has higher entropy than the initial sum.  Finally, we argue again that although the trader could decrease the entropy of the CD system by suitable use of an external system, if the external system satisfies the axioms and the CD change can be achieved by arbitrarily small change in the external system then the CD change must be zero.
\item[A4] A system $A$ can be scaled by any factor $\lambda>0$, and $X \preceq Y$ for $A$, $\lambda>0$ implies $\lambda X \preceq \lambda Y$ for $\lambda A$:~Here we have the problem that we can only scale a CD economy by factors that leave the number of agents an integer.  Furthermore, we have to specify what the encounter graph is for the scaled version, for which there is a lot of freedom. On the other hand, the equilibrium for a CD economy does not depend on the encounter graph as long as it is connected (see (\ref{eq:CDdensity}) in section~\ref{sec:toy}).
Furthermore, all agents are equivalent (at least in the simplest CD economies considered here) so scaling the set of agents by any factor that creates an integer number of agents is unambiguously achievable.  From from the correspondence between $G^\alpha M^\eta$ and $\preceq$ in a CD economy, obtained under A14, we see that $X \preceq Y$ does imply $\lambda X \preceq \lambda Y$.
\item[A5] $\forall \lambda \in (0,1)$, $X \sim (\lambda X, (1-\lambda X)$:~again, even if the number of agents is divided into integers, there is a question of how the encounter graph is cut.  But assuming these are settled,  the direction $(\lambda X, (1-\lambda)X) \preceq X$ is straightforward.  The direction $X\preceq (\lambda X, (1-\lambda)X)$ is not obvious, because cutting $X$ will in general produce an order $1/\sqrt{N}$ imbalance between the amounts of money and goods in the two parts.  At the macro-level, however, this is irrelevant.
\item[A6] $(X,\eps_k Z_0) \preceq (Y,\eps_k Z_1)$ for a sequence $\eps_k \to 0$ implies $X \preceq Y$:~This is satisfied by the interpretation of $\preceq$ that allows the trader to end up with arbitrarily small change in the external system.
\item[A7] $\forall t \in (0,1)$ and states $X,Y$, $(tX,(1-t)Y) \preceq tX+(1-t)Y$:~trivial (if scaling is well defined).
\item[A9] The forward sector $A_X$ has unique support plane and it varies Lipschitz continuously with $X$:~From the discussion under A14, we have the explicit formula $G^\alpha M^\eta \ge G_0^\alpha M_0^\eta$ for the forward sector of $(G_0,M_0)$, so it has unique supporting plane $\eta(M-M_0)/M_0 + \alpha (G-G_0)/G_0=0$,  which varies Lipschitzly with $(G_0,M_0)$.
\item[A10] $\partial A_X$ is connected:~It is the curve $G^\alpha M^\eta = G_0^\alpha M_0^\eta$.
\end{itemize}

\section{Derivation of entropy for Cobb-Douglas economies}
\label{app:CDentropy}

Here we derive the formula (\ref{eq:SCD}) for the entropy, \begin{equation}
S = N \log \bar{g}^\alpha \bar{m}^\eta,
\label{eq:CDentropy}
\end{equation}
of the toy economy of Section~\ref{sec:toy}, via the construction in the proof of Theorem 1 in [LY], where $N$ is the number of agents, and $\bar{g} = G/N$ and $\bar{m}=M/N$ are the amounts of goods and money per agent. We also show that it is calibrated {between CD economies.}

We have to find which states $(G,M)$ are reversibly accessible from a subdivision of the agents in the ratio $\lambda_0:\lambda_1$ (with $\lambda_i > 0$ and $\lambda_0+\lambda_1 = 1$) into two parts with given initial states of the form $\lambda_i(G_i,M_i), i=0,1$ with $(G_0,M_0) \prec (G_1,M_1)$. 
Then $\lambda_1$ (up to an affine transformation to allow for the freedom of choice of initial states) will be the entropy of state $(G,M)$.

First, as shown in section~\ref{sec:toy}, the trader can move each subsystem reversibly to any state $(\tilde{G},\tilde{M})$ with the same value of $\tilde{G}^\alpha \tilde{M}^\eta$ as initially, by offering to buy and sell goods at a price $\mu$ just above or below $$\mu_c = \frac{\alpha}{\eta} \frac{\tilde{M}}{\tilde{G}}.$$
The result is nett purchase of goods by the subsystem if $\mu < \mu_c$, sale if $\mu > \mu_c$.
In the case of equality, we have reversible accessibility of the states infinitesimally close along the price line $\Delta \tilde{M} + \mu_c \Delta \tilde{G} = 0$.
Integrating $d\tilde{M} + \mu_c\, d\tilde{G} =0$ produces $\tilde{G}^\alpha \tilde{M}^\eta=$ cst. 

In the above process, the trader has to buy or sell goods, but we ask the trader to buy exactly the same quantity of goods from one subsystem as it sells to the other.  That way it may gain or lose money but will not change its quantity of goods. 

We choose the movement by the trader of the two subsystems along their curves $\tilde{G}^\alpha \tilde{M}^\eta=$ cst to end with the money per agent $\bar{m}_i = \tilde{M}_i/N_i$ in each subsystem $i=0,1$ becoming the same.  This is feasible because the prices are positive and the changes in amounts of goods in the two subsystems have opposite signs, so the changes in amounts of money in the two subsystems have opposite signs and vary continuously between extremes with opposite differences in money per agent.  Thus, at the end of this phase, we have \begin{equation}
\bar{g}_i^\alpha \bar{m}^\eta = (G/N)^\alpha (M_i/N)^\eta,
\label{eq:A1}
\end{equation}
where $\bar{g}_i = \tilde{G}_i/N_i$ and $\bar{m}$ is the common mean money per agent.

Having achieved equal money per agent, the two subsystems are now in financial equilibrium, because the  money is distributed between the two subsystems in proportion to their number of agents  (see {the discussion of} A13 in the previous Appendix, noting that the two subsystems have the same $\eta$).
We now put the two subsystems into financial contact as well as contact with the trader.  The trader buys goods from one at its market price and sells them to the other at its market price, in the direction to make the goods per agent become the same in each, and in addition agents transfer money according to their utility for money, which keeps $\bar{m}_0 = \bar{m}_1$ at all times.   This results in
$N d\bar{m}=-\mu_0 d\tilde{G}_0 - \mu_1 d\tilde{G}_1$ and $\tilde{G}_0+\tilde{G}_1 = G$ constant.  Then, using $\mu_i = \frac{\alpha}{\eta}\frac{\tilde{M}_i}{\tilde{G}_i}$, we obtain
$$\frac{d\bar{m}}{\bar{m}} = -\frac{\alpha}{\eta}\left(\lambda_0 \frac{d\tilde{G}_0}{\tilde{G}_0} + \lambda_1 \frac{d\tilde{G}_1}{\tilde{G}_1}\right).$$  As a consequence,
$$\eta \log \bar{m} + \alpha (\lambda_0 \log \bar{g}_0 + \lambda_1 \log \bar{g}_1)$$ is conserved.
We can write $\log \bar{m} = \lambda_0 \log \bar{m} + \lambda_1 \log \bar{m}$.
Thus the conserved quantity in this second phase can be written as $$\lambda_0 \log \bar{g}_0^\alpha \bar{m}^\eta + \lambda_1 \log \bar{g}_1^\alpha \bar{m}^\eta.$$
Using (\ref{eq:A1}), it starts at $\lambda_0 k_0 + \lambda_1 k_1$, where $k_i = \log (G/N)^\alpha (M_i/N)^\eta$.  {Note that by the assumption that} $(G_0,M_0)\prec (G_1,M_1)$, then $k_1>k_0$.

The process ends when both subsystems have the same goods per agent. They already have the same money per agent and so can then be merged reversibly.  Using $\lambda_0+\lambda_1=1$, it follows that for the final state
\begin{equation}\log \bar{g}^\alpha \bar{m}^\eta = \lambda_0 k_0 + \lambda_1 k_1.
\label{eq:result}
\end{equation}
We see that 
$$\lambda_1=\frac{1}{k_1-k_0}(\log \bar{g}^\alpha \bar{m}^\eta -k_0).$$ 
Hence the entropy function is $\log \bar{g}^\alpha \bar{m}^\eta$ up to an \rev{orientation-preserving} affine transformation.  To respect scaling symmetry, we insert a factor of $N$.
Hence, up to addition of a possible constant per agent (that is irrelevant for fixed numbers of agents), we obtain the formula (\ref{eq:CDentropy}).

Note that it is not necessary that the trader apply this particular two-stage process, nor any process at all.  Contact between the two parts under an ``exchange rate'', as under A14, will move the system arbitrarily close to reversibly in the space where $M_A+M_B$ and $G_A+G_B$ are conserved if the exchange rate satisfies (\ref{eq:tgt}). But this equation is precisely the equation of the tangent to the curve of constant  $\sum_i \lambda_i \log \bar{g}_i^\alpha \bar{m}_i^\eta$.  So $\lambda_1$ is determined by (\ref{eq:result}) again.

Finally we have to check whether the formula needs calibration for different CD systems.  On putting two CD economies in financial contact, we already obtained under A13 that there is no nett money flow iff they have the same value of $\frac{M}{N\eta}$, equivalently of $\frac{N\eta}{M}$, but using (\ref{eq:CDentropy}), this is the value of $\frac{\partial S}{\partial M}$.  So it is the condition for no direction of entropy increase with respect to movement of money, and we deduce that their entropies are already calibrated.

\end{appendix}

\end{document}